\documentclass[final]{siamltex}
\usepackage{graphicx,psfrag}
\usepackage{hyperref}
\usepackage{enumerate}
\usepackage{amsmath}
\usepackage{amssymb}

\graphicspath{{Pictures/}}

\begin{document}

\title{Flooding in urban drainage systems: Coupling hyperbolic conservation laws  for sewer systems and surface flow}

\author{R. Borsche \footnotemark[1] \and A. Klar \footnotemark[1]
  \footnotemark[2] }
  
\footnotetext[1]{Technische Universit\"at Kaiserslautern, Department of Mathematics, Erwin-Schr\"odinger-Stra{\ss}e, 67663 Kaiserslautern, Germany 
  (\{borsche,klar\}@mathematik.uni-kl.de)}
\footnotetext[2]{Fraunhofer ITWM, Fraunhoferplatz 1, 67663 Kaiserslautern, Germany} 

\maketitle

\begin{abstract}
  \noindent
  In this paper we propose a  model for a sewer network coupled to  surface flow and investigate it numerically. In particular, we present a new model for the manholes in storm sewer systems.
  It is derived using the balance of the total energy in the complete network. 
  The resulting system of equations contains, aside from hyperbolic conservation laws for the sewer network  and algebraic relations for the coupling conditions, a  system of ODEs governing the flow in the manholes. 
  The manholes provide natural points for the interaction of the sewer system and the run off on the urban surface modelled by shallow water equations.
  Finally, a numerical method for the coupled system is presented. In several numerical tests we study the influence of the manhole model on the sewer system and the coupling with 2D surface flow.   

\end{abstract}

\section{Introduction}
Many mathematical models have been developed to study the flow in sewer systems.
Most of todays models choose separate equations for the dynamics in the horizontal tubes and for the flow at a junction or manhole \cite{Extran,Hamoka,Mouse,NISN}.
For the flow in a single nearly horizontal pipe the Saint Venant equations are well established.
These can be extended to handle also pressurized flows by the concept of the Preissmann Slot \cite{TT001704700,CungeHolly,Leon}.
Other approaches are e.g. two phase models in \cite{BourdariasGerbi,LeonGhidaouiSchmidtGarcia,Mouse} or models additionally tracking the air pressure inside the tube \cite{BourdariasErsoyGerbi,TrappedAir}.

For the description of the flow at a junction or manhole a variety of different models has been developed. 
Models for the flow at a junction without vertical extension are compared in detail in \cite{AbdallahThesis,KesserwaniGhostineVazquezMoseAbdallahGhenaim}. 
Mathematical investigations of such coupling conditions can be found in \cite{ColomboHertySachers,LeugeringSchmidt}.
Models of junctions including the flow in the manholes are considered in e.g. \cite{Extran,LeonLiuGhidaouiSchmidtGarcia,Hamoka,Mouse,NISN}.
In \cite{GuoSong,leon:705} models including a separate ODE for the flow in the manhole are derived using conservation of mass and momentum. 

One important aspect of the manholes is the linking of the sewer network to the urban surface.
The runoff from the surface into the sewer is mainly ducted by inlets, which can be associated to their nearest manhole. 
In contrast an eventual surcharge of the sewer directly occurs at the manholes. 
The flow on the surface can be modeled by the shallow water equations \cite{Olympic,AbderrezzakPaquierMignot} or one of their simplifications \cite{EttrichSchmittThomas,EttrichRotheSteinerThomas}. 
A coupling between sewer system and the surface flow can be realized by suitable source terms in the manhole model and in the 2D surface description respectively, e.g. \cite{EttrichRotheSteinerThomas,NOVATECH}.

The purpose of the present paper
is to develop and investigate numerically  a new consistent coupled model for sewer system and surface flow
based on several  ingredients.
First, flow in the network is modeled by a conservation law, in this case the  one-dimensional St. Venant equations. Second, the  coupling conditions at the junctions are consistently modeled using the solution of Riemann problems. 
Third, the surface flow is again modelled by a conservation law, in the present case the two-dimensional shallow water equations.
Fourth, the coupling between surface flow and sewer network is obtained via a 
consistent model for the  manhole based on ordinary differential equations and derived from energy considerations for the complete network.
Finally, the coupled model is investigated numerically using  well-balanced Riemann solvers for all components.

This paper contains 6 sections.
In section \ref{Models:sec:Sewer} a mathematical model for the flow in sewer system is presented. 
Among many standard components the focus is given to a new model describing the states inside the manholes. 
This model can be connected to the shallow water equations described in section
\ref{Models:sec:surface}, establishing the interaction of surface run off and sewer network.
In section \ref{Analysis:sec:ConservationOfEnergy} we analyze the balance of the total energy in this coupled system.
Equipped with the new manhole model the total energy in the sewer network can be conserved and due to the interplay with the surface only minor variations occur.
In the last section we present a numerical method for the coupled system and several numerical test cases, which elucidate the behavior of the manhole model and the interaction of surface flow with the sewer system.

\section{The sewer network}\label{Models:sec:Sewer}
Many mathematical models for the flow in sewer systems have been developed in the past \cite{Extran,Hamoka,Leon,Mouse,NISN}.
Most of them describe the dynamics inside each single component separately and link these models by suitable coupling conditions.
We will follow this approach, as it allows to model easily networks of any dimension.

For the description of the network the following notations will be used.
$\mathcal{E}_{edges}\subset \mathbb{N}$ and $\mathcal{N}_{nodes}\subset \mathbb{N}$ are the sets of all indices of the edges and junctions respectively.
The set of all indices of the tubes connected to the junction $j\in\mathcal{N}_{nodes}$ is $\mathcal{E}^j_{edges}\subset \mathcal{E}_{edges}$ and
the total number of tubes connected to the node $j$ is denoted by $n^j_{edges}=\left|\mathcal{E}^j_{edges}\right|$. 
The orientation of an edge $i$ respective to a node $j$ is labeled by $\delta_i^j$, where $\delta_i^j=1$ if the conduit begins a the node and $\delta_i^j=-1$ if it ends there.
The manholes inherit the indicies of the corresponding junctions.

\subsection{The Saint Venant equations}
The free surface flow in a single conduct can be described by the Saint Venant equations \cite{KesserwaniGhostineVazquezMoseAbdallahGhenaim,CungeHolly,Leon,BourdariasGerbi,Mouse,NISN}.
These are derived for the free surface flow inside a nearly horizontal tube
and can be extended to the case of pressurized flow by a slight modification of the corresponding pressure law, the so called Preissmann Slot \cite{CungeHolly}.
For a tube with index $i$ the Saint Venant equations read
\begin{align}\label{Models:eq:StVE}
\begin{array}{rl}
\partial_tA_i+\partial_xQ_i&=0\\
\partial_tQ_i+\partial_x(\frac{Q_i^2}{A_i}+p_i(x,A_i))&=-gA_i\partial_x z_i - \left(S_f\right)_i\,.
\end{array}
\end{align}
Here $A_i(t,x)$ denotes the wetted cross sectional area and $Q_i(t,x)$ is the flow of water into $x$-direction at a given time $t\in \mathbb{R}^+$ and location $x\in \mathbb{R}$.
The averaged hydrostatic pressure law $p_i(\cdot,\cdot):\mathbb{R}\times\mathbb{R}^+\rightarrow \mathbb{R}^+$,
\begin{align} \label{Models:eq:pressurelaw}
p_i(x,A_i)=g\int^{A_i}_0 \left(h_i(x,A_i)-h_i(x,a)\right) da\,
\end{align}
depends on the geometry of the tube.
$g$ is the gravitational acceleration, $h_i(\cdot,\cdot):\mathbb{R}\times\mathbb{R}^+\rightarrow\mathbb{R}^+$ is the relative height of water corresponding to the wetted area
and $z_i(t,x)$ is the bottom elevation.
For the friction term $S_f$ the formula of Manning can be used \cite{CungeHolly}
\begin{align}\label{Model:eq:frictionStVE}
\left(S_f\right)_i=g\left(n_{f}\right)_i^2\frac{Q_i\left|Q_i\right|}{A_ir_{hy}^{4/3}}\,.
\end{align}
$n_f$ is the Manning friction coefficient and $r_{hy}=\frac{A_i}{U_i}$ is the hydraulic radius, where $U_i$ is the wetted perimeter corresponding to $A_i$.

Altogether the equations \eqref{Models:eq:StVE} state the conservation of mass and the balance of linear momentum.
For suitable choices of the initial conditions, sufficiently bounded source terms and appropriate boundary conditions on finite domains, the system \eqref{Models:eq:StVE} is well posed, \cite{GuerraMarcelliniSchleper}.

\subsection{Coupling conditions without manhole}
In order to connect several tubes to a network suitable coupling conditions have to be imposed.
This is usually managed by a set of algebraic equations at the nodes, e.g. \cite{GuoSong,KesserwaniGhostineVazquezMoseAbdallahGhenaim,LeonLiuGhidaouiSchmidtGarcia}.
A detailed comparison of the most common coupling conditions for sewer networks can be found in \cite{KesserwaniGhostineVazquezMoseAbdallahGhenaim}.
In the following we consider the coupling of $n^j_{edges}$ tubes at the junction $j$, similar to \cite{LeugeringSchmidt}.

In order to assure a constant number of coupling conditions at the junction we assume that the flow is subsonic all times, i.e. $\left|\frac{Q_i}{A_i}\right|<\sqrt{\partial_A p(A_i)}\quad \forall \ i\in \mathcal{E}^j_{edges}$.
The case of purely supersonic flow is discussed in \cite{LeonLiuGhidaouiSchmidtGarcia,NISN}.

For a node $j$ the first equation of the coupling conditions states the conservation of mass by balancing the flows at the junction
\begin{align}\label{Models:eq:junction_mass}
\sum_{ i\in \mathcal{E}^j_{edges}} \delta^j_iQ_i(t,\chi_i^j)=0\,.
\end{align}
The orientation of the tubes is adjusted by the functions $\chi_i^j$  and $\delta^j_i$, where $\chi_i^j$ denotes the end or starting point of the conduit.

For a well defined set of coupling conditions \cite{ColomboHertySachers} further $n^j_{edges}-1$ equations are needed.
Here, we impose the equality of the hydraulic heads $\bar{h}$
\begin{align}\label{Models:eq:equalhydraulicheads}
\bar{h}_i(\chi_i^j,A_i(t,\chi_i^j)=\bar{h}_k(\chi_i^j,A_i(t,\chi_i^j)) \qquad i\neq k \quad i,k\in \mathcal{E}^j_{edges}\ ,
\end{align}
where the hydraulic heads or energy levels are defined as
\begin{align}\label{Models:eq:hh}
\bar{h}_i(x,A_i,Q_i)=\frac{1}{2g}\frac{Q_i^2}{A_i^2}+h_i(x,A_i) 
+z_i
\qquad i \in \mathcal{E}_{edges}\,.
\end{align}
This set of equations assures the conservation of the total energy at the junction, as shown in \cite{BorscheDiss}.
Since at a junction turbulences can occur, the total energy is in general not conserved. 
Therefore we add terms $\Delta L^j_{i}$, which can model the loss of the total energy, i.e. the equations \eqref{Models:eq:equalhydraulicheads} are modified to
\begin{align}\label{Models:eq:equalhydraulicheadslosses}
\bar{h}_i(\chi_i^j,A_i(t,\chi_i^j)+\Delta L^j_{i}
=\bar{h}_k(\chi_i^j,A_i(t,\chi_i^j))+\Delta L^j_{k} \qquad i\neq k \quad i,k\in \mathcal{E}^j_{edges}
\,.
\end{align}
In the following we will denote this common hydraulic head by $\bar{h}^j_{node}$, as it is independent of the choice of the conduit. 
The terms $\Delta L^j_{i}$ may depend on the current states in the conduits at the junction.
A possible choice $\Delta L^j_{i}=l_i\frac{Q_i|Q_i|}{2g A^2_i}$ is proposed in \cite{LeonLiuGhidaouiSchmidtGarcia}, where $l_i$ is a local head loss coefficient.
These terms can also be used to incorporate the geometrical structure of the manhole into the equations.
Thus the final set of coupling conditions is \eqref{Models:eq:junction_mass} together with \eqref{Models:eq:equalhydraulicheadslosses}.

\subsection{A Manhole}
A manhole is a vertical tube, which is usually located at the junction between horizontal pipes. 
In the following we will describe the state within a manhole $j$ by two variables, the water level $h^j_M$ and the inflow into the manhole from the bottom $Q^j_M$.
As we assume the inflow to enter at the bottom this induces a movement of the complete column of water with the speed $\frac{Q^j_M}{A^j_M}$, where $A_M^j$ is the horizontal cross sectional area of the manhole. 
\begin{figure}[h]
\centering
\includegraphics[width=5cm]{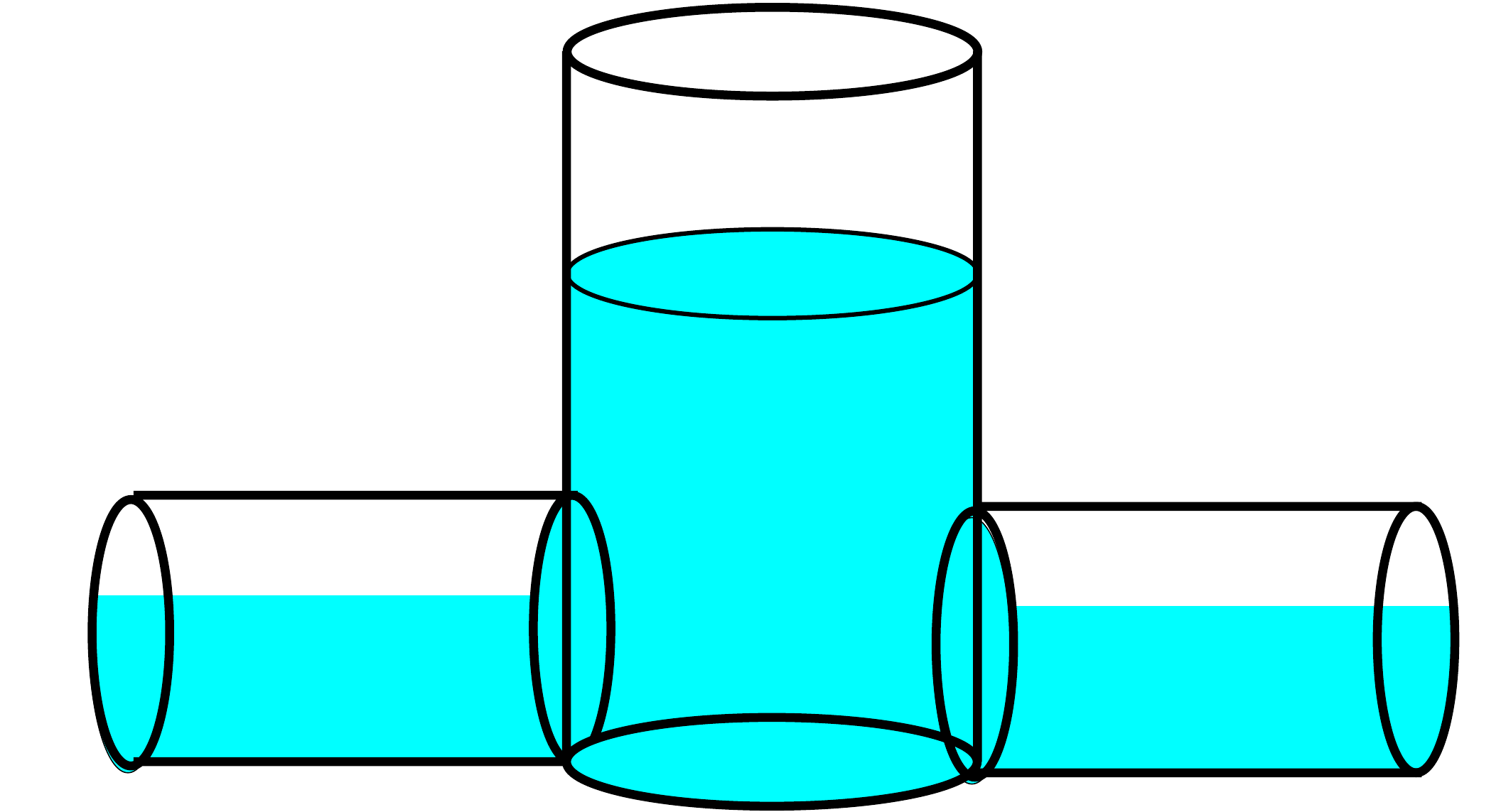}
\caption{A manhole and two horizontal conduits}
\end{figure}
Thus the water level inside the manhole changes according to the following ODE
\begin{align}\label{Models:eq:ManholeMass}
\partial_t h_M^j=\frac{Q_M^j+Q^j_{ext}}{A_M^j}\,.
\end{align}
$Q_{ext}^j(t)$ represents any external inflows from the top, e.g. it will include the water coming from the surface.

In order to maintain the global conservation of mass in the entire network, the mass entering the manhole from below has to be subtracted from the connected tubes.
Thus,  we modify the mass balance at the junction \eqref{Models:eq:junction_mass} to 
\begin{align}\label{Model:eq:junction_mass_manhole}
\sum_{ i\in \mathcal{E}^j_{edges}} \delta^j_iQ_i(t,\chi_i^j)+Q_M^j(t)=0\,.
\end{align}
The remaining equations of the coupling conditions \eqref{Models:eq:equalhydraulicheadslosses} are not influenced directly by the presence of the manhole.

\subsubsection{The full manhole model}
The following model can be derived by regarding the balance of the total energy in the complete network.
The corresponding calculations are shown in section \ref{Analysis:sec:ConservationOfEnergy}.
Finally we obtain the following ODE describing the evolution of $Q_M$
\begin{align}\label{Models:eq:ManholeQM}
\partial_t Q^j_M=\frac{gA^j_M}{h_M^j}\left(\bar{h}_{node}^j-\bar{h}^j_M\right)-\Delta L_M^j\,
\end{align}
with $\bar{h}^j_M$ defined similarly to \eqref{Models:eq:hh}.
It is easy to see, that the change in $Q^j_M$ tends to balance differences of the hydraulic head of the manhole with the hydraulic head of the junction.
This change is inversely proportional to the volume stored within the drop shaft, since the inflow at the bottom always moves the complete mass in the manhole.
The term $\Delta L_M^j$ summarizes different losses of energy acting on the flow.
One possible choice is a version of the Darcy Weisbach formula
\begin{align}\label{Models:eq:LossFullManhole}
\Delta L_M^j= \frac{\lambda_{DW}^j}{8} U_M^j Q_M^j\frac{|Q_M^j|}{\left(A_M^j\right)^2}\,,
\end{align}
modeling the friction along the vertical walls.
$U_M^j$ denotes the perimeter of the manhole and $\lambda_{DW}^j$ is the Darcy Weisbach friction factor.
Further losses might be added to include the change of direction of the flow.
The total energy in the system decreases, if $\Delta L_M^j$ has the same sign as the flow $Q^j_M$.

Thus we can compose \eqref{Models:eq:ManholeMass} and \eqref{Models:eq:ManholeQM} to a single system of ODEs describing the states in the manhole
\begin{align}\label{Models:eq:FullManhole}
\partial_t
\left(
\begin{array}{c}
 h_M^j\\ Q^j_M
 \end{array}
 \right)
 =
 \left(\begin{array}{c}
 \frac{Q_M^j+Q^j_{ext}}{A_M^j}\\
\frac{gA^j_M}{h_M^j}\left(\bar{h}_{node}^j-\bar{h}^j_M\right)-\Delta L_M^j
\end{array}
\right)\,.
\end{align}
In the following, this will be referred to as the full manhole model, together with the coupling conditions \eqref{Model:eq:junction_mass_manhole} and \eqref{Models:eq:equalhydraulicheadslosses}.

{\it Remark:}
It is important to note, that $\bar{h}^j_M$ should not be replaced by simply $h^j_M$, as then the manhole model would be similar to the undamped harmonic oscillator, which leads to oscillating states in the manhole.

\subsubsection{A simplification of the manhole model}
The above full manhole model \eqref{Models:eq:FullManhole} can be simplified by assuming the change of the flow to be small, such that the term $\partial_t Q^j_M$ can be neglected.
Thus  in \eqref{Models:eq:ManholeQM} $Q^j_M$ appears only in the term $\bar{h}^j_M$.
By solving for $Q^j_M$ we obtain the following expression 
\begin{align} \label{Model:eq:QM_simple}
Q_M^j(t)=sign\left(\bar{h}_{node}^j-h_M^j-\frac{h^j_M}{gA^j_M}\Delta L_M^j\right)A_M^j\sqrt{\left|2g\left(\bar{h}_{node}^j-h_M^j\right)-\frac{2h^j_M}{A^j_M}\Delta L_M^j\right|}\,.
\end{align} 
This equation can also be directly motivated by applying Bernoulli's equation to the involved flows, which is equivalent to focus only on stationary flows.

Inserting \eqref{Model:eq:QM_simple} into \eqref{Models:eq:ManholeMass} results into an ODE with a square root on the right hand side. 
As this is not useful for analytical as well as numerical reasons, we replace $Q^j_M$ in \eqref{Models:eq:ManholeMass} by the expression of coupling conditions \eqref{Models:eq:junction_mass} and obtain as ODE for the manhole
\begin{align} \label{Model:eq:Manhole_simple_ODE}
\partial_t h_M^j(t)=\frac{1}{A_M^j}\left(-\sum_{ i\in \mathcal{E}^j_{edges}} \delta^j_iQ_i(t,\chi_i^j)+Q^j_{ext}(t)\right)\,.
\end{align} 
As missing coupling condition we replace \eqref{Models:eq:ManholeMass} by 
\begin{align} \label{Model:eq:Manhole_simple}
h_M(t)=\bar{h}^j_{node}(t)-\frac{h^j_M(t)}{gA^j_M}\Delta L_M^j-\frac{1}{2g\ \left(A_M^j\right)^2}\left(\sum_{ i\in \mathcal{E}^j_{edges}} \delta^j_iQ_i(t,\chi_i^j)\right)\left| \sum_{ i\in \mathcal{E}^j_{edges}} \delta^j_iQ_i(t,\chi_i^j)\right|\,.
\end{align} 
Thus the simplified manhole model consists of the ODE \eqref{Model:eq:Manhole_simple_ODE} and the coupling conditions \eqref{Models:eq:equalhydraulicheadslosses}, \eqref{Model:eq:Manhole_simple}.

Both models, the full model and the simplified one, fit into the framework of \cite{0951-7715-23-11-002,MixedODEPDE,BorscheDiss}, i.e with suitable initial conditions they form a well posed system of equations.

\section{The Surface}\label{Models:sec:surface}

The flow on the urban surface can be modeled by the $2D$ shallow water equations \cite{EttrichSchmittThomas,AbderrezzakPaquierMignot,HsuChenChang}.
The water level $h$ above the surface $z$ and the flow in $x-$ resp. $y-$direction $hu$ resp. $hv$ are described by
\begin{align}\label{Models:eq:ShallowWater}
\partial_t h +\partial_x\left(hu\right)+\partial_y\left(hv\right)&=S^1_{sewer}
\nonumber\\
\partial_t\left(hu\right)+\partial_x\left(hu^2+\frac{g}{2}h^2\right)+\partial_y\left(hu v\right)&=-gh\partial_x z-S^2_f+S^2_{sewer}\\
\partial_t\left(hv\right)+\partial_x\left(hu v\right)+\partial_y\left(hv^2+\frac{g}{2}h^2\right)&=-gh\partial_y z-S_f^3+S^3_{sewer}
\nonumber\,.
\end{align}
Expressions for the friction terms $S^2_f,S^3_f$ can be found, for example,  in  \cite{CungeHolly}.

For the interaction of a sewer system with the surface flow different models have been proposed \cite{HsuChenChang,EttrichSchmittThomas,LeandroChen}.
The following construction of  $S^i_{sewer}, \quad i=1,\dots,3$ is closely related to the ideas in \cite{LeandroChen}.

The flow from the surface into the sewer is usually led through inlets.
These inlets can be directly assigned to the nearest manhole.
In the following we will assume, that the water level in the inlets is equal to the one in the manhole.
To avoid any indexing we will discuss the modeling at a single manhole. 
We denote by $A_{inlet}\in \mathbb{R}^2$ the area on the surface where all the inlets of the manhole are located. 
It is not necessarily a connected set, since it can contain several inlets and the top of the manhole itself.
At each point $(x,y)\in A_{inlet}$ the inflow $Q_S$ into the sewer system mainly depends on the difference of the considered water levels 
\begin{align}\label{Models:eq:deltahinlet}
\Delta h_{inlet}(t,x,y)=\min\left(h(t,x,y),h(t,x,y)+z(x,y)-\left(h_M(t)+z_M\right)\right) 
 \qquad
 (x,y)\in A_{inlet}\,.
 \end{align}
Here we require the natural property $z(x,y)>z_M$.
In order to compute the exchange of water, we use a classical weir formula \cite{CungeHolly}
\begin{align}\label{Model:eq:SurfaceSewer}
Q_{S}(t,x,y)
=sign(\Delta h_{inlet}(t,x,y))\gamma_{inlet}(x,y)
\sqrt{2g\left|\Delta h_{inlet}(t,x,y)\right|}
 \qquad
(x,y)\in A_{inlet}\,
\end{align}
with the weir constant $\gamma_{inlet}$, which can depend on the shape and size of the corresponding inlet.

For the sources in the momentum equation we further assume, that the momentum is reduced proportional to the amount  exiting the surface.
Considering the opposite direction, the water exiting the sewer system has no significant horizontal momentum.

Altogether, this leads to the following definition of the source terms for \eqref{Models:eq:ShallowWater}
\begin{align}\label{Models:eq:S_sewer}
S_{sewer}
=
\left(
\begin{array}{c}
S^1_{sewer}\\ 
S^2_{sewer}\\
S^3_{sewer}
\end{array} 
\right)
=
\left(
\begin{array}{c}
-Q_S\\ 
-\max(0,Q_S)u\\
-\max(0,Q_S)v
\end{array} 
\right)\,.
\end{align}
The water, which is added to or subtracted from the surface, has to vanish or reappear in the manhole respectively.
Thus, the external inflow from the surface to the sewer is defined by
\begin{align}\label{Models:eq:Q_ext}
Q_{ext}(t)=\int_{A_{inlet}}Q_S(t,x,y)dx dy\,,
\end{align}
where $Q_{ext}$ is the function introduced in \eqref{Models:eq:ManholeMass}.

At this point we have to mention, that the capacity of the manhole has no upper bound.
This extension is made to ensure, that the pressure of the height of water at the surface is correctly transfered to the sewer system.
It can be interpreted as a Preissmann Slot of the size $A_M$ above the manhole.
The capacity of the resulting ghost storage at the surface is tolerable, if the water level above the ground is small compared to the depth of the manhole. 
 
 \section{Conservation of Energy}\label{Analysis:sec:ConservationOfEnergy}
In this section we study the evolution of the total energy in the coupled system of sewer and surface flow. 
As in the above construction we will begin with the components separately and merge their results for the general perspective.
Throughout the computations all the states within the system are assumed to be regular enough for the underlying differential operations.

\subsection{The total energy in a single tube}
Consider a single tube $i$ extending from  $x=0$ to $x=+\infty$. 
The total energy $\mathcal{E}_i(t)$ within this tube $\mathcal{E}_i(t)=\int^{\infty}_{0}E_i(x,t)dx$
is described by the energy density
\begin{align}\label{Analysis:eq:EnergyTube}
E_i(t,x)=\frac{1}{2}\frac{Q_i^2}{A_i}+g\int^{A_i}_0h_i(x,a) da+gA_iz_i
\,,
\end{align}
which is composed of the kinetic and potential energy densities.
The change of the energy density can be described by \cite{BourdariasGerbi}
\begin{align*}
\partial_tE_i(t,x)+\partial_xF_i(t,x)=\frac{Q_i}{A_i}\left(S_f\right)_i
\,,
\end{align*}
where $F_i(t,x)=
\frac{1}{2}\frac{Q_i^3}{A_i^2}
+gh_i(x,A_i)Q_i
+gz_iQ_i =gQ_i\bar{h}_i$
is the energy density flux.
Thus the total energy within the tube evolves according to 
\begin{align*}
\frac{d}{dt}\mathcal{E}_i(t)
&=\int^{\infty}_{0}\partial_tE_i(x,t)dx
=-F_i(t,0)-\int^{\infty}_0\frac{Q_i}{A_i}\left(S_f\right)_i dx
\,,
\end{align*}
if the energy density at $+\infty$ does not change. 
This implies that in a single tube the energy decreases and this only due to the friction, as $-\frac{Q}{A}S_f$ is always negative by \eqref{Model:eq:frictionStVE}.
Furthermore energy can be added or subtracted by an in- or outflow at the boundary at $x=0$.

\subsection{The total energy at a junction with a manhole}
The total energy $\mathcal{E}_N$ in a network of one junction connecting several tubes and a manhole is the sum of the energy in its components
\begin{align*}
\mathcal{E}_N(t)=
\mathcal{E}_M(t)+
\sum^n_{i=1} \mathcal{E}_i(t,x)dx
\,.
\end{align*}
The energy in a manhole $\mathcal{E}_M$ is composed of potential and kinetic energy 
\begin{align}\label{eq:Energy_Manhole}
\mathcal{E}_M(t)=
\frac{1}{2}\frac{Q^2_M}{A_M}h_M
+\frac{g}{2}A_Mh_M^2+gA_Mh_Mz_M
\,.
\end{align}
For the kinetic part we have to recall that the inflow $Q_M$ is located at the bottom, such that the whole amount of water in the manhole moves with the speed $\frac{Q_M}{A_M}$.

The change in time of $\mathcal{E}_N(t)$ can be computed by
\begin{align*}
\frac{d}{dt}\mathcal{E}_N(t)
&=
\frac{d}{dt}\mathcal{E}_M(t)
+\sum^n_{i=1}\left[F_i(t,0)-\int^{\infty}_0\frac{Q}{A}\left(S_f\right)_i dx\right]\\
&=
gA_M\left(h_M+z_M\right)\frac{d}{dt}h_M
+\frac{Q_M}{A_M}h_M\frac{d}{dt}Q_M
+\frac{1}{2}\frac{Q^2_M}{A_M}\frac{d}{dt}h_M
\\ & \qquad
+\sum^n_{i=1}\left[F_i(t,0)-\int^{\infty}_0\frac{Q}{A}\left(S_f\right)_i dx \right]\\
&=
\frac{Q_M}{A_M}h_M\frac{d}{dt}Q_M
+\left(Q_M+Q_{ext}\right)\bar{h}_M
+\sum^n_{i=1}\left[
Q_i\bar h_i-\Sigma_f\right]
\,.
\end{align*}
Here, we have used \eqref{Models:eq:ManholeMass} to replace $\frac{d}{dt}h_M$ and defined $\Sigma_f=\sum^{n}_{i=1}\int^{\infty}_0\frac{Q_i}{A_i}\left(S_f\right)_i\geq 0$ for the friction losses in the conduits.
By using the coupling conditions \eqref{Models:eq:junction_mass} and \eqref{Models:eq:equalhydraulicheadslosses} we obtain
\begin{align*}
\frac{d}{dt}\mathcal{E}_N(t)
&=
\frac{Q_M}{A_M}h_M\frac{d}{dt}Q_M
+g Q_M
\left(
\bar{h}_M-\bar{h}_{node}
\right)
+g Q_{ext}
\bar{h}_M
-\Sigma_{\Delta L}
-\Sigma_f
\,,
\end{align*}
where $\Sigma_{\Delta L}=\sum^n_{i=1}Q_i\Delta L_i$ collects the losses included in the coupling conditions.
If formula \eqref{Models:eq:equalhydraulicheadslosses} is utilized, the term $\Sigma_{\Delta L}\geq 0$ decreases the total energy in the system.

If the full model for the manhole \eqref{Models:eq:FullManhole} is considered, $\mathcal{E}_N$ evolves as
\begin{align}\label{Analysis:eq:SewerEnergyConservation}
\frac{d}{dt}\mathcal{E}_N(t)
&=
g Q_{ext}\bar{h}_M
-\Sigma_{\Delta L_M}
-\Sigma_{\Delta L}
-\Sigma_f
\,,
\end{align}
where $\Sigma_{\Delta L_M}=\frac{Q_M}{A_M}h_M\Delta L_M$ is positive, if $\Delta L_M$ is of the form \eqref{Models:eq:LossFullManhole}. 
This relation implies that the total energy in the system of a manhole and connected tubes is exactly reduced by the amount specified by the loss terms in the corresponding equations.
The energy can only increase due to external inflows $Q_{ext}$.

If we use the simplified model \eqref{Model:eq:QM_simple} to describe the dynamics in the manhole, the total energy changes as
\begin{align*}
\frac{d}{dt}\mathcal{E}_N(t)
=\frac{Q_M}{A_M}h_M\frac{d}{dt}Q_M
+g Q_{ext}\bar{h}_M
-\Sigma_{L_M}
-\Sigma_{\Delta L}
-\Sigma_f
\,.
\end{align*}
In general the term $\frac{d}{dt}Q_M$ has no fixed sign, i.e. the total energy does not necessarily decrease.
But if $\frac{d}{dt}Q_M$ is small enough, the gain of energy might be compensated by the sum of the considered losses.

\subsection{The total energy on the surface}
Analogous to the $1D$ case we define the total energy on a surface $\Omega$ as $\mathcal{E}_S(t)=\int_{\Omega} E_S(t,x,y) dx dy$ with the energy density 
\begin{align*}
E_S(t,x,y) &=\frac{1}{2}hu^2+\frac{1}{2}hv^2+\frac{g}{2}h^2+g hz\,,
&
&(x,y)\in \Omega
\,.
\end{align*}
We can also  find a balance law for the evolution of the energy density, namely
\begin{align*}
\partial_t E_S+\partial_x F_S+\partial_y G_S
=
-\sigma_{Sf}+\sigma_{S}
\,.
\end{align*}
The energy density flux  in $x$-direction $F_S$ and in $y$-direction $G_S$, are defined as
\begin{align*}
F_S=
hu\left(
\frac{1}{2}u^2+\frac{1}{2}v^2
+gh+gz
\right)
\quad,\qquad
G_S=
hv\left(
\frac{1}{2}u^2+\frac{1}{2}v^2
+gh+gz
\right)
\,
\end{align*}
and the friction term is
$ \sigma_{Sf}= uS^2_f +vS^3_f$ .
The exchange with the sewer system has the representation
\begin{align}\label{Analysis:eq:SigmaS}
\sigma_{S}=
\left(-\frac{1}{2}\left(u^2 
+v^2\right)
+g\left(h+z\right)
\right)
 S^1_{sewer}
+uS^2_{sewer}+vS^3_{sewer}
\,.
  \end{align}
Thus, the evolution of $\mathcal{E}_S$ is governed by
\begin{align}\label{Analysis:eq:SurfaceEnergyConservation}
\nonumber
\frac{d}{dt}
\mathcal{E}_S(t)
&=\int_{\Omega} \frac{d}{dt} E_S(t,x,y) dx dy
\\
&=
\int_{\partial\Omega} 
\left(
\begin{array}{c}
F_S
 \\
G_S
\end{array}
\right)
\vec{n}
\ ds
-\Sigma_{Sf}
+\Sigma_{S}
\,,
\end{align}
where the source terms $\Sigma_{Sf}$ and $\Sigma_{S}$ are 
\begin{align*}
\Sigma_{Sf}=
\int_{\Omega} 
\sigma_{Sf}
\ dx dy
\quad,\qquad
\Sigma_{S}=
\int_{A_{inlet}} 
\sigma_{S}
\ dx dy
\,.
\end{align*}
The integration of $\sigma_S$ is only over $A_{inlet}$ as it is zero elsewhere and $-\Sigma_{Sf}$ is negative, if standard formulas, e.g. \cite{EttrichSchmittThomas}, are applied.

If there are no fluxes across the boundary the total energy on the surface decreases according to the friction terms and energy can be exchanged with the sewer system.

\subsection{The total energy in the sewer and on the surface}
In order to estimate the evolution of the total energy in the coupled system, we combine the results of the previous sections.
Thus the total energy in the coupled system
\begin{align*}
\mathcal{E}(t)=\mathcal{E}_S(t)+\mathcal{E}_N(t)
\,
\end{align*}
evolves due to \eqref{Analysis:eq:SewerEnergyConservation} and \eqref{Analysis:eq:SurfaceEnergyConservation} as
\begin{align*}
\frac{d}{dt}\mathcal{E}(t)
&=
\frac{d}{dt}\mathcal{E}_S(t)
+\frac{d}{dt}\mathcal{E}_N(t)\\
&=
-\Sigma_{Sf}
+\Sigma_{S}
+g Q_{ext}\bar{h}_M
-\Sigma_{\Delta L_M}
-\Sigma_{\Delta L}
-\Sigma_f
\,.
\end{align*}
As the friction terms only reduce $\mathcal{E}$ we focus on the balance between sewer and surface.
These are governed by 
\begin{align}\label{Analysis:eq:EnergyTransfer}
\nonumber
\Sigma_{S}+g Q_{ext}\bar{h}_M
&=
g Q_{ext}\bar{h}_M
+
\int_{A_{inlet}}
\frac{1}{2}\left(u^2 
+v^2\right)
Q_S
-g\left(h+z\right)
Q_S
 +u  S^2_S
 +v  S^3_S
   \ dx dy \\
\nonumber
&=
\int_{A_{inlet}}
g Q_{S}\bar{h}_M
-g\left(h+z\right)
Q_S
+\frac{1}{2}\left(u^2 
+v^2\right)
Q_S
 +u  S^2_S
  +v  S^3_S
   \ dx dy
   \\
\nonumber
&=
\int_{A_{inlet}}
g\left(h_M+z_M-h-z\right)
Q_S
+\frac{1}{2}\frac{Q_M^2}{A_M^2}Q_S
\\
\nonumber
&\qquad
+u^2\left( \frac{1}{2}Q_S
-\max(0,Q_S)\right)
+v^2\left(\frac{1}{2}Q_S
-\max(0,Q_S)\right)
   \ dx dy
\\
&=
\int_{A_{inlet}}
g\left(h_M+z_M-h-z\right)
Q_S
+\frac{1}{2}\frac{Q_M^2}{A_M^2}Q_S
-\frac{1}{2}\left(u^2+v^2\right)|Q_S|
   \ dx dy
   \,.
\end{align}
Here we applied the coupling conditions \eqref{Models:eq:S_sewer} and \eqref{Models:eq:Q_ext}.

It is important to note, that in general the above expression \eqref{Analysis:eq:EnergyTransfer} is not zero.
The first part containing the water levels is always negative, due to the orientation of $Q_S$.
This is the potential energy, which is lost if water drops from the surface into the manhole or from the Preissman Slot of the manhole on the surface.

The last term in \eqref{Analysis:eq:EnergyTransfer} containing the surface velocities is also negative.
If the water enters the sewer, $Q_S>0$, it loses all its horizontal momentum, e.g. by hitting the vertical walls of the manhole.
If $Q_S<0$ then the inertia of the added water slows down the flow passing by on the surface.

The only term which might be positive is $\frac{1}{2}\frac{Q_M^2}{A_M^2}Q_S$.
It is negative if $Q_S<0$, i.e. water enters the surface.
This is the kinetic energy of the vertical movement of the transfered water.
It is considered as loss, since the vertical velocities are neglected on the surface. 
The only case in which energy is added, is if $Q_S>0$, that means   water from the surface enters the sewer system.
All the water inside the manhole is assumed to move with the velocity $\frac{Q_M}{A_M}$.
The energy generated in this part is exactly the kinetic energy, which the incoming water needs to move with the water already inside the manhole.

This drawback can be cured by substituting in equation \eqref{Models:eq:deltahinlet} the term $h_M+z_M$ by $\bar{h}_M$.
This modification can be  omitted in the current model, since  the movement inside the manhole is assumed not to be very large.

\section{Numerical Methods}
In this section we describe a second order numerical method to solve the system of sewer network and surface flow. We use a  second order splitting scheme. The components of the scheme are described in the following. 
A special focus is given on the numerical incorporation of the coupling procedures.
\subsection{A single tube}
For the Saint Venant equation there exist many powerful solvers. 
Here we use the Augmented Riemann Solver described in \cite{Borsche}.
It is well-balanced w.r.t. the bottom slope and can accurately capture wet-dry interfaces.
For the well-balancing a special approximation of the pressure law \eqref{Models:eq:pressurelaw} has to be used.
At the boundaries of the domain we use ghost cells \cite{LeVequeBook}, i.e. no modification of the solver is required at these points.  
 
\subsubsection{Approximation of the pressure law}
The pressure law \eqref{Models:eq:pressurelaw} strongly depends on the geometry of the conduit.
In sewer systems mainly circular tubes are used.
For tubes of circular shape the expression \eqref{Models:eq:pressurelaw} can not be further simplified, i.e. it has to be approximated in a suitable way.

In the following we estimate the curve of the pressure law by approximating the shape of the tube, instead of approximating \eqref{Models:eq:pressurelaw} directly as e.g. in \cite{Leon}.
For a tube with a hexagonal profile we obtain the following relations for the width $w$ and the height $h$
\begin{align}\label{Numerics:eq:HexagonalShapeh}
 w=
\left\{
\begin{array}{ll}
\frac{2}{c}h & 0\leq h\leq h_1\\
2r & h_1<h\leq h_2\\
\frac{2}{c}(2r-h) & h_2<h\leq h_P\\
w_P & h_P <h\\
\end{array}
\right.
\, ,\ 
h=
\left\{
\begin{array}{ll}
\sqrt{cA} & 0\leq A\leq A_1\\
\frac{1}{2r} A+\frac{c}{2}r & A_1<A\leq A_2\\
2r-\sqrt{c\pi r^2-cA} & A_2<A\leq A_P\\
\frac{1}{w_P}A-\frac{\pi}{w_P}r^2-\frac{c}{4}w_P+2r &A_P<A
\end{array}
\right.
\end{align}
This ansatz also allows us to incorporate directly the Preissmann Slot \cite{CungeHolly,BourdariasGerbi,Leon}. 
In order to closely approximate the volumes of a circular tube with radius $r$ we choose the constants as $c=2-\frac{\pi}{2}$, $h_1=cr$, $h_2=\frac{\pi}{2}r$ 
and for a Preissmann Slot of width $w_P$ the corresponding critical levels are $h_P=2r-\frac{c}{2}w_P$ and $A_P=\pi r^2-\frac{c}{4}w_P^2$.

\begin{figure}[htpb]
\includegraphics[width=7cm]{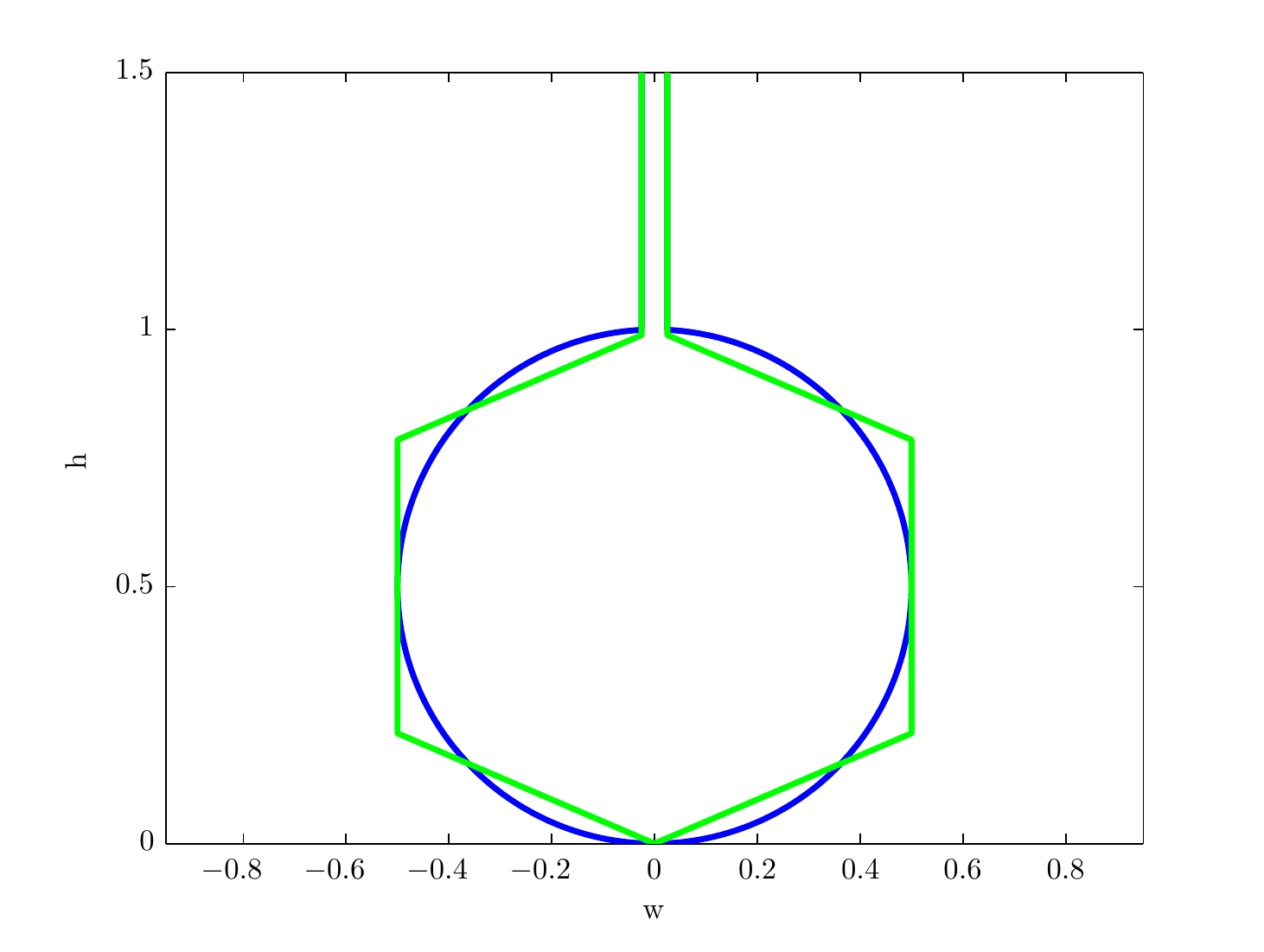}
\includegraphics[width=7cm]{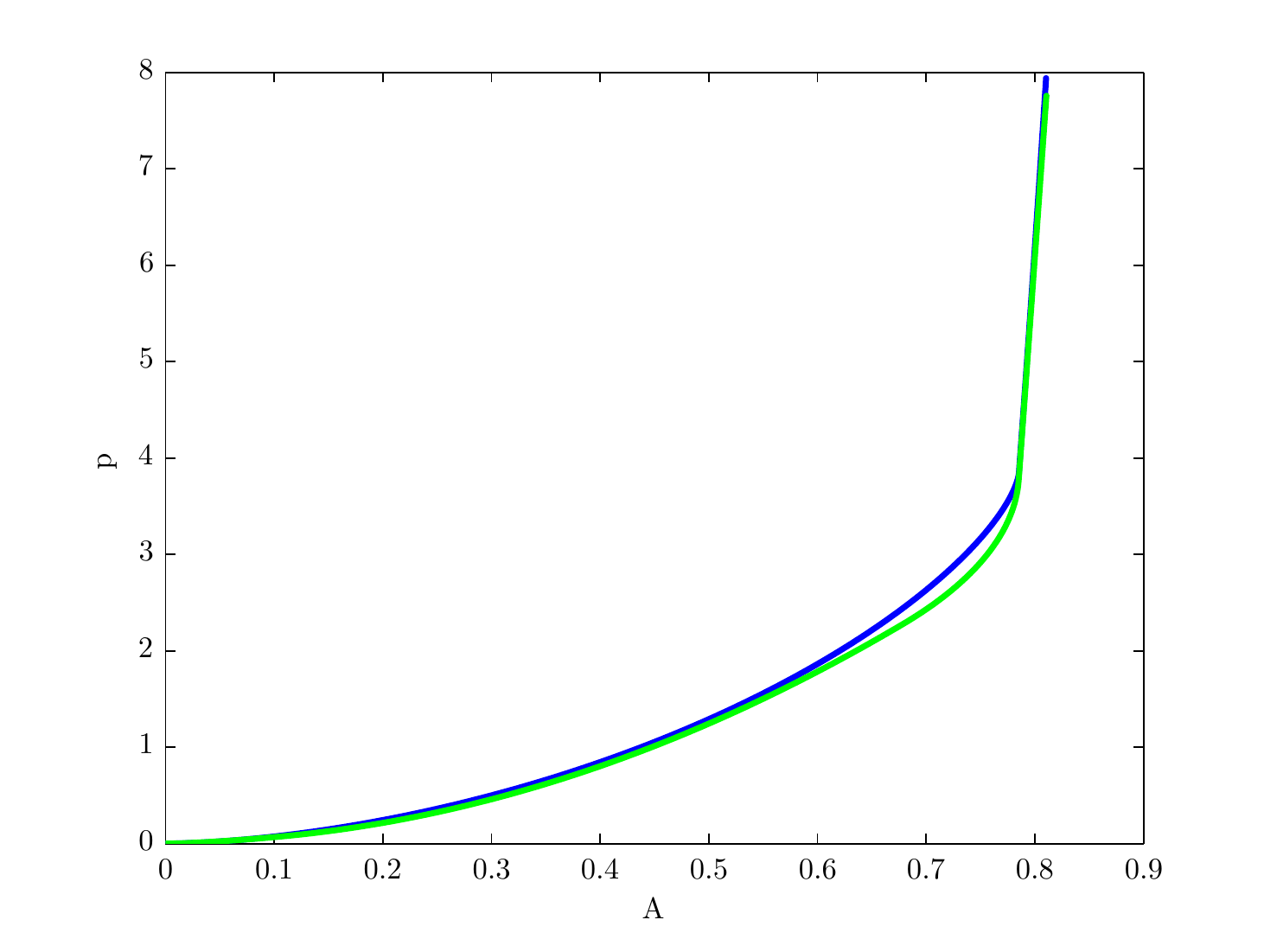}
\caption{The cross section and the corresponding pressure law of a circular (blue) and a hexagonal (green) tube with Preissmann Slot.
\label{Numerics:fig:PreissmannCrossection}
}
\end{figure}
In figure \ref{Numerics:fig:PreissmannCrossection} (left) we show the cross section of tube with a circular profile and one with a hexagonal profile.
The corresponding pressure laws (right) show a very similar behavior.

For the geometry of such a hexagonal tube the pressure law \eqref{Models:eq:pressurelaw} can be simplified to the following piecewise polynomial expression
\begin{align}\label{Numerics:eq:HexagonalShapeP}
p=g
\left\{
\begin{array}{ll}
\frac{1}{3c}h^3 & 0\leq h\leq h_1\\
rh^2-cr^2h+\frac{1}{3}c^2r^3 & h_1<h\leq h_2\\
-\frac{1}{3c}h^3+\frac{2}{c}rh^2+\left(\pi-\frac{4}{c}\right)r^2h+
\left(\frac{1}{3}c^2+\frac{\pi^3}{24 c}\right)r^3 
& h_2<h\leq h_P\\
\frac{w_P}{2}h^2+(\pi r^2-2 rw_P+\frac{c}{4}w_P^2)h-\pi r^3+2r^2w_P-\frac{c}{2}rw_P^2+\frac{c^2}{4}w_P^3 
& h_P<h\ .
\end{array}
\right.
\end{align}
In the following we will refer to tubes with this pressure law as hexagonal conduits of radius $r$. 

\subsection{Coupling at a junction}\label{Numerics:sec:CouplingConditions}
For the solving of the coupling at a junction we first consider the case of a junction without manhole.
In a second step the drop shaft can be included into the coupling procedure.

\subsubsection{Coupling of tubes}
The coupling of hyperbolic conservation laws is needed in many different applications, e.g. supply chains \cite{FugenschuhGoettlichHertyKlarMartin,HertyRinghofer}, traffic flow \cite{HertyKlar,HertyKirchnerKlar} and gas networks \cite{HertySachers,HertySeaid}
Here we follow an approach closely related to \cite{ColomboGaravelloRPJunction}.

For the coupling of $n$ tubes governed by the Saint Venant equations \eqref{Models:eq:StVE} we have the $n$ equation \eqref{Models:eq:junction_mass} and \eqref{Models:eq:equalhydraulicheadslosses} as coupling conditions.
But in the ghost cells of the connected edges we have $2n$ unknowns.
The additional equations we obtain due to the following considerations. 
The states in the node should only allow waves to travel into the domain, i.e. no waves or information is lost at the junction.
All such states lie on the so called Lax-Curves. 
If all edges are oriented in such a way, that they start at the node, we have to use the reversed Lax-Curves \cite{ColomboGaravelloRPJunction}.
The reversed Lax-Curves for the Saint-Venant equations \eqref{Models:eq:StVE} are
\begin{align*}
Q_m=L_2^-(A_m;A_R,Q_R)=
\begin{cases}
\frac{Q_R}{A_R}A_m+(A_m-A_R)\sqrt{\frac{A_m}{A_R}\frac{p_R(A_m)-p_R(A_R)}{A_m-A_R}}
\qquad & if\  A^R_m>A_R\\
\frac{Q_R}{A_R}A^R_m+A_m\int^{A_m}_{A_R}\frac{\sqrt{\partial_A p_R(a)}}{a}da
& else\,,
\end{cases}
\end{align*}
with $A_R$, $Q_R$ the first states within the domain and $A_m$, $Q_m$ the states in the node.
Since we have to apply these relations for each edge, the $n$ reversed Lax-Curves complete the set of coupling conditions.

Altogether we have $2n$ nonlinear equations for $2n$ unknowns.
This system of nonlinear equations we solve with a Newton type method.
In general the values of the previous time step provide a good starting value, such that only few iterations are needed.

The resulting states can now be used to fill the ghost cells of the conduits and the next time step therein can be computed.
This procedures requires, that all connected tubes are synchronized, i.e. the time steps have to be chosen according to the most restrictive CFL condition,
\begin{align*}
\Delta t\  \cdot \max_{i\in \mathcal{E}^j_{edges}}\left(\frac{\lambda_{max,i}}{\Delta x_i}\right)<1\,,
\end{align*}
where $\lambda_{max,i}$ is the maximum of the absolute value of the speed of propagation in the $i$-th tube.

\subsubsection{A junction with a manhole}
As a manhole is always located at a junction of at least two tubes we incorporate the manhole model into the coupling conditions of the node. 
In the following we describe a procedure for the full model \eqref{Models:eq:FullManhole}, the simplified model \eqref{Model:eq:Manhole_simple} can be treated in a similar way.

Additionally to the $2n$ unknowns in the ghost cells, we now have to consider the two unknowns of the ODE of the manhole.
There exist different approaches to solve DAE systems \cite{MR2450187}.
Here we incorporate an implicit two stage Runge-Kutta method directly into the solver of the algebraic coupling conditions.
This has the advantage, that the mass added to the manhole can be exactly balanced with the mass subtracted from the tubes.
The method is chosen to be implicit, such that no further restriction on the time step can arise.
As the manhole models are quite simple ODEs, the computational effort is small compared to the rest of the system.

\subsection{The surface}
The flow on the surface is governed by the shallow water equations \eqref{Models:eq:ShallowWater}. 
For these a huge variety of solvers have been developed, e.g. \cite{LeVeque-wavepropagation,NoelleXingShu,Seaid,ToroSWE}. 
In the following we use the Augmented Riemann solver presented in \cite{GeorgeDiss,GeorgeAugmented}, which is of the same type as the one for the conduits \cite{Borsche}.
For all details of the 2D realization we refer to these publications. 

\subsection{Coupling surface flows and sewer system}
For the coupling of hyperbolic equations several approaches exist \cite{LeVequeBook}.
Nevertheless there are some particular aspects of the coupling between sewer system and surface flow, which we will address in this section.

As general coupling procedure we use the Strang splitting \cite{LeVequeBook}. 
We first solve the sewer system for $\frac{\Delta t}{2}$ with constant states on the surface,
followed by a step $\Delta t$ of the surface with constant states in the manholes
and finally $\frac{\Delta t}{2}$ of the sewer system again.
Since the time step restrictions in both models can vary significantly the $\Delta t$ of the coupling is not a global time step for the complete system.
Instead we choose for the coupling a local time step restriction.
On the surface in the area around $A_{inlet}$ and in the sewer parts of the connected tubes a local time step for the coupling of both systems is determined by  local CFL conditions. 
If the time step e.g. in the sewer system is smaller than the timestep $\Delta t$ of the coupling, several iterations of the network are computed before the surface update starts.
This has the advantage, that a surcharged conduit, which is not connected to the surface, does not slow down the computation of the complete system. 

Another important aspect for the coupling is the well balancing between both models.
The coupling is governed on the surface by the equation \eqref{Models:eq:S_sewer} and in the sewer by \eqref{Models:eq:Q_ext}.
For the balancing of both models the artificial storage capacity of the manhole model is of major importance. 
Below the surface the manhole has a constant cross sectional area $A_M$.
On the surface we extend the capacity of the manhole with an cross section $|A_{inlet}|$.

In order to motivate this choice we consider the following example, where the extension of the manhole has a cross section smaller than $|A_{inlet}|$.
Assume we have a constant water level on the surface and the water inside the sewer system is at the same hight.
If now a small wave passes the surface, the water level on the surface rises. 
Thus according to \eqref{Models:eq:S_sewer} the inflow into the connected manhole is computed.
The incoming water \eqref{Models:eq:Q_ext} is distributed on the area of the extended manhole, i.e. the water level of the manhole can rise above the water level of the surface.
As we want to keep the extended storage as small as possible we prolong the manhole with the area $|A_{inlet}|$.

\section{Numerical examples} 
In this section we present numerical examples elucidating the behavior of the models described above.
If not specified differently the following settings are used.

The length of a tube is $10$ and the pressure law of the hexagonal profile \eqref{Numerics:eq:HexagonalShapeP} and \eqref{Numerics:eq:HexagonalShapeh} for a radius $r=0.5$ is used.
At the boundary free outflow conditions are prescribed.
The gravitational acceleration is set to $g=9.81$ and the friction term is neglected, i.e. $n_{f}=0$.
The computational grid is $100$ cells per tube and the time step is chosen variable according to a CFL bound of $0.99$.
All figures show the height of water above the bottom $h(A)+z$.
For the manholes we choose a cross sectional area of $A_M=0.25\, \pi$ and the additional losses are omitted.
The initial filling is orientated on the connected tubes.
The surface is initialized without water and free outflow conditions are imposed at the boundary.
For the friction formula \eqref{Model:eq:frictionStVE} with  $n_{f}=0.025$ is used.
The area associated to the inlets is $A_{inlet}=0.25\, \pi$.
 
\subsection{The manhole models}\label{Numerics:sec:ManholeTests}
Here we investigate the behavior of the manhole models \eqref{Models:eq:FullManhole} and \eqref{Model:eq:Manhole_simple}, including their influence on the flow of the connected tubes.
In these examples we consider two tubes of length $25$, connected by a manhole in between.

\begin{figure}[htpb]
\includegraphics[width=6cm]{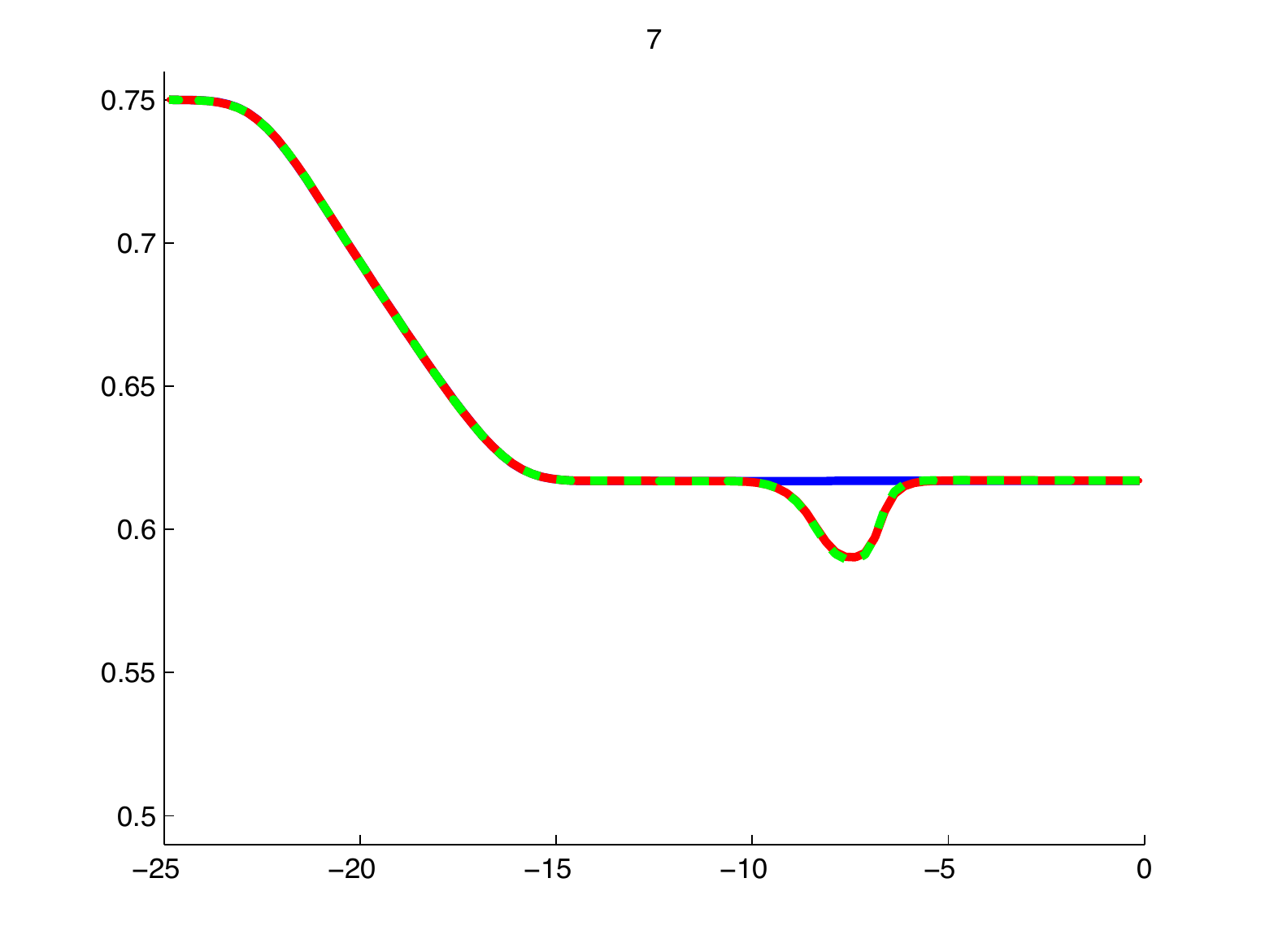}
\includegraphics[width=6cm]{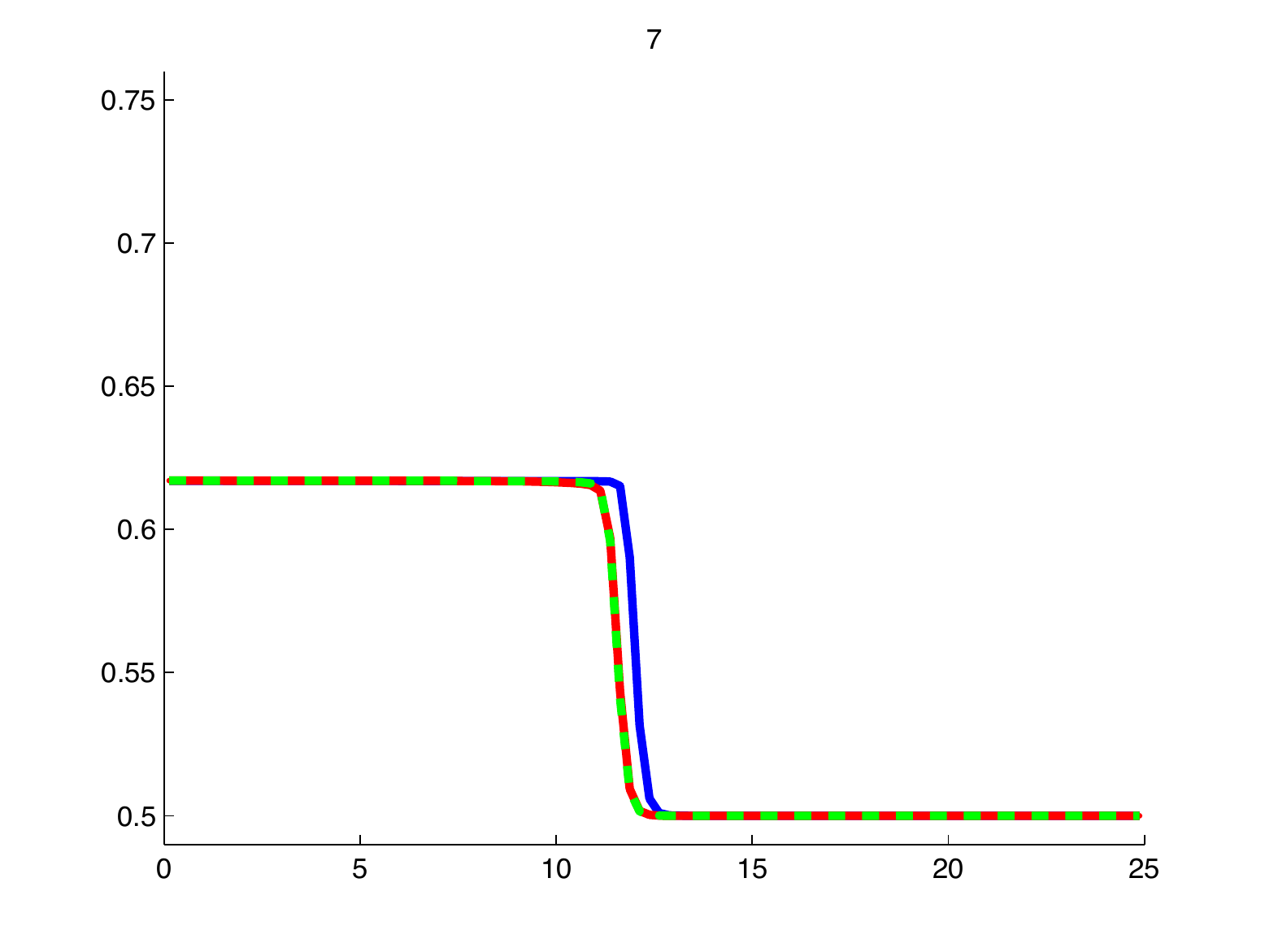}
\caption{The height of water in the first tube (left) and the second tube (right) at t=$7$. Compared are the full model (dashed, green), the reduced one (dash dot, red) and the evolution without a manhole (solid, blue).
\label{Numerics:fig:ManholeModels_Shock_Tubes}}
\end{figure}

\begin{figure}[htpb]
\includegraphics[width=6cm]{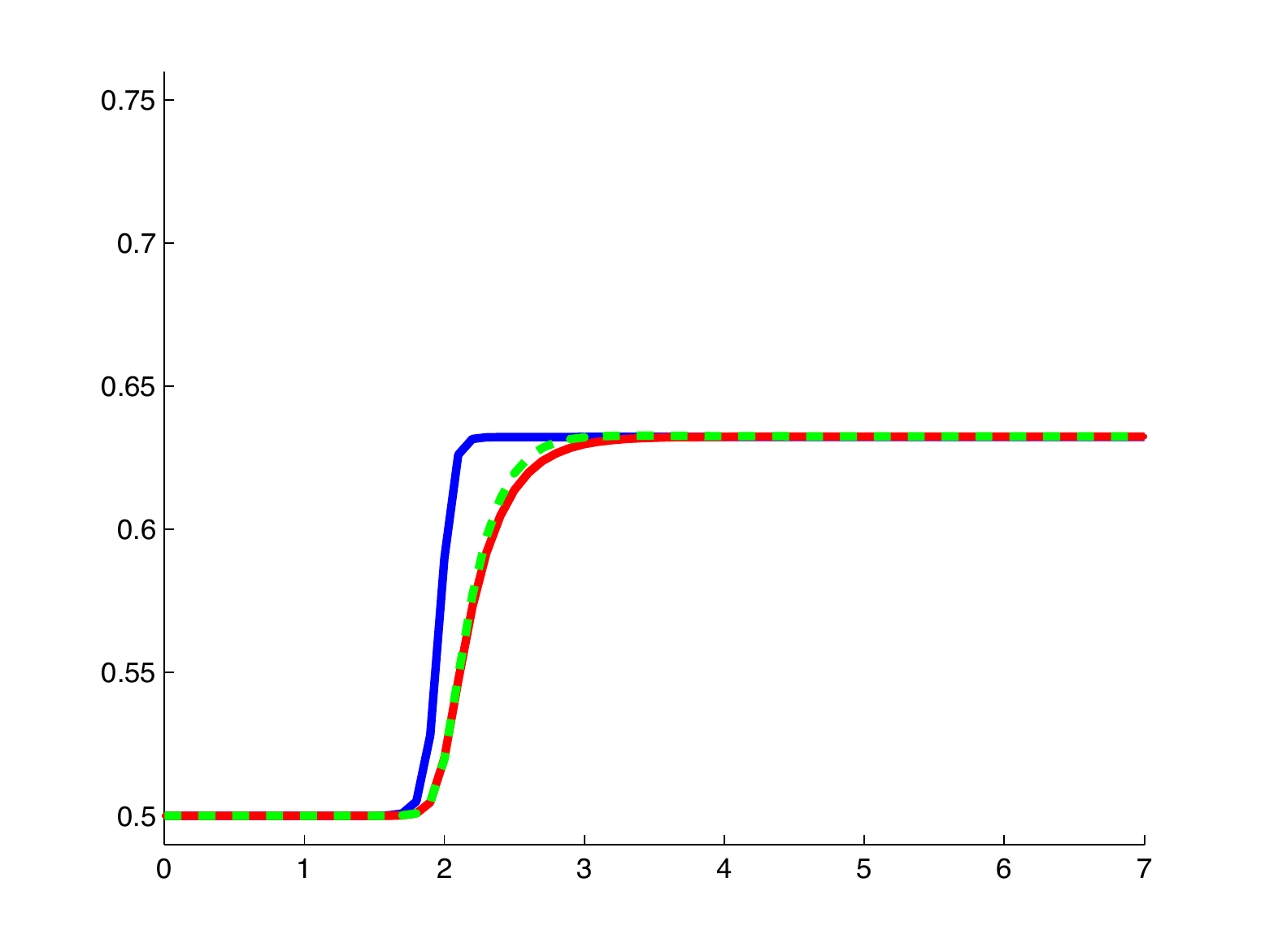}
\includegraphics[width=6cm]{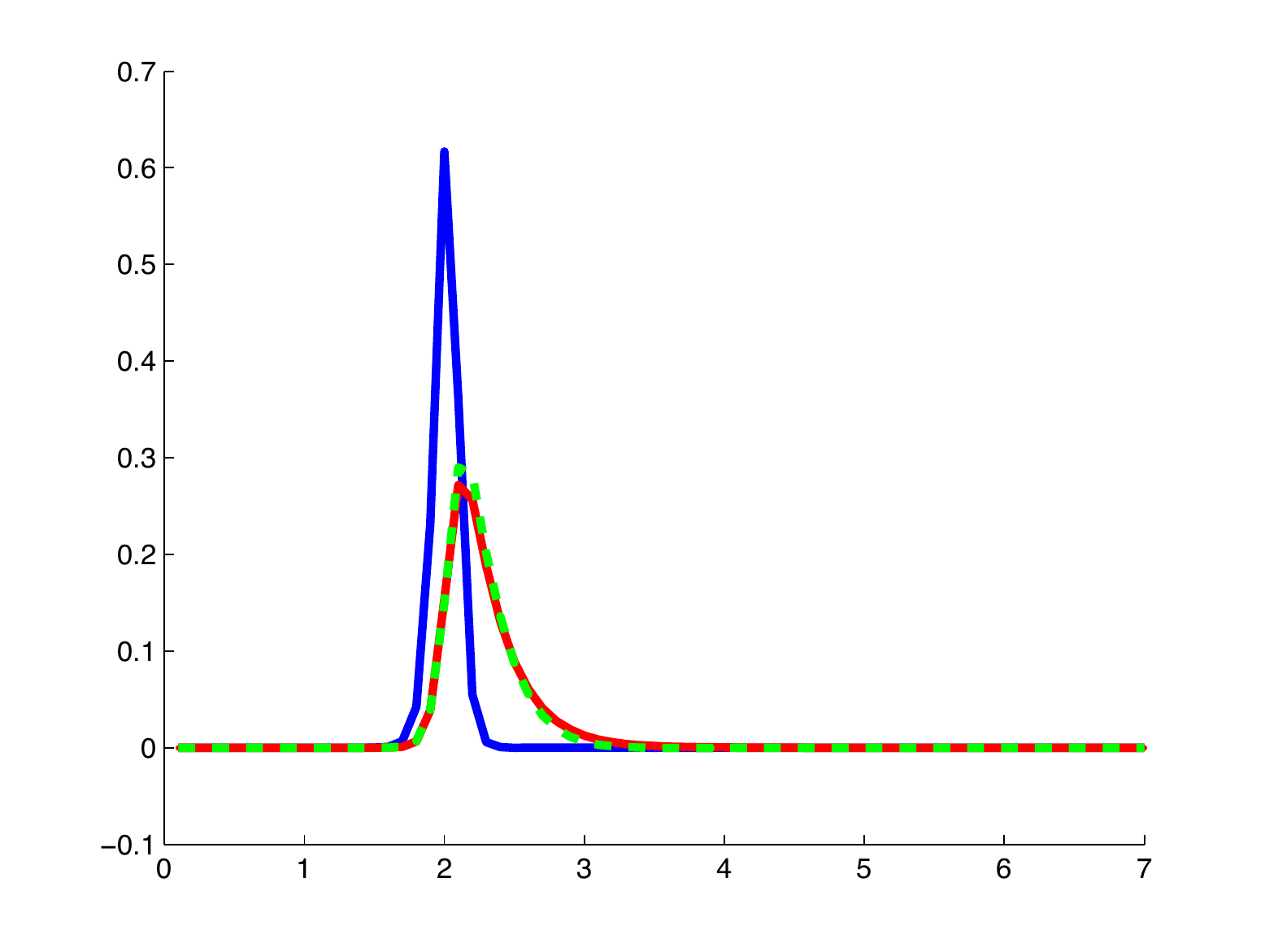}
\caption{The height of water in the manhole (left) and the flow into the manhole (right) from $t=0$ to $t=10$. Compared are the full model(dashed, green), the reduced one (dash dot, red) and the situation without (solid, blue).
\label{Numerics:fig:ManholeModels_Shock_Manhole}}
\end{figure}

The first test case shows a shock passing the manhole.
The corresponding initial conditions are $h_2(A_2(0,x))\equiv 0.5$ and $Q_2\equiv 0$ in the second conduit, while a Riemann Problem with $h_1(A_1(0,x_L))\equiv 0.75$, $h_1(A_1(0,x_R))\equiv 0.5$ and $Q_1\equiv 0$ is located at $x_L<-5<x_R$.
Thus, a shock wave travels to the right passing the tube.
The rarefaction wave moving in the opposite direction is of no further interest.
At about $t=2$ the shock hits the manhole and fills it according to the surrounding states.
As shown in figure \ref{Numerics:fig:ManholeModels_Shock_Manhole}, both models behave almost identical and the inflow is damped compared to the case without manhole (blue).
In figure \ref{Numerics:fig:ManholeModels_Shock_Tubes} the depth of water in the conduits at $t=7$ is plotted.
The shock is slightly retarded and a small kink travels backwards in the first tube.
The new middle state is the same in all cases.

Similar observations can be made for a rarefaction wave passing the manhole.
We consider the initial conditions $h_1(A_1(0,x_L))\equiv 0.5$, $h_1(A_1(0,x_R))\equiv 0.75$ and $Q_1\equiv 0$ with $x_L<-5<x_R$, respectively $h_2(A_2(0,x))\equiv 0.75$ and $Q_2\equiv 0$.
The shock wave is moving to the left, while the rarefaction wave travels through the junction.
\begin{figure}[htpb]
\includegraphics[width=6cm]{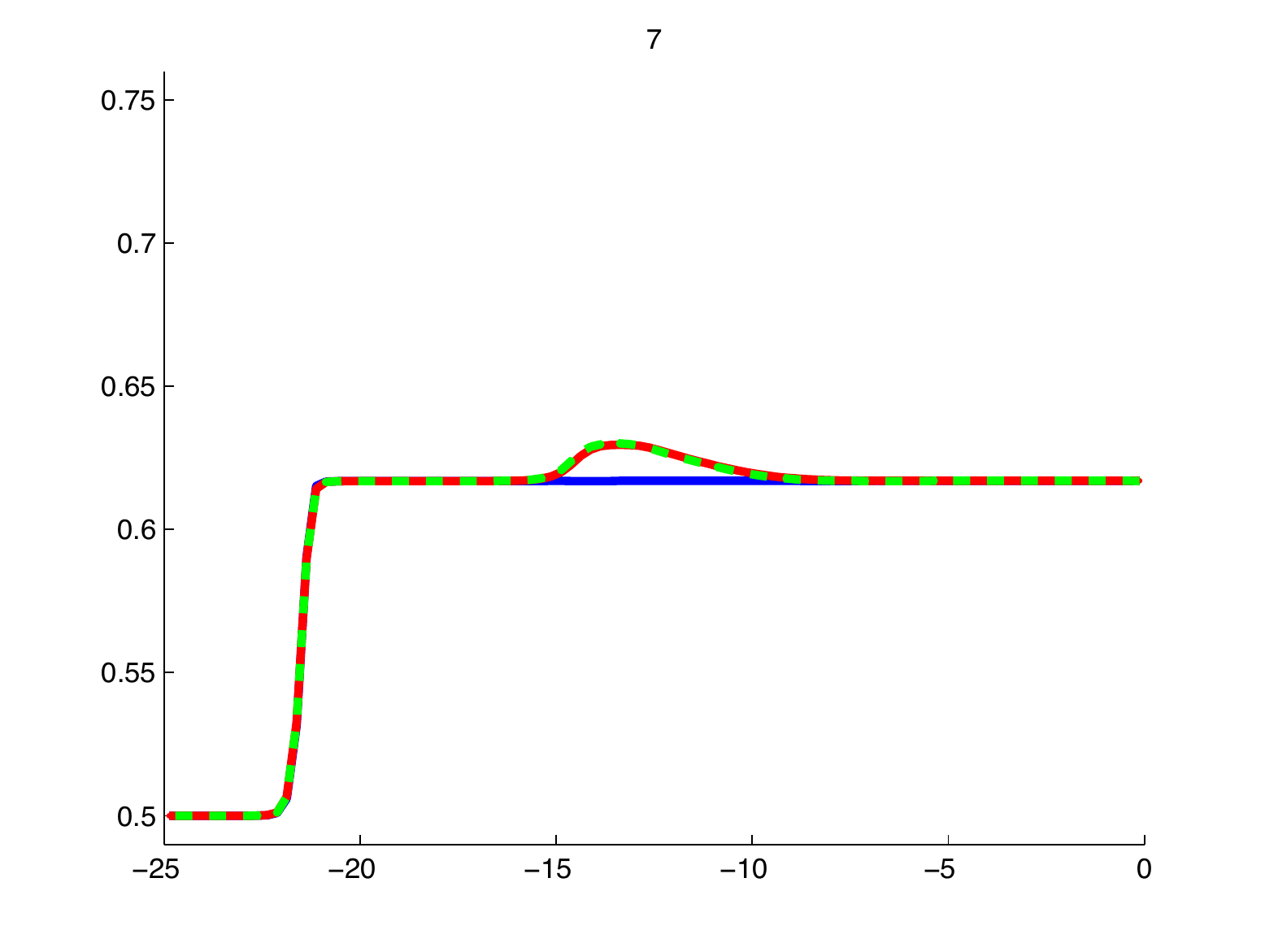}
\includegraphics[width=6cm]{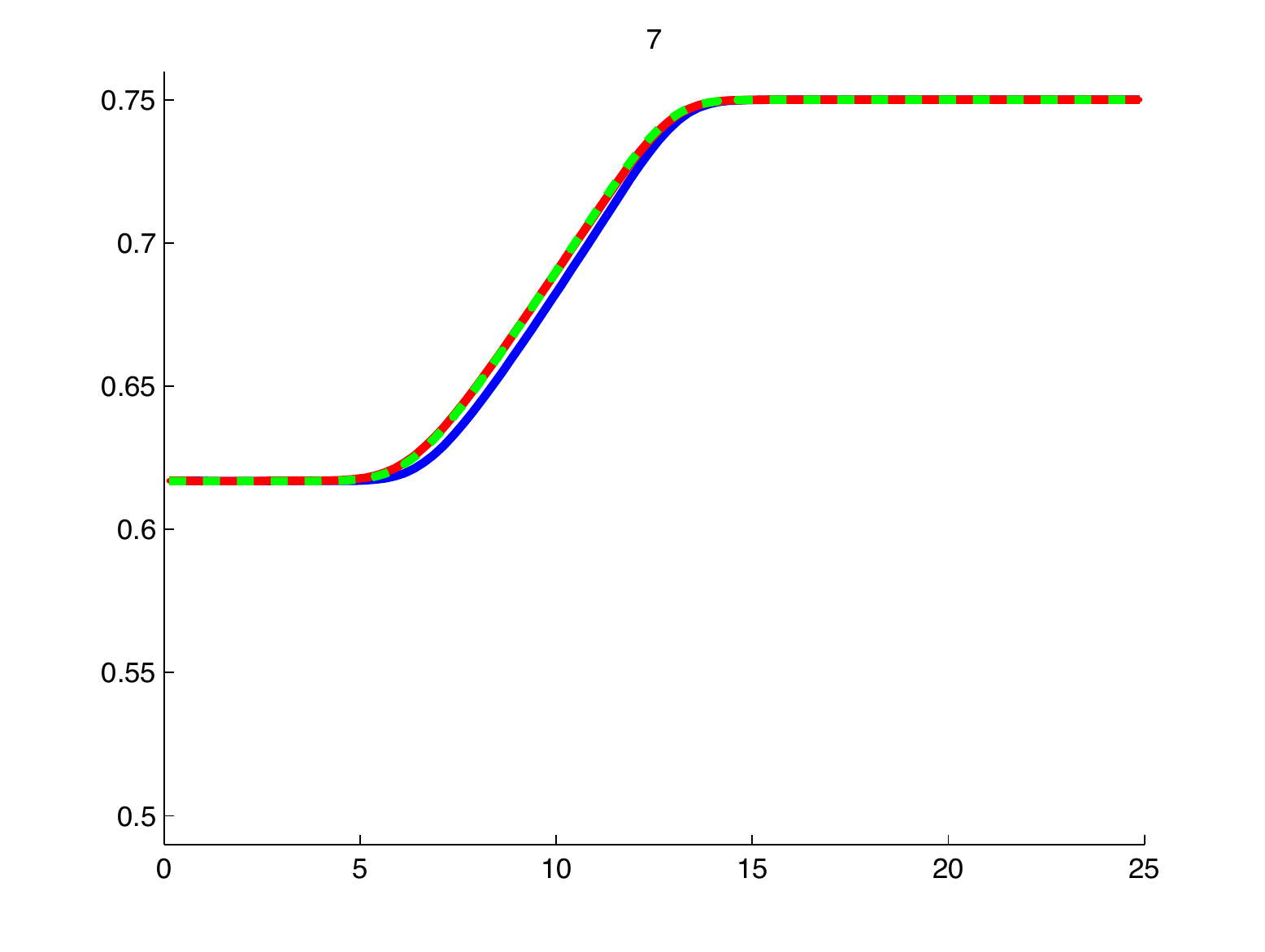}
\caption{The height of water in the first tube (left) and the second tube (right) at t=$7$. Compared are the full model(dashed), the reduced one (dash dot) and the evolution without a manhole (solid line).
\label{Numerics:fig:ManholeModels_Rare_Tubes}}
\end{figure}
As shown in figure \ref{Numerics:fig:ManholeModels_Rare_Tubes}, a small hump is formed moving towards the left.
The shape of the rarefaction wave remains almost unchanged.
As before the influence of the manhole is distributed equally to both tubes.
\begin{figure}[htpb]
\includegraphics[width=6cm]{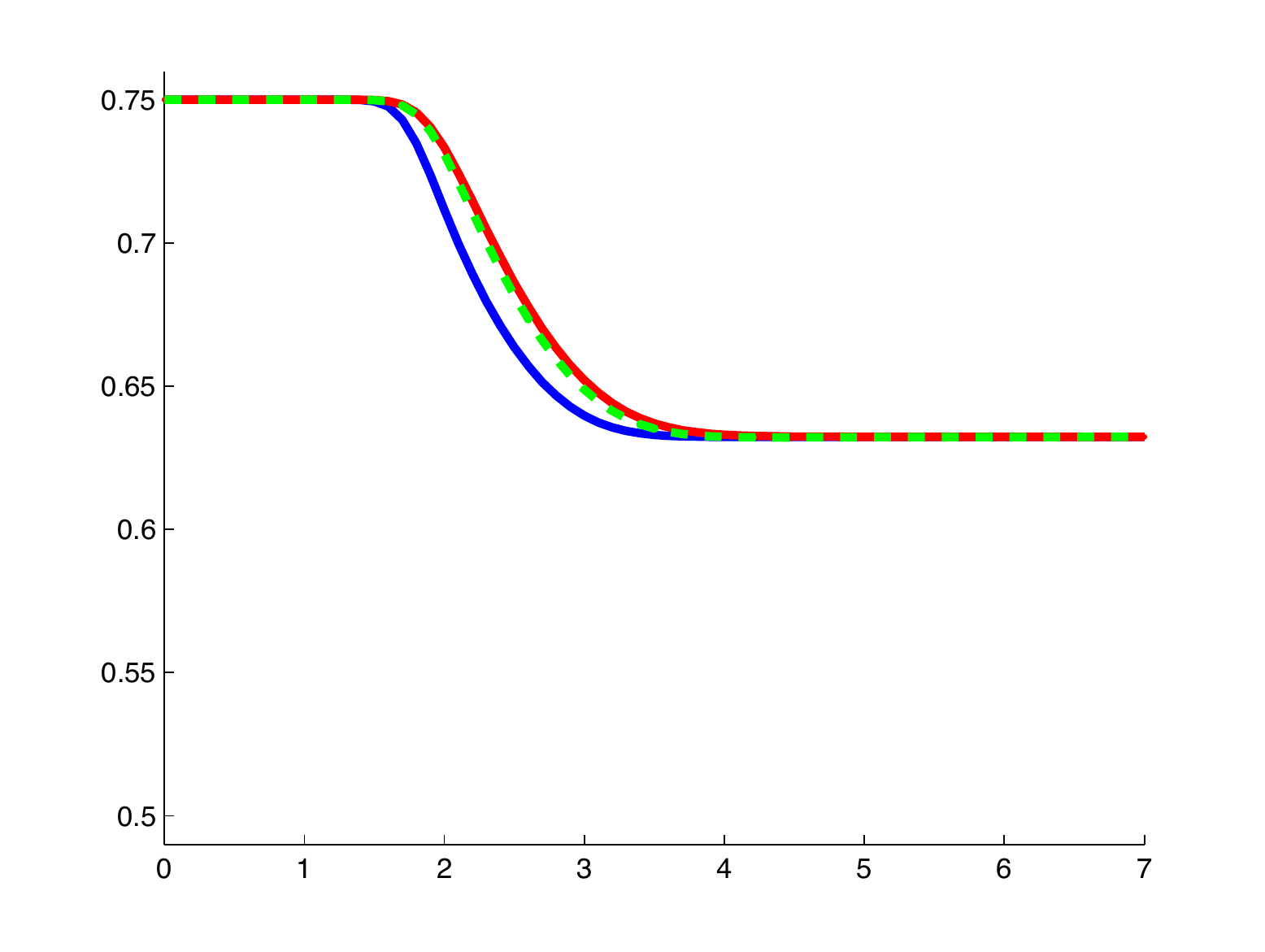}
\includegraphics[width=6cm]{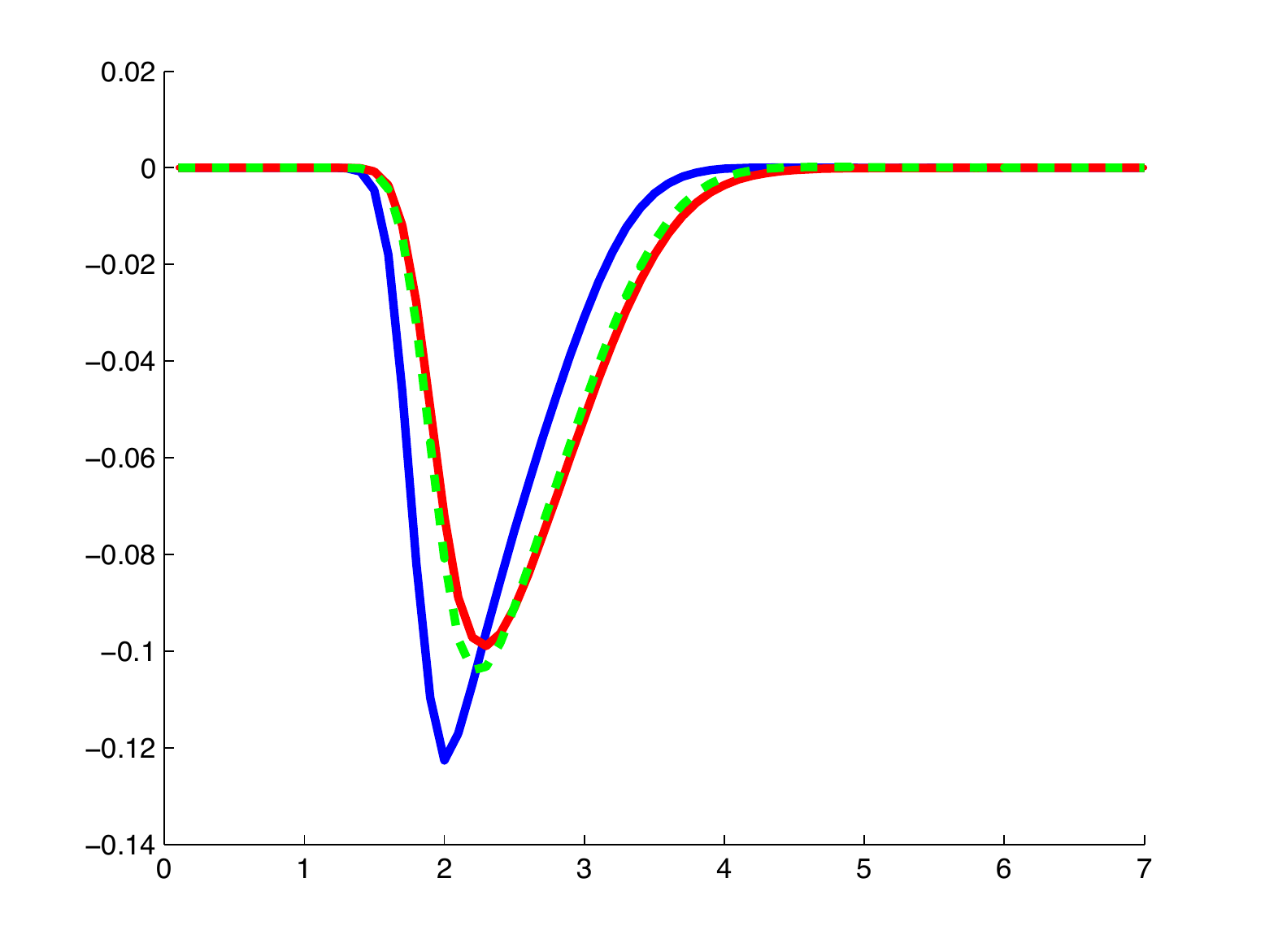}
\caption{The height of water in the manhole (left) and the flow into the manhole (right) from $t=0$ to $t=10$ time units. Compared are the full model(dashed), the reduced one (dash dot) and the situation without (solid line) 
\label{Numerics:fig:Manhole_CompareModels_Rare}}
\end{figure}
The behavior inside the manhole is plotted in figure \ref{Numerics:fig:Manhole_CompareModels_Rare}.
The height of water adapts to the depth of the passing wave.
As it can be seen in comparison to the situation without a manhole (blue) this happens slightly retarded.
The same can be observed for the flow $Q_M$, where the manhole smooths the outflow.
Again no significant difference in the two models for the manhole can be observed.

\begin{figure}[htpb]
\includegraphics[width=6cm]{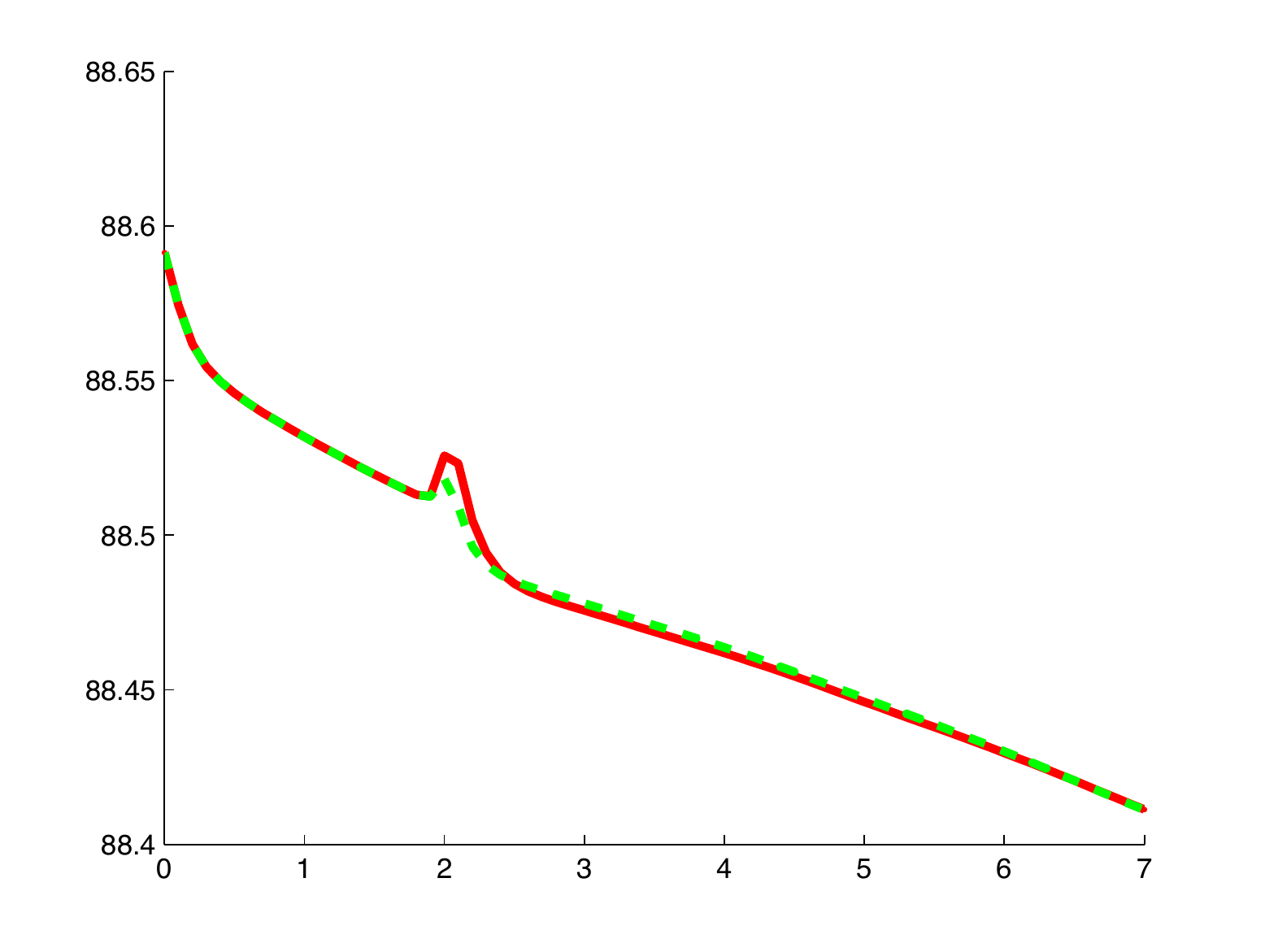}
\includegraphics[width=6cm]{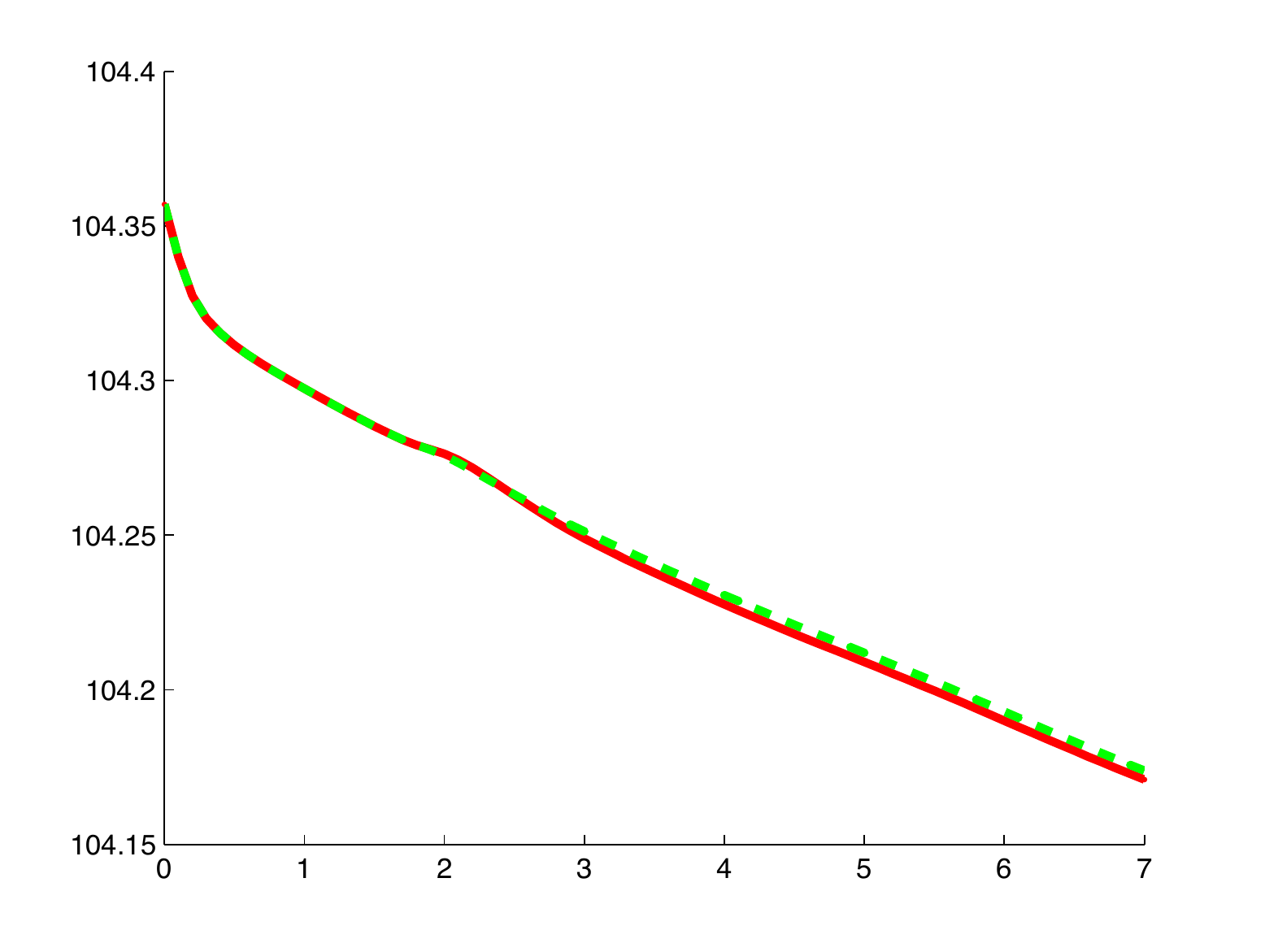}
\caption{The total energy $\mathcal{E}$ in the whole system from $t=0$ to $t=7$ time units. Compared are the full model(red, solid) and the reduced (green, dashed) one.
\label{Numerics:fig:Manhole_energy}}
\end{figure}

Finally, we look at the total energy in the system.
In figure \ref{Numerics:fig:Manhole_energy}, for both previous test cases the evolution of the total energy in the system is shown.
As the flow at the boundary is zero and no external inflow at the manhole is considered, no energy is subtracted from or added to the network.
In both situations the total energy is changing during time.
This is not in contrast to the calculation made in section \ref{Analysis:sec:ConservationOfEnergy}, since these are only valid for smooth solutions.
Since in both examples a shock wave occurs, the entropy admissibility condition causes the energy to decrease.
The only moment when energy is added to the system is, when the shock hits the junction.
But compared to the losses, due to the normal propagation of the shock, this is of negligible magnitude.
When the rarefaction wave passes the junction the total energy does not grow.
The scaling in both figures is relatively fine, but the contribution of the manhole to the energy of the full system is small even though.
As we already observed before, there is no significant difference between both models in the pictures.

In the following we investigate the influence of the size of the manhole.
Therefore, we repeat the test case of the shock passing the manhole for different cross sectional areas of the manhole.
Consider the full manhole model with cross sectional areas are $A_{M_1}=0.125\ \pi$, $A_{M_2}=0.25\ \pi$ and $A_{M_3}=0.5\ \pi$.
\begin{figure}[htpb]
\includegraphics[width=6cm]{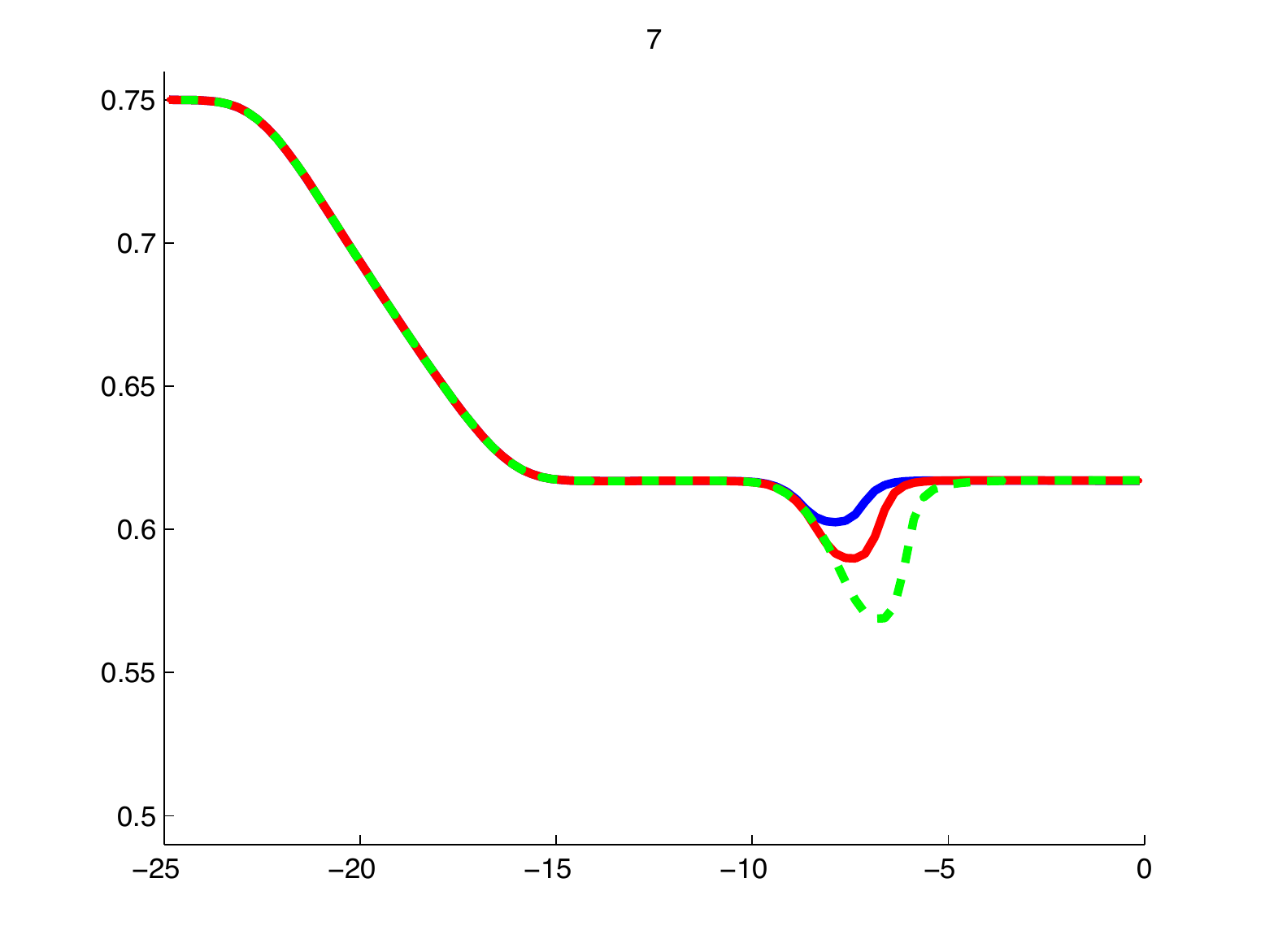}
\includegraphics[width=6cm]{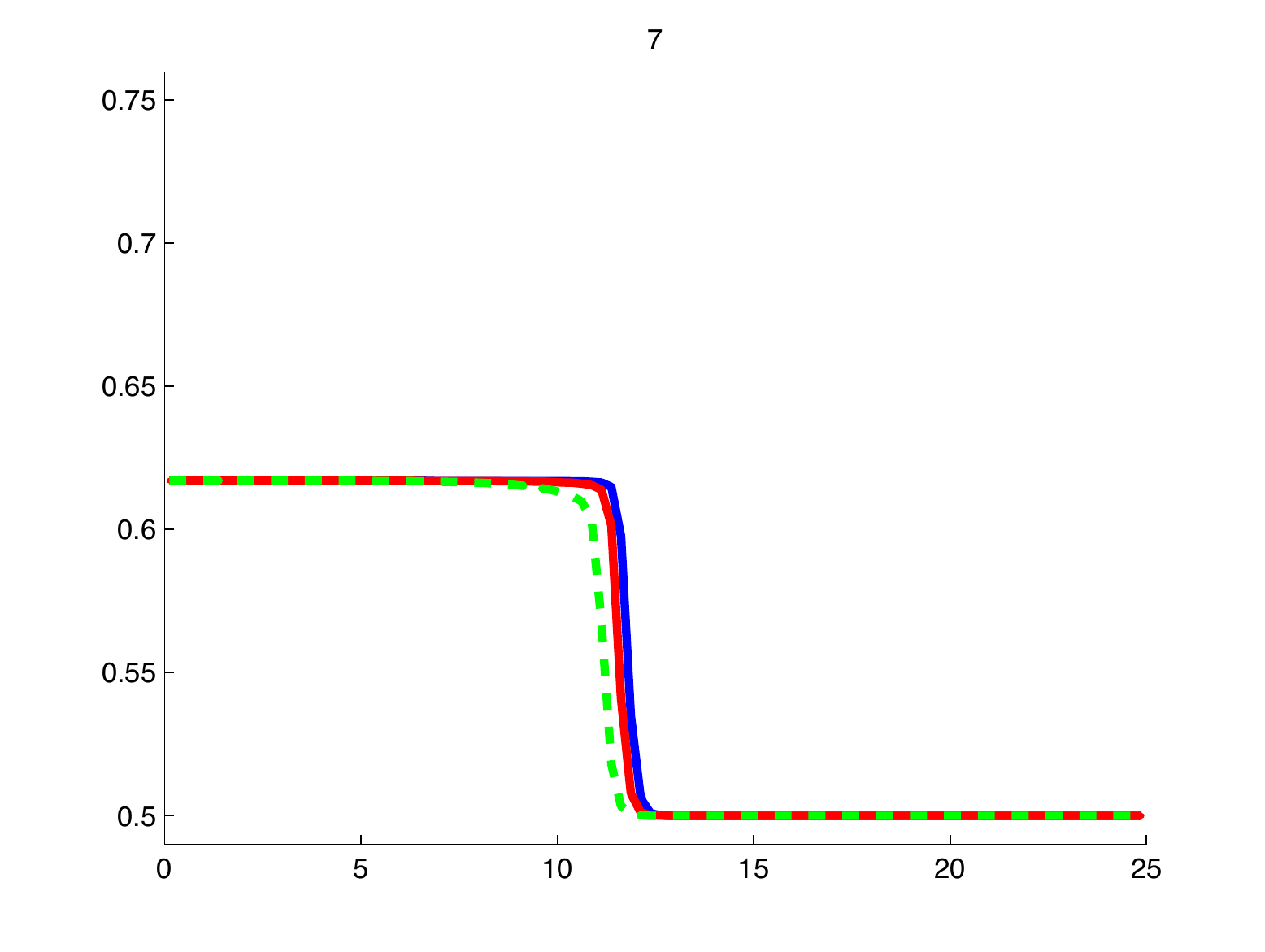}
\caption{The water level in the first tube (left) and the second tube (right) at t=$7$ for different cross sectional areas of the manhole, $A_M=0.125\ \pi$ (solid, blue), $A_M=0.25\ \pi$ (dash dot, red) and $A_M= 0.5\ \pi$ (dashed, green).
\label{Numerics:fig:ManholeAreas_tubes}}
\end{figure}
In figure \ref{Numerics:fig:ManholeAreas_tubes} the depth of water in the conduits is shown.
There is a direct dependence on the size of the manhole, i.e. the larger the area of the manhole is the larger the kink becomes and the more the shock is retarded.
\begin{figure}[htpb]
\includegraphics[width=6cm]{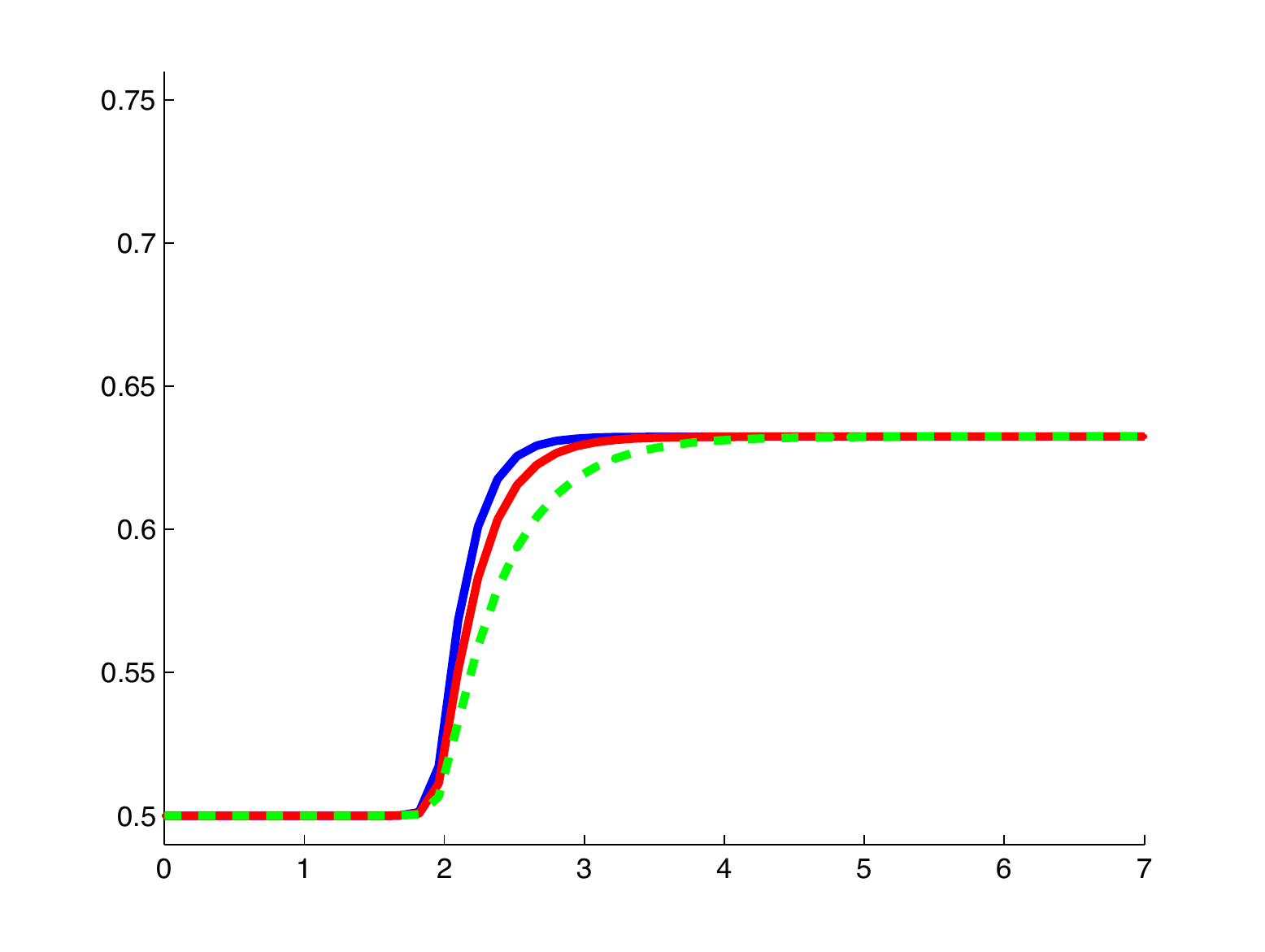}
\includegraphics[width=6cm]{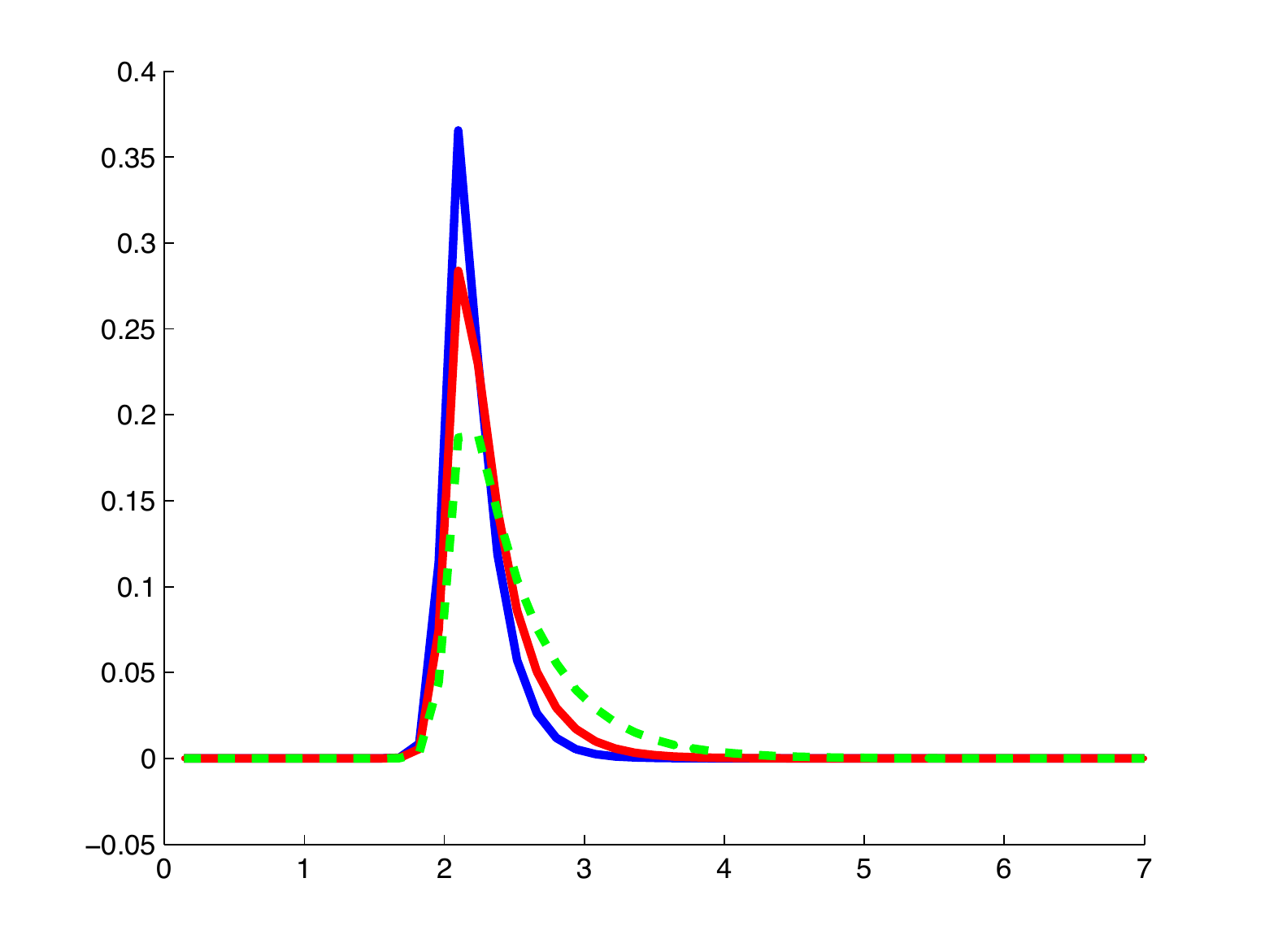}
\caption{The water level in the manhole (left) and the flow into the manhole (right) from $t=0$ to $t=7$ for different cross sectional areas of the manhole, $A_M=0.125\ \pi$ (solid, blue), $A_M=0.25\ \pi$ (dash dot, red) and $A_M= 0.5\ \pi$ (dashed, green)
\label{Numerics:fig:ManholeAreas_Manhole}}
\end{figure}
The states in the manhole are shown in figure \ref{Numerics:fig:ManholeAreas_Manhole}.
For smaller cross sectional areas the new steady heights are reached faster, whereas larger storage capacities need more time to adapt to the surrounding flows.

In total the influence of the manhole is moderate.
It can store water according to its size, but does not modify the actual structures of the waves.

\subsection{Coupling sewer and surface flow}
In this section we investigate the interplay of the surface flow and the sewer system.
A special focus is given to the well balancing of the coupling of both models.

\subsubsection{A well balanced test}
\begin{figure}[htpb]
\includegraphics[width=6cm]{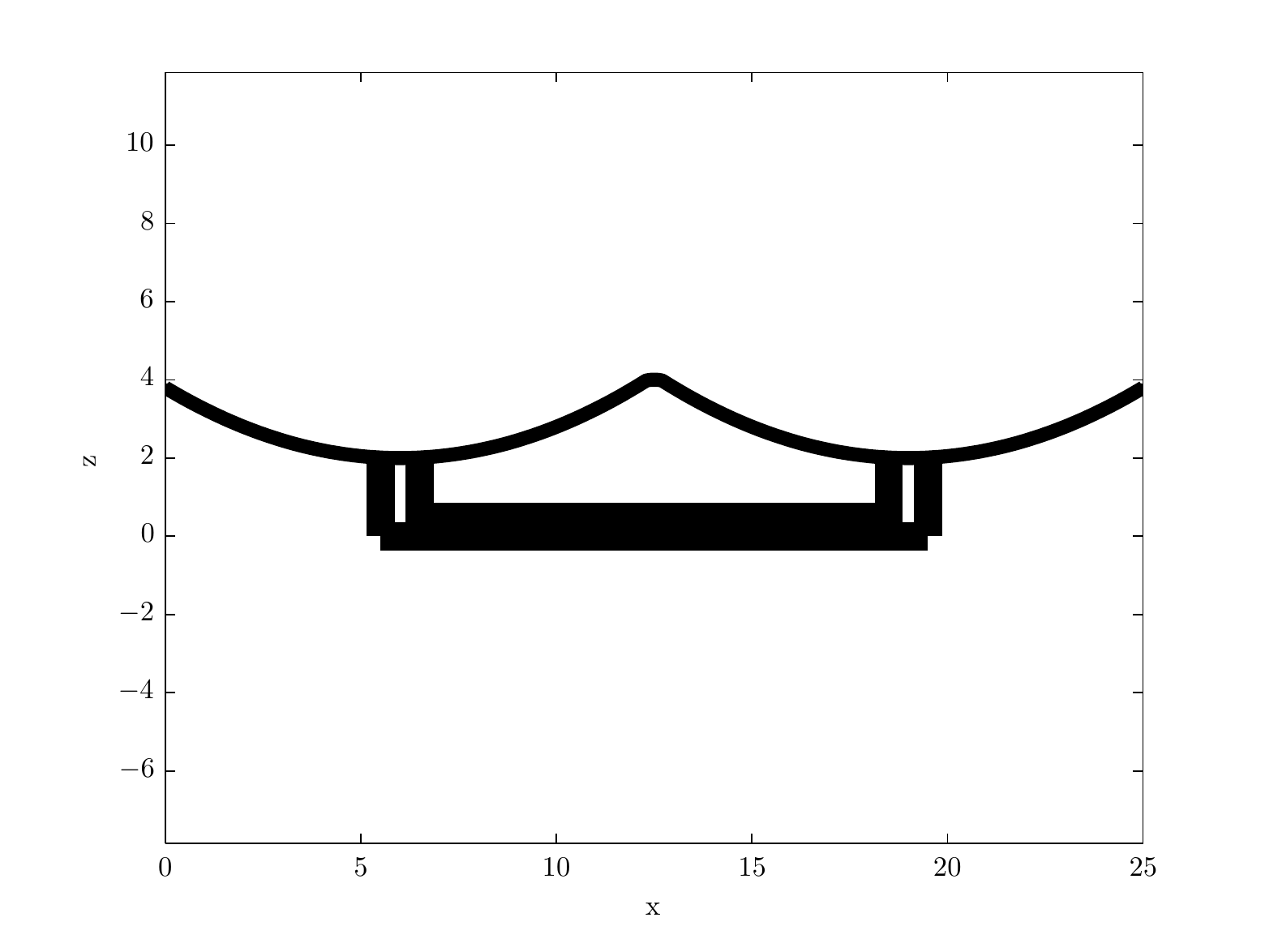}
\includegraphics[width=6cm]{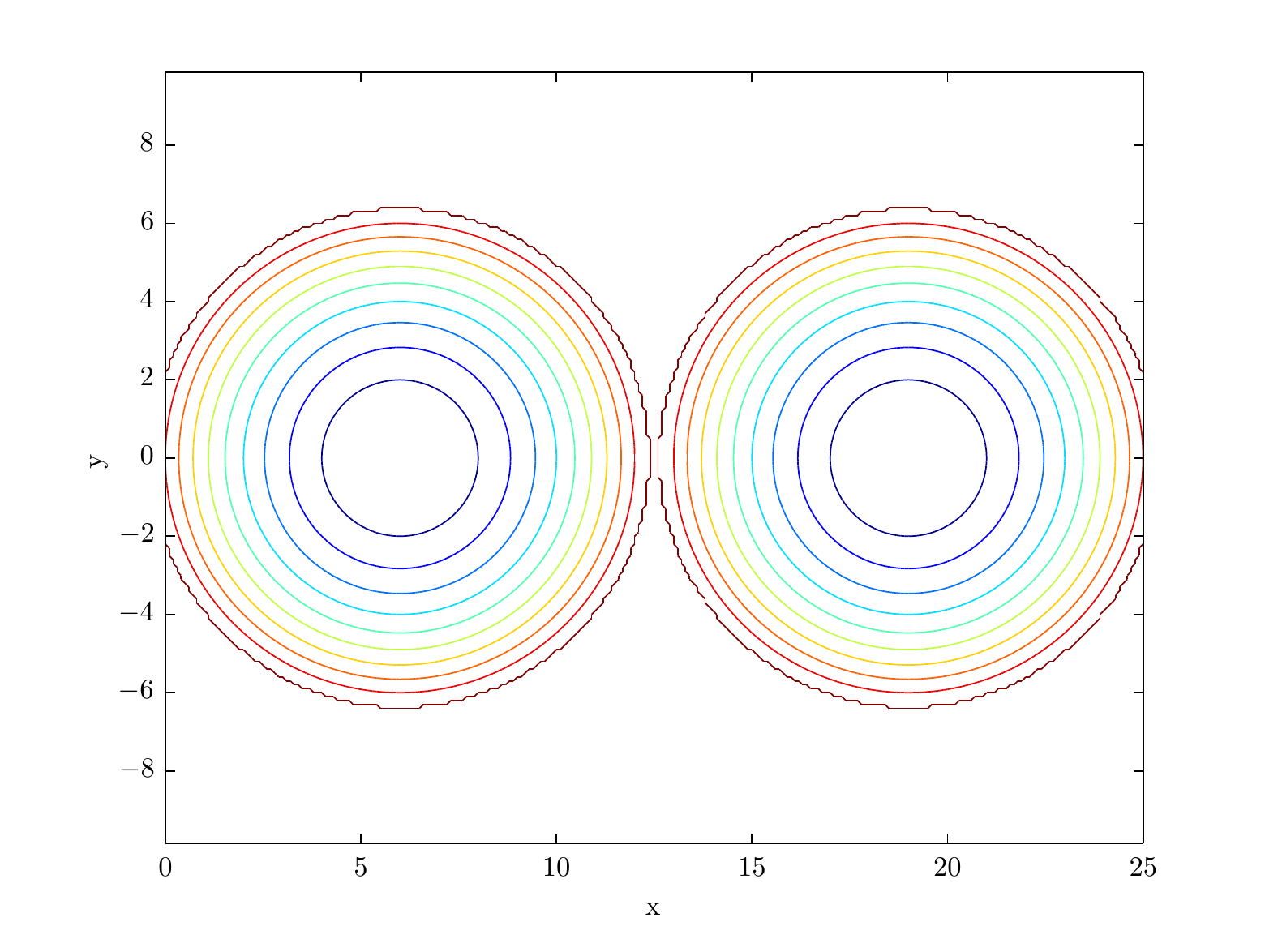}
\caption{A conduit connecting two hollows on the surface (left: vertical cut, right: top view).
\label{Numerics:fig:SNI_WellBalanced_draw}}
\end{figure}

First, we consider a simple test scenario with a minimal network and a small surface.
The sewer system is represented by a single tube of radius $r=0.25$ ranging from $\vec{x}_1=(6,0)$ to $\vec{x}_2=(19,0)$ and two manholes each at one end of the conduit.
At the top of the drop shafts two circular hollows on a surface of $[0,25]\times[-7,7]$ are placed
\begin{align*}
z(\vec{x})=\min\left(\min\left(2+0.05*\|\vec{x}-\vec{x}_1\|^2_2,2+0.05*\|\vec{x}-\vec{x}_1\|^2_2\right),4\right)
\,.
\end{align*}
Thus the sewer connects both hollows with tubes forming an 'U', as shown in figure \ref{Numerics:fig:SNI_WellBalanced_draw}.

{\bf In- and out-flow}\\
{\noindent} 
The initial states in the sewer system have a water level of $h(A(0,x))=h^1_M=h^2_M=0.5$, on the surface the right hollow is filled up to $h(\vec{x})+z(\vec{x})=3$ and the left one is empty.
For the computation we used $100$ grid cells in the conduit and $100\times 56$ cells on the surface.

\begin{figure}[htpb]
\includegraphics[width=6cm]{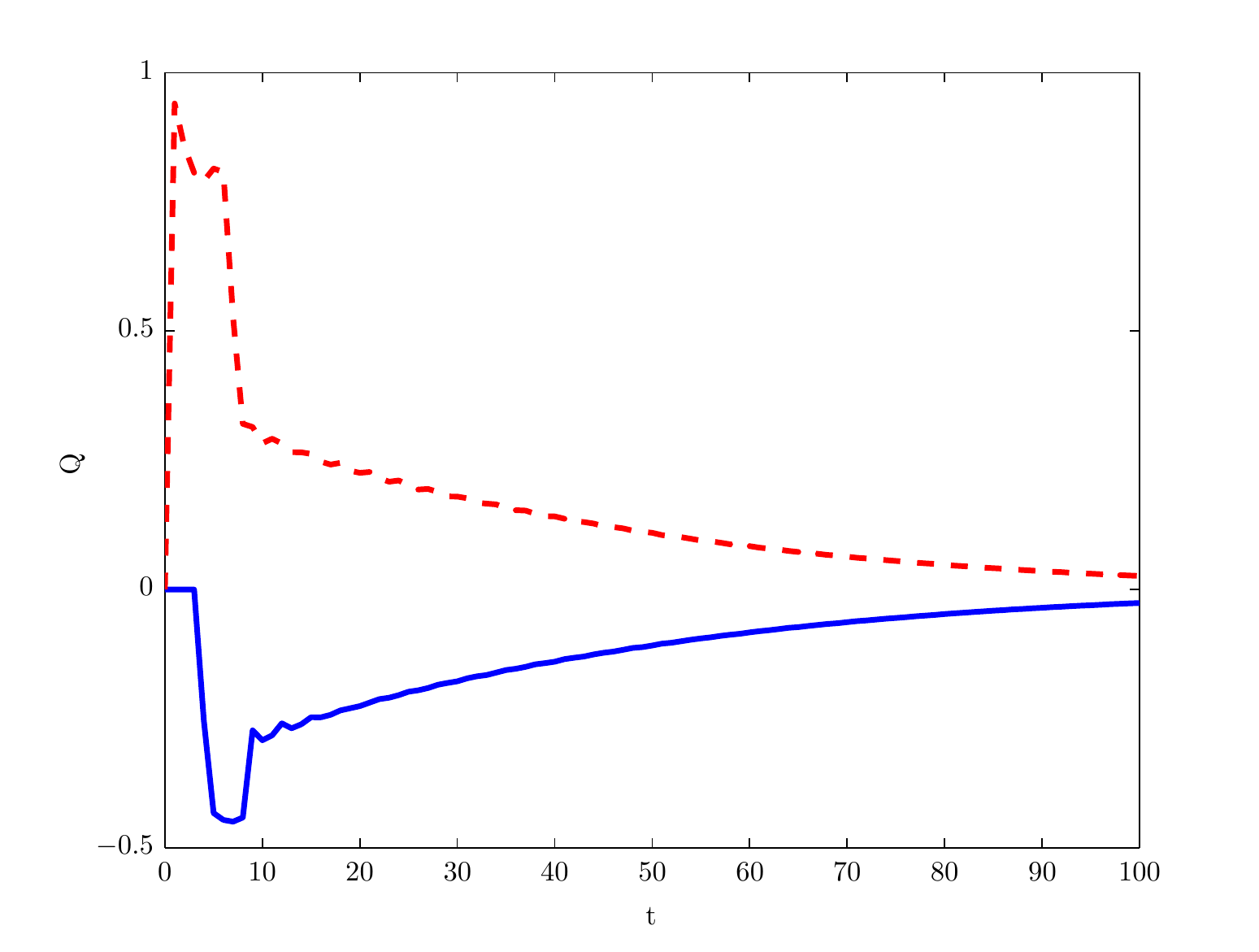}
\includegraphics[width=6cm]{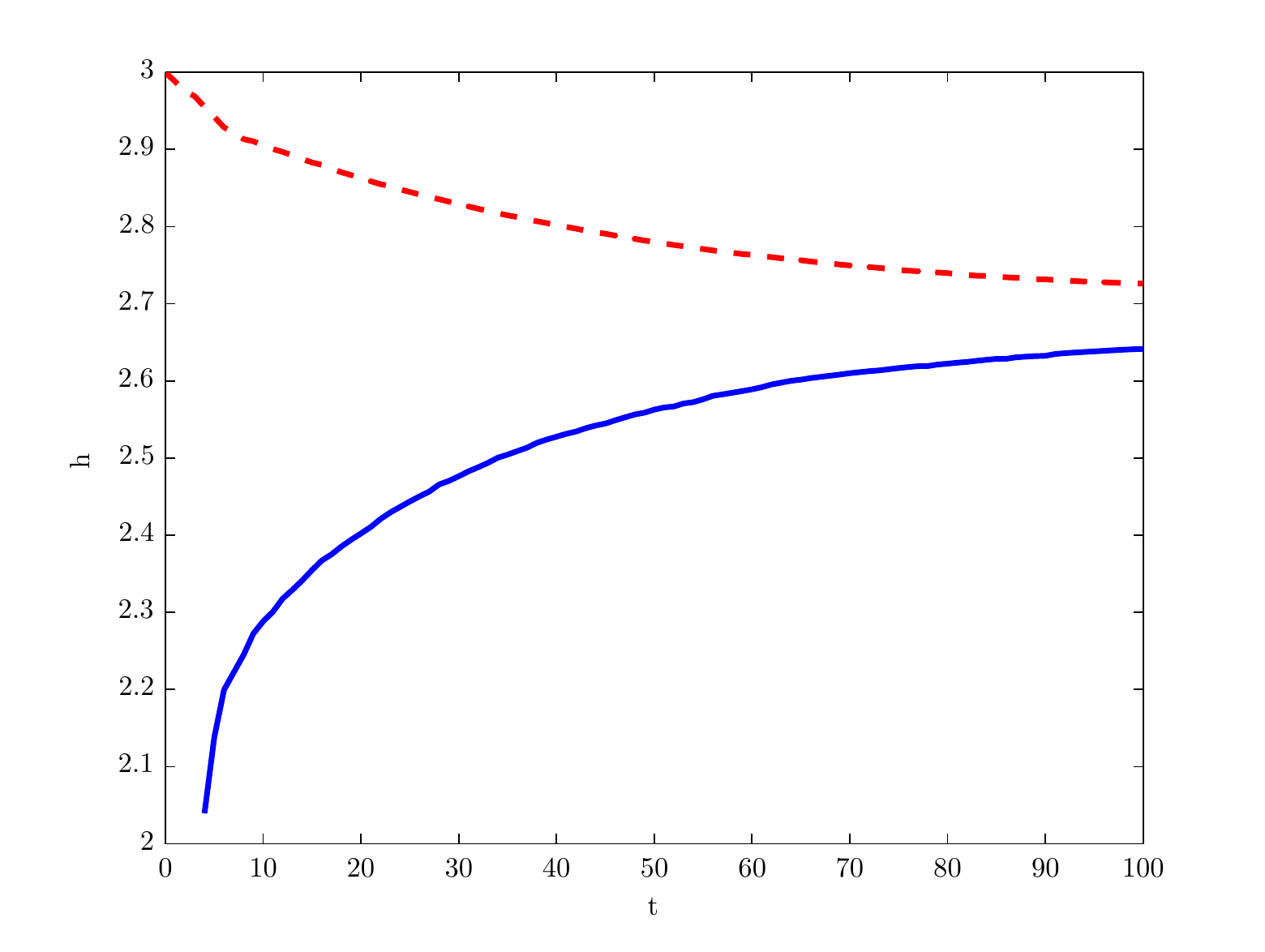}
\caption{The flow (left) into manhole 1 (blue solid line) and manhole 2 (red dashed line). The averaged water level (right) in the left (blue solid line) and right (red dashed line) hollow.
\label{Numerics:fig:SNI_WellBalanced_Qext}}
\end{figure}

\begin{figure}[htpb]
\includegraphics[width=7cm]{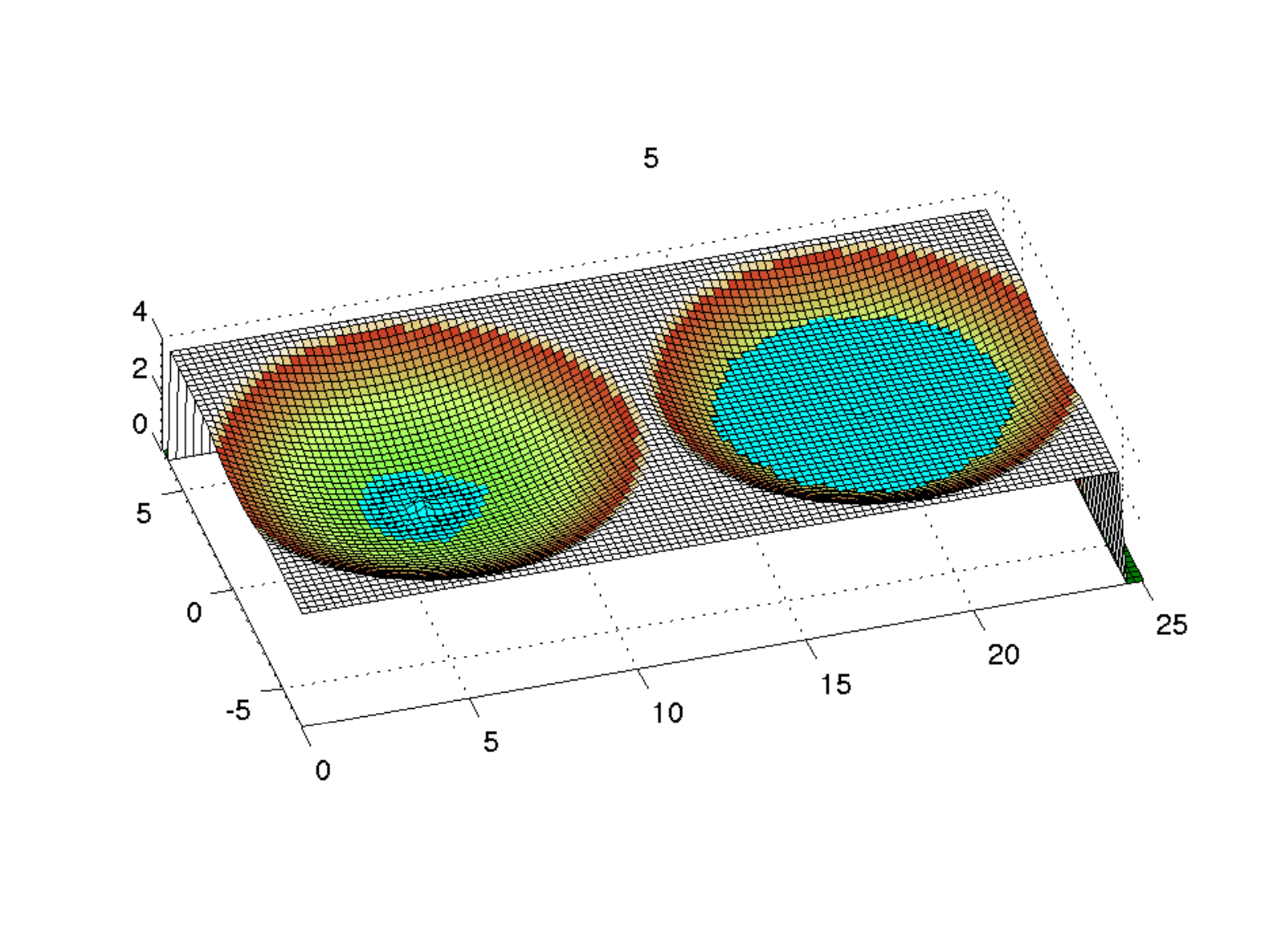}
\includegraphics[width=7cm]{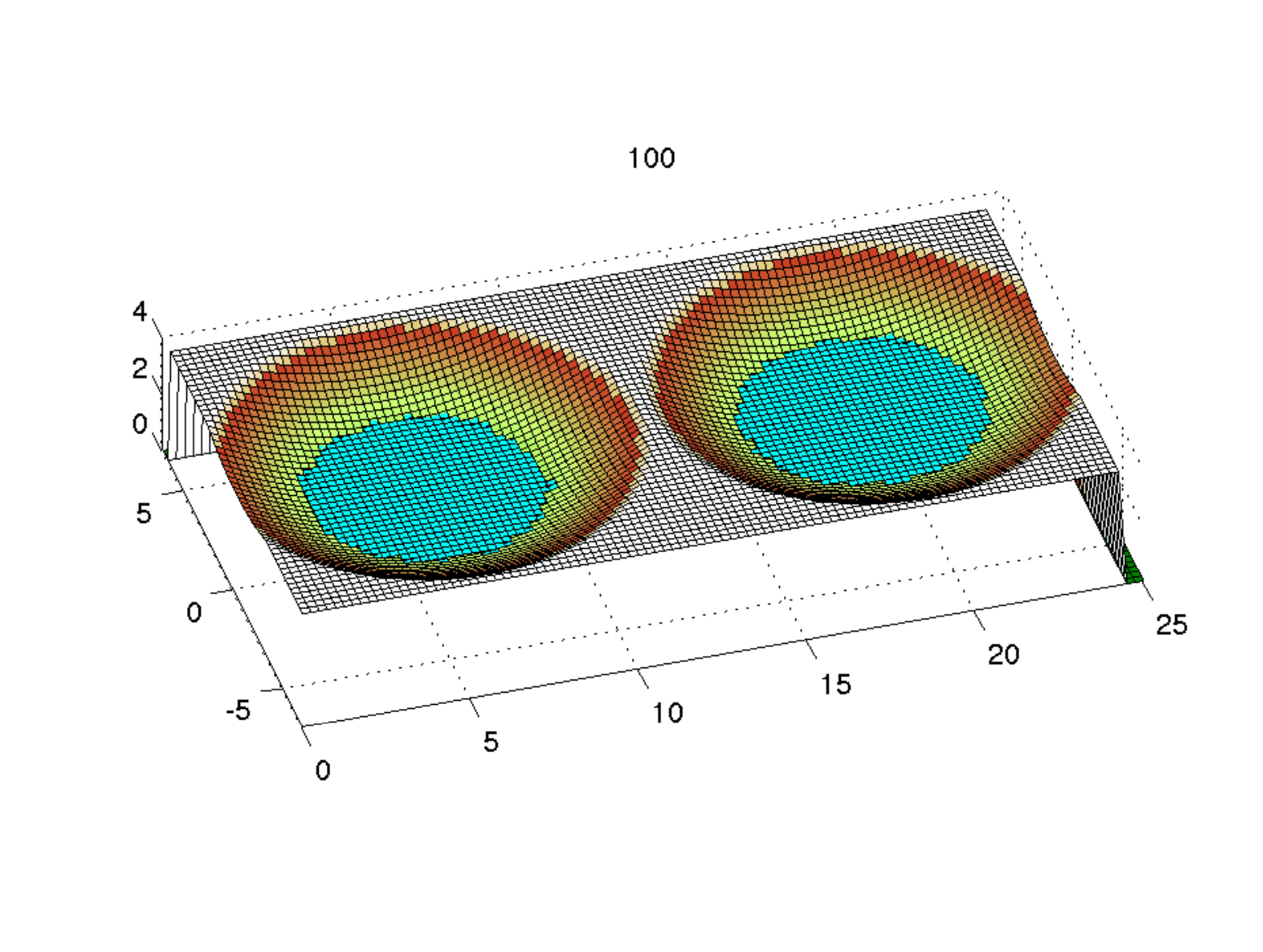}
\caption{Well balanced test: The water level on the surface at t=5 (left) and t=100 (right).
\label{Numerics:fig:SNI_WellBalanced_Surface}}
\end{figure}

At the beginning, the water in the right hollow flushes into the manhole.
This sudden inflow is transported by a shock wave through the conduit.
At about $t=4$ the shock reaches the left manhole and the water enters the surface, see figure \ref{Numerics:fig:SNI_WellBalanced_Surface}.
For the next few seconds  fast waves travel through the tube, which can be guessed by the small variations in the outflow of the left manhole in figure \ref{Numerics:fig:SNI_WellBalanced_Qext}. 
As these activities decay fast, a constant flow from the right to the left hollow is established.
This causes the water level at the feeding hollow to decrease.
Since the depth of water in the receiving hollow increases, the exchange is  slowing down.
The longer the computation runs the more both water levels and the hydraulic head in the sewer system approach a  constant balancing state.

In figure \ref{Numerics:fig:SNI_WellBalanced_Qext} the inflow into the manholes and the averaged water level on the surface are plotted.
At the first manhole (blue solid line) the inflow is at its maximum, since all the water fits into the tubes. 
This drain is abruptly reduced, when the sewer is filled.
Afterwards the smooth exchange is established.
The dynamics of the averaged water levels on the surface is more regular due to the smoothing of the averaging process.
But, as we can see in figure \ref{Numerics:fig:SNI_WellBalanced_Surface}, the surface is almost planar and only at the first outflow in the left hollow a small well can be observed.

{\bf Well-balancing}\\
{\noindent} 
With the identical configuration of surface and sewer we perform a second well balanced test by modifying the initial conditions.
Consider the sewer filled up to a height of $h_M=h_M=h(A_1)=3$ and on the surface the left hollow is filled up to $h(\vec{x})+z(\vec{x})=3$. 
In the right hollow we place a Riemann Problem consisting of a left part, which contains the inlet, with $h(\vec{x})+z(\vec{x})=3\ ,x_1<20$ and on the right the water level is slightly higher $h(\vec{x})+z(\vec{x})=3.1\ ,x_1\geq 20$.

\begin{figure}[htpb]
\includegraphics[width=6cm]{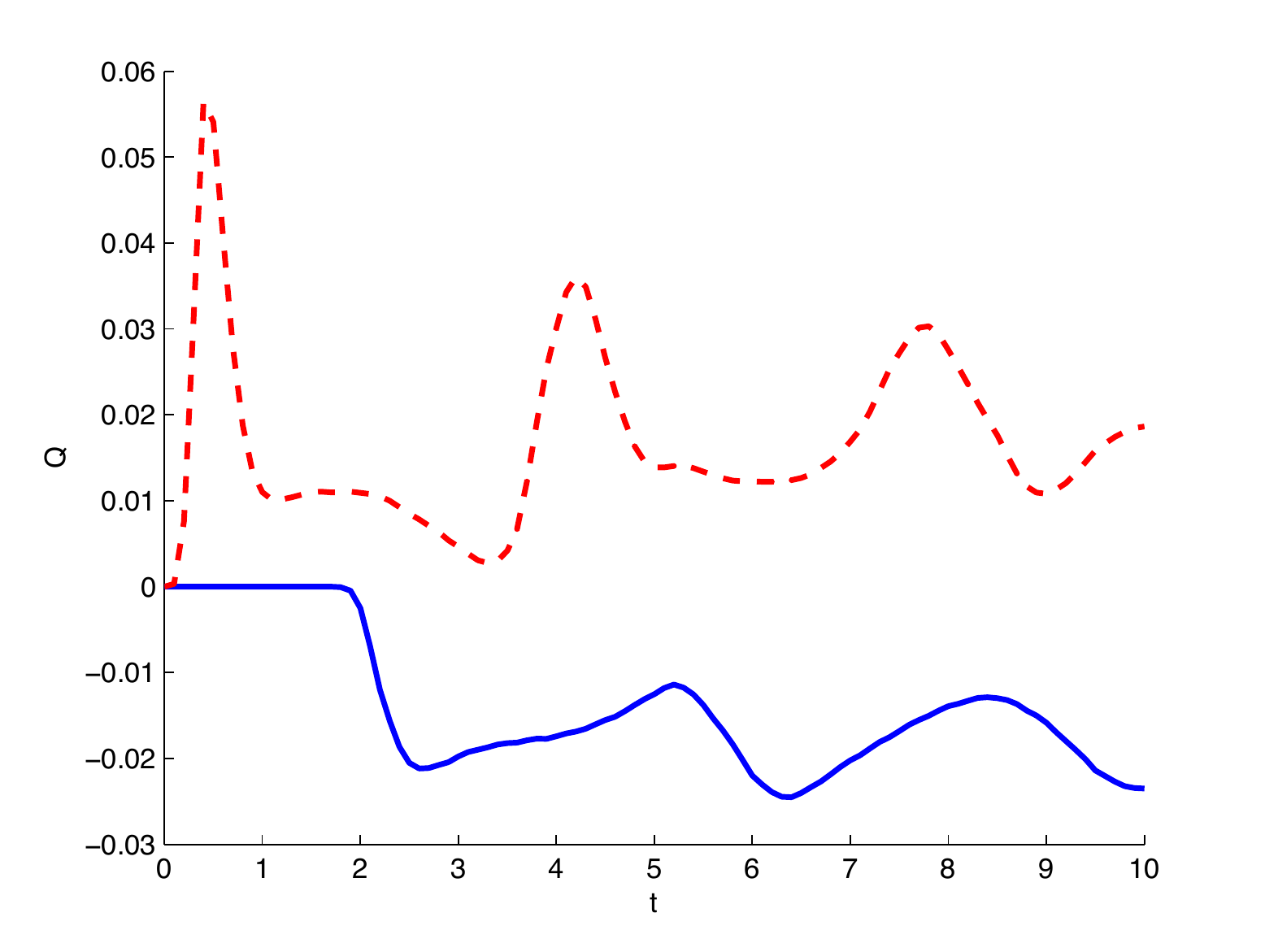}
\caption{The flow int the manholes on the left (solid blue) and on the right (dashed red).
\label{Numerics:fig:SNI_WellBalanced2}}
\end{figure}

From this Riemann Problem a shock wave passes the manhole and water is pressed into the sewer.
As shown in figure \ref{Numerics:fig:SNI_WellBalanced2}, the water level in the left hollow and the left manhole remains constant up to $t=2$ until the wave reaches the left side.
This wave keeps traveling between both ends, but with decreasing amplitude.
In the long run a constant water level in the whole system will establish.
      
\subsubsection{An interceptor sewer}
\begin{figure}[htpb]
\includegraphics[width=6cm]{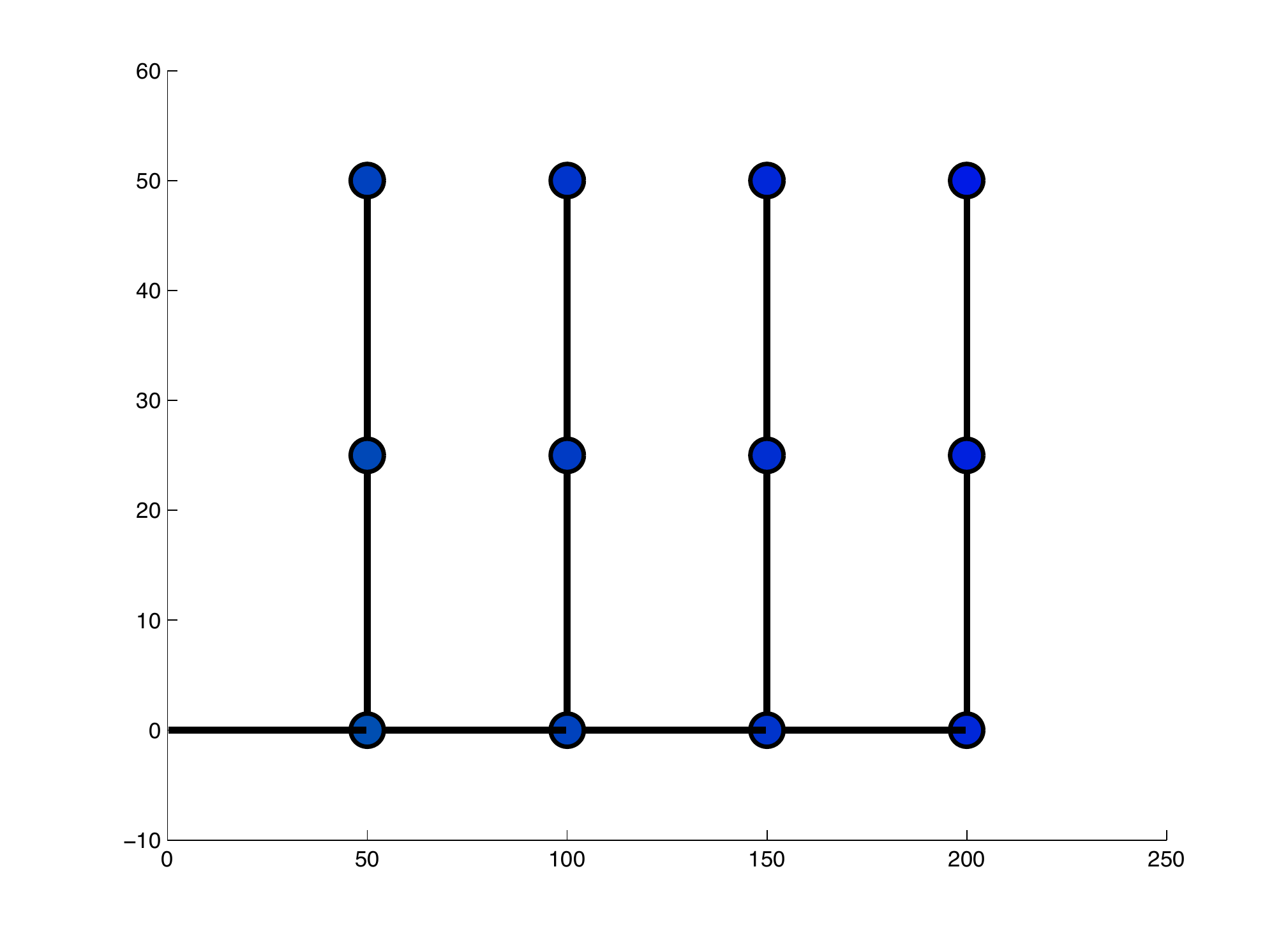}
\includegraphics[width=6cm]{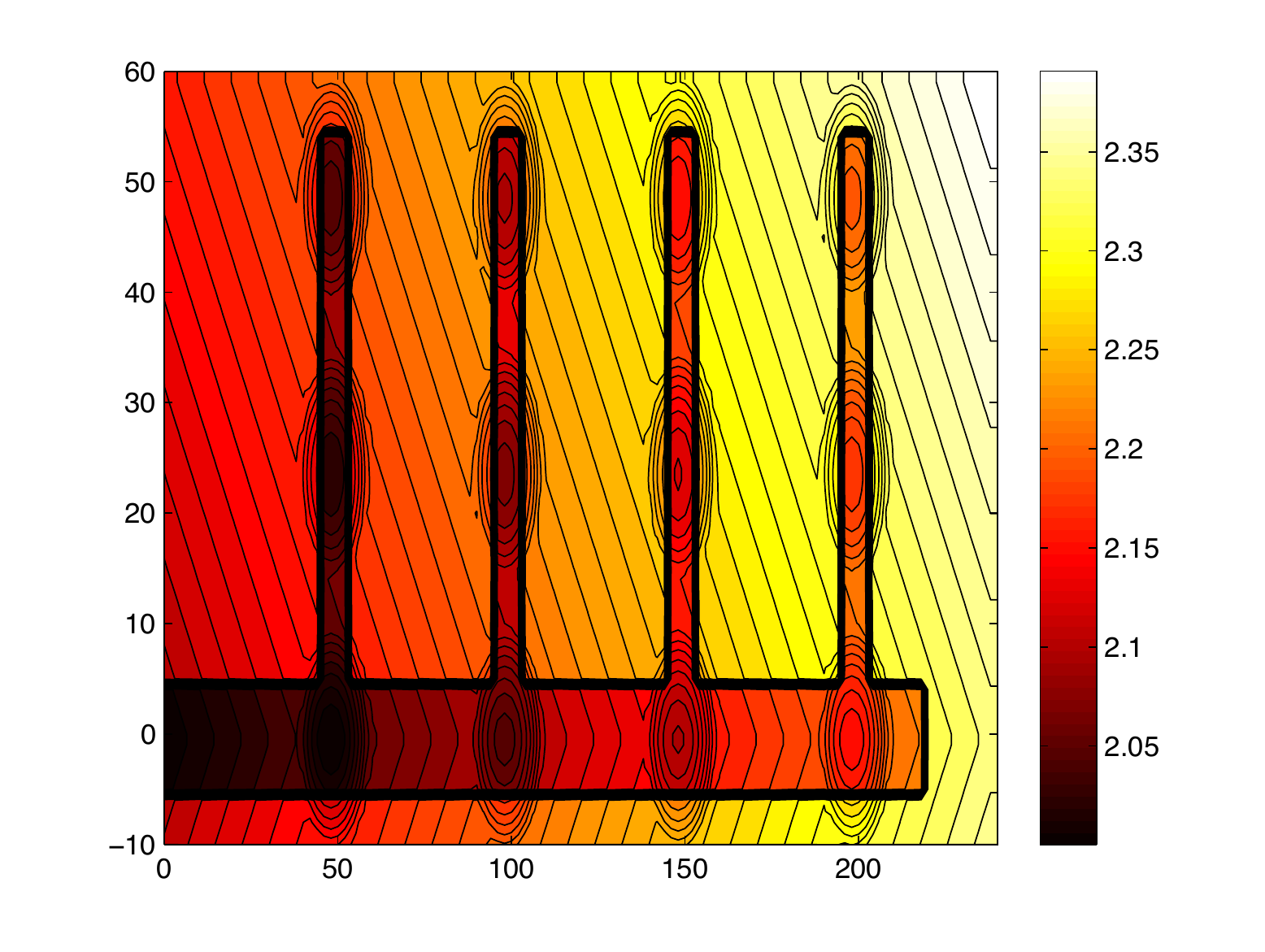}
\caption{The sewer network (left) and the surface elevation (right) of an interceptor sewer with four branches.
\label{Numerics:fig:Trunk_DarwSewer}}
\end{figure}

The next examples are two tests with a larger network as shown in figure \ref{Numerics:fig:Trunk_DarwSewer}.
As inflow of water we consider a short but intense rain falling on the surface.  

The sewer consists of four conduits forming an interceptor sewer and four branches of two tubes each.
At each junction a manhole is placed.
The conduits of the interceptor have a length of $50$, the tubes of the branches are $25$ long.
All tubes of the network have a constant slope, orientating at  $\partial_x z(x,y)=\partial_y z(x,y)=1.e^{-3}$.
The  surface is $2$ length units above the basis of the tubes, following the same slope.
On the surface there are streets, which are lowered by $0.1$ compared to the surrounding elevation.
For simplicity the accurate placement of inlets is omitted and the water is lead to the manholes by lowering the bottom up to $0.05$ circularly around the drop shafts.
The effective area of exchange has the size $A_{inlet}=0.5$, the manholes have a cross-sectional area of  $A_M=0.25\, \pi$.
The rain has the constant intensity $I=1.e-5$ and lasts from $t=0$ up to $t=1200$.
As boundary conditions on the surface non transmissive conditions are applied to the upper ends, while a free outflow condition is posed at the left end in $x$-direction.
This combination is chosen to guarantee no inflow of water along the boundaries and no damming at the lower part.
For the free end in the sewer system a free outflow is considered.
The tubes are discretized with $100$ points each, while the surface has a grid of $120 \times 70$ points.
The computation is done up to $T=2400$.

{\bf The overload}\\
{\noindent} 
In the first test case all tubes have a diameter of $0.2$.
They are initially filled up to a height of $h_i(A_i)+z_i=0.45$, with flow at rest $Q_i=0$, $i=1,\dots , 12$.

When the rain starts falling, water runs towards the manholes.
This constant drain into the sewer system exceeds the outflow at its free end, i.e. the water level in the network begins to rise.
At about $t=800$ the first manhole, at the entrance of the first branch, is completely filled.
The next manhole of the interceptor, as well as both ones of the first branch, follow immediately.
\begin{figure}[htpb]
\includegraphics[width=6cm]{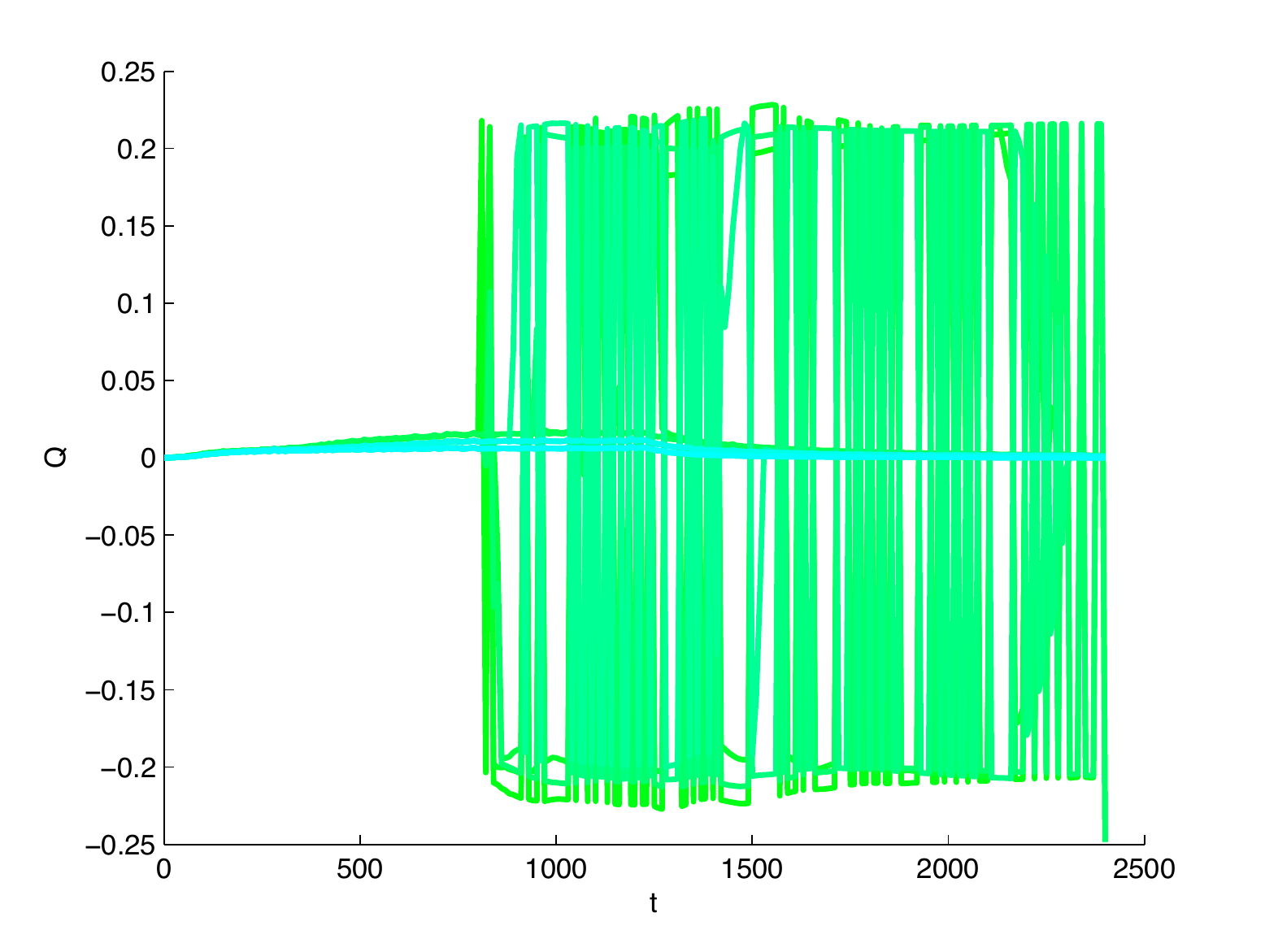}
\includegraphics[width=6cm]{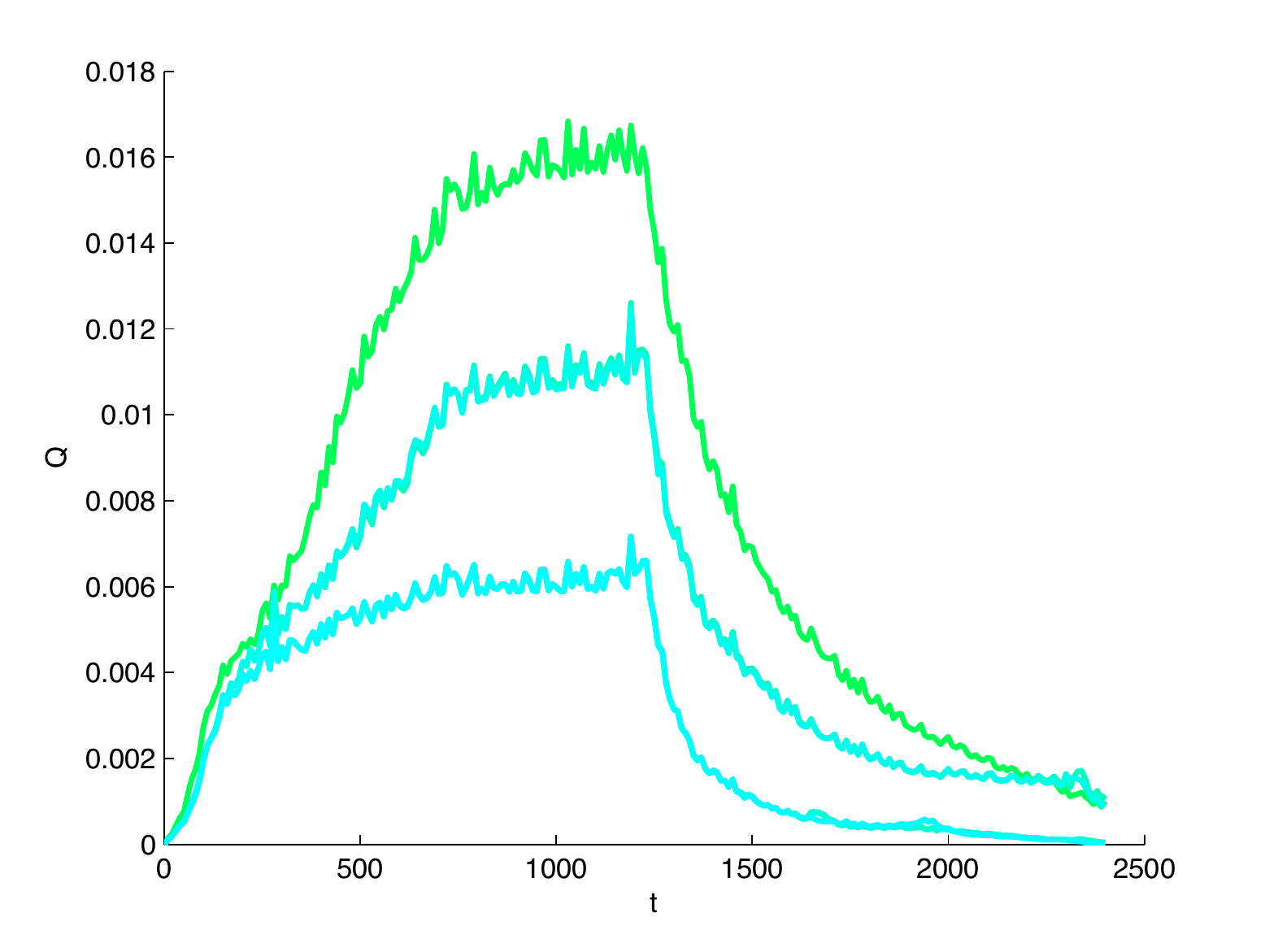}
\caption{The flow from the surface to the manholes: all manholes (left) and only the non-surcharging ones (right).
\label{Numerics:fig:Trunk_FlowManholes}}
\end{figure}
In the left picture of figure \ref{Numerics:fig:Trunk_FlowManholes} the inflow of all $12$ manholes is shown, on the right only the not surcharging ones are depicted.
The smallest inflow occurs at the manholes at the heads of the branches (blue), since only the direct surroundings serve as catchment areas.
For the manholes in the middle of the branches (blue/green) the streets collect some additional water. 
The largest inflow have the inlets of the main street (green), above the interceptor sewer. 
Here, all the rain, which missed the previous entrances, is accumulated.
Through the manhole at the end of the street a bit less water passes, since there is no further street following.
The oscillations observed on the left picture are caused by small waves in the sewer and on the surface, which force the coupling to switch between in- and outflow. 
The changing appears to be very abrupt, which is mainly due to  the output-discretization of the data.
\begin{figure}[htpb]
\includegraphics[width=7cm]{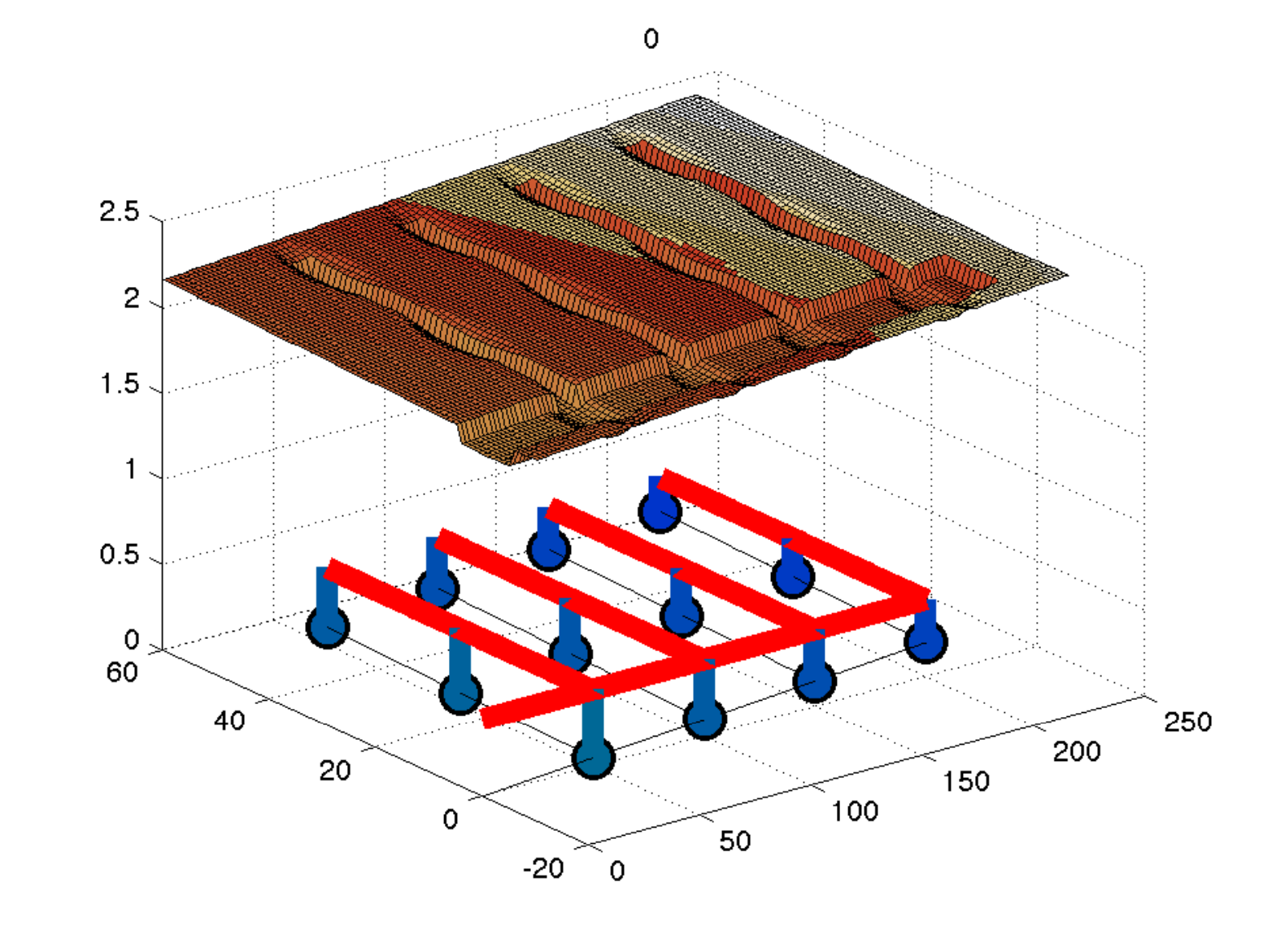}
\includegraphics[width=7cm]{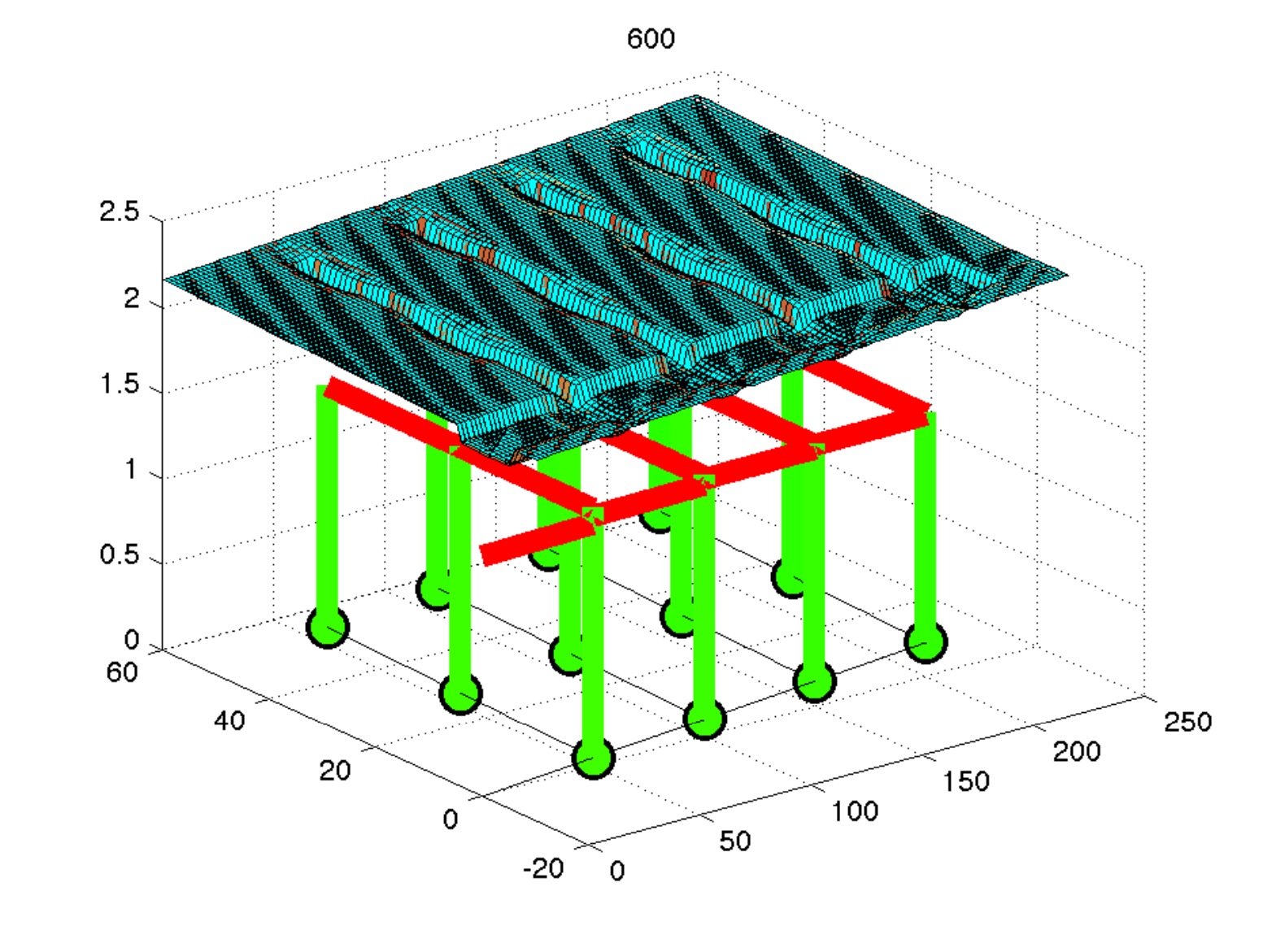}\\
\includegraphics[width=7cm]{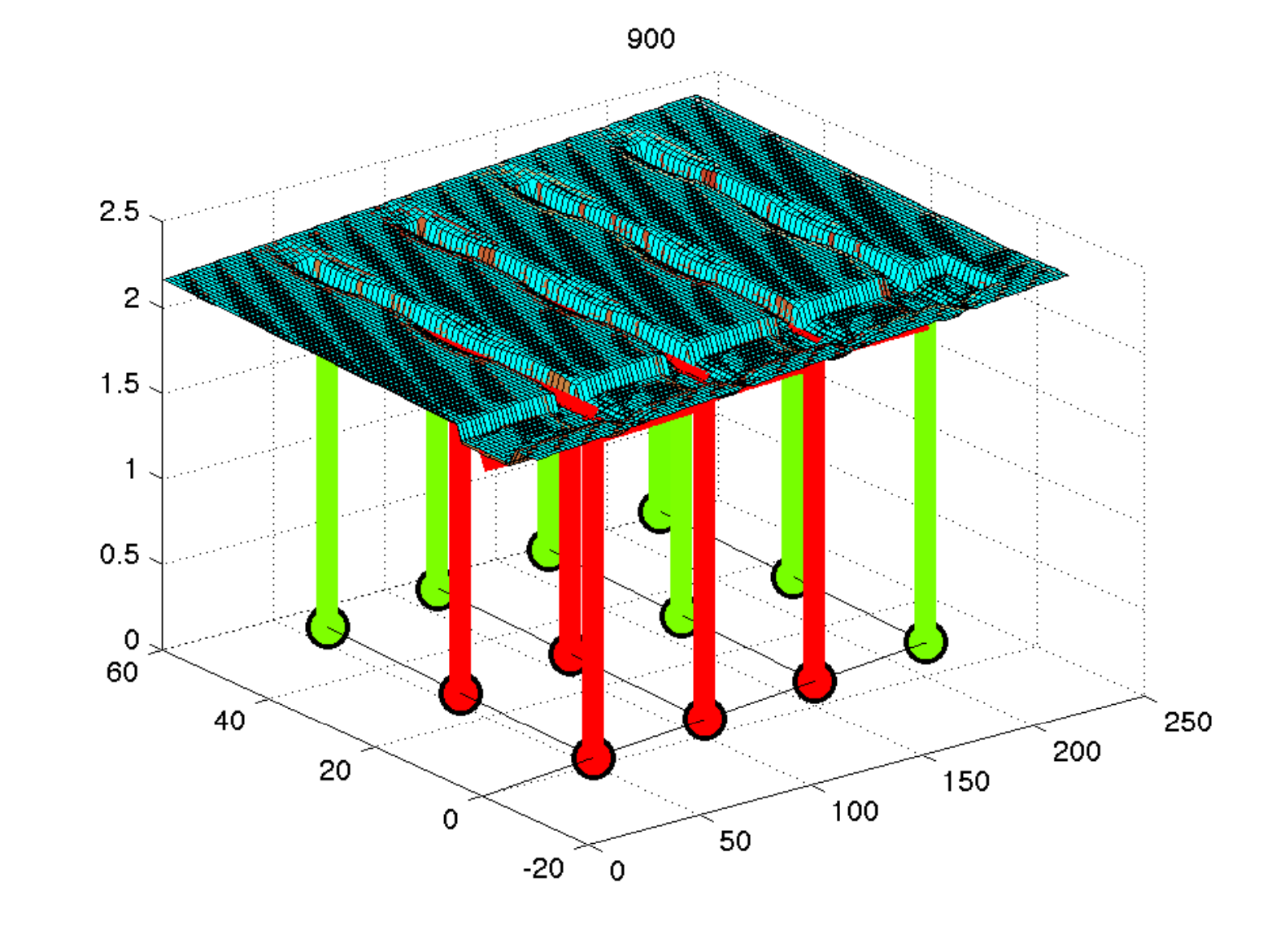}
\includegraphics[width=7cm]{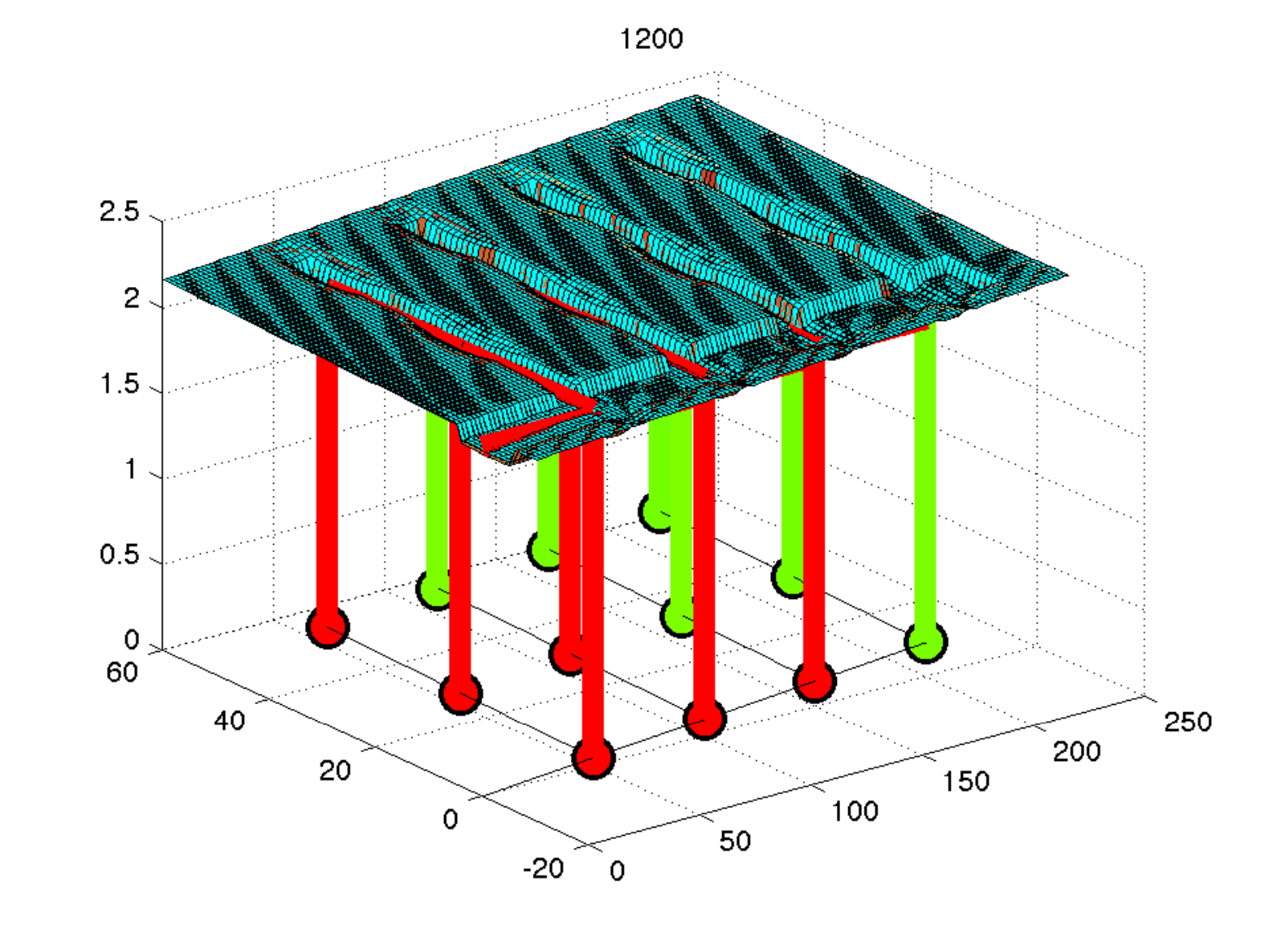}\\
\includegraphics[width=7cm]{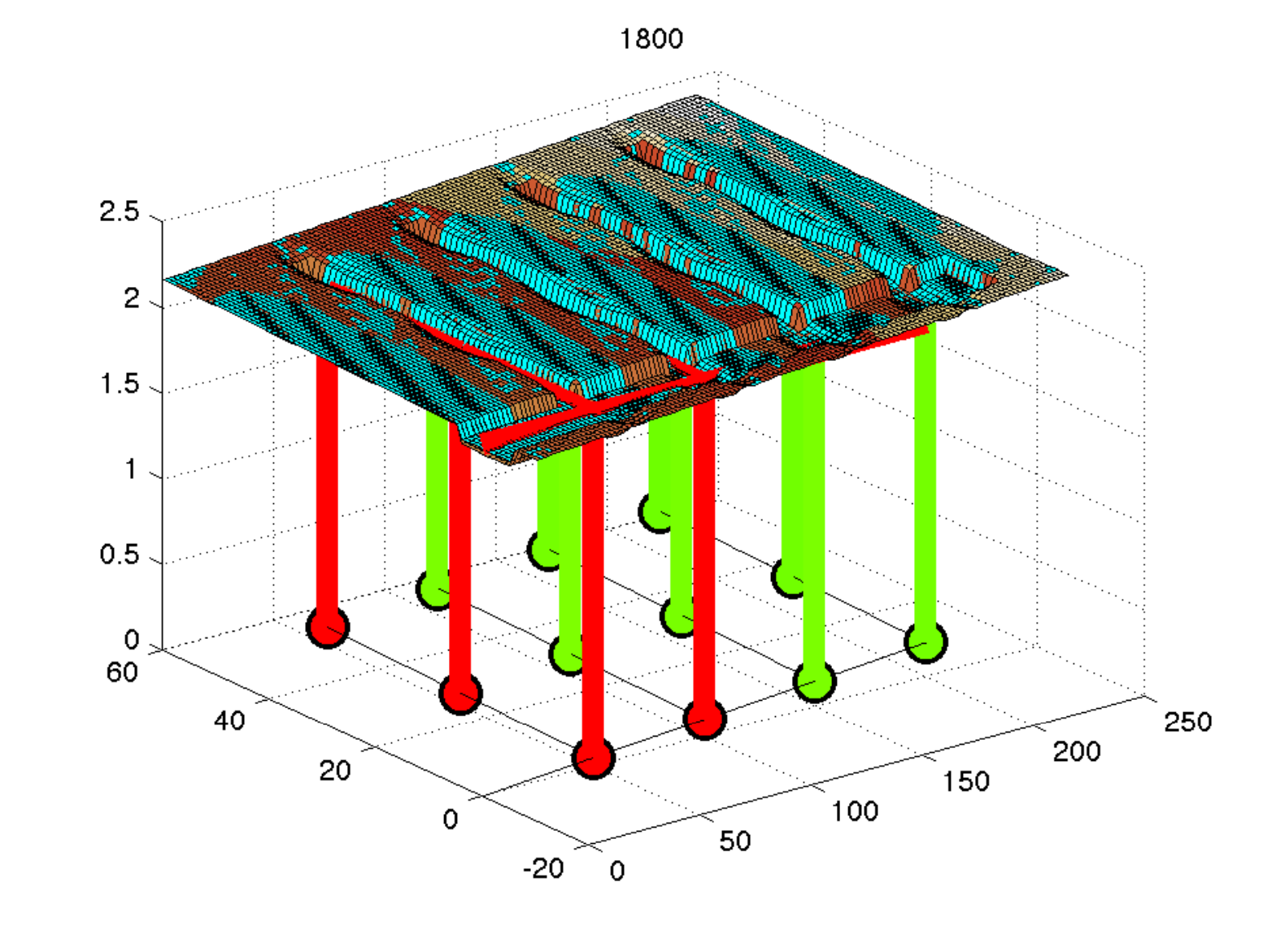}
\includegraphics[width=7cm]{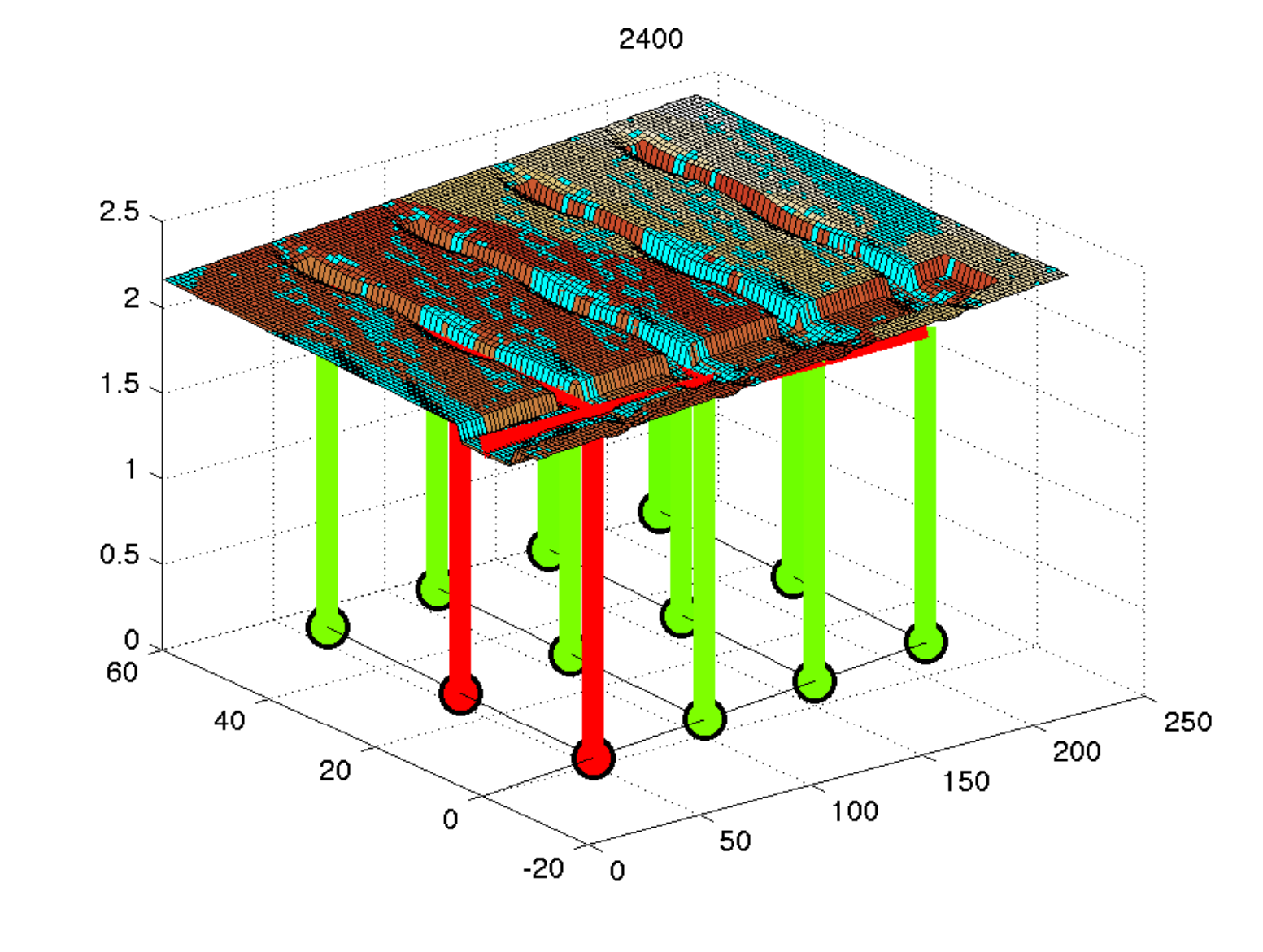}
\caption{The full system at the times $t=0\,,600\,,900\,,1200\,,1800\,,2400$.
\label{Numerics:fig:Trunk_System}}
\end{figure}
When the rains stops, the inflow at the branches is decaying.
The manholes of the interceptor sewer still suffer the load of the sewer system.
Here, the decay is significantly delayed in comparison to the end of the rainfall.

Similar observations can be made for the surface flow.
In figure \ref{Numerics:fig:Trunk_System} the coupled system at the times $t=0\,,600\,,900\,,1200\,,1800\,,2400$ is shown.
The colors in the network symbolize the percentage of the actual filling, where red indicates pressurized flow in the conduits and surcharge of the manholes.
At the beginning the surface is dry.
The rain immediately covers the whole surface with a thin film of water.
At a sufficient depth of surface water, a flow following the bottom topography establishes and the water is conducted into the sewer system.
Shortly after the rainfall stops and the faraway areas begin to dry.
In the streets water still remains and at the surcharged manholes water is even added to the surface.
The longer the computation runs, the more parts dry out, until only at the head of the main street water is left on the surface.

\begin{figure}[htpb]
\includegraphics[width=6cm]{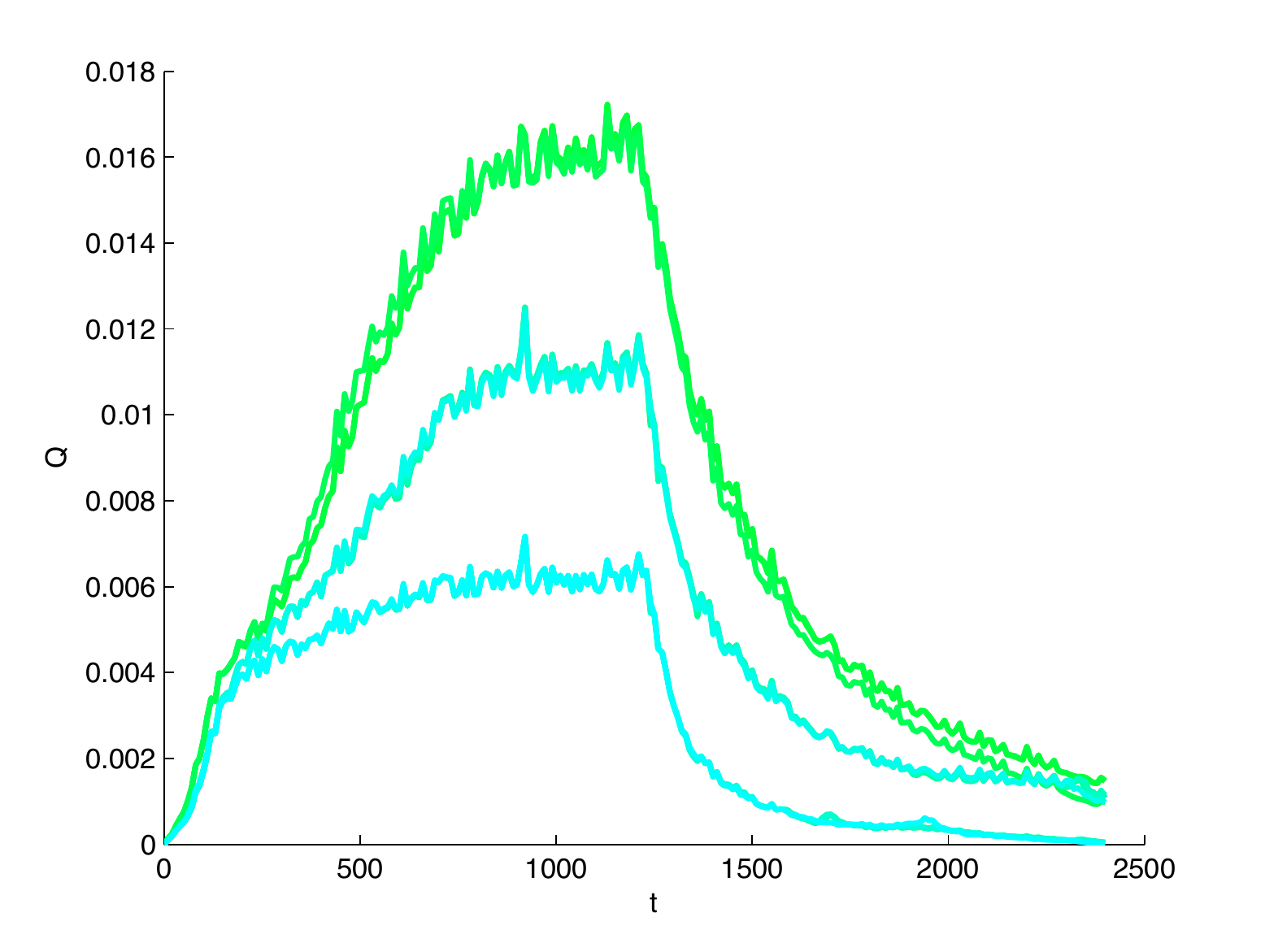}
\includegraphics[width=6cm]{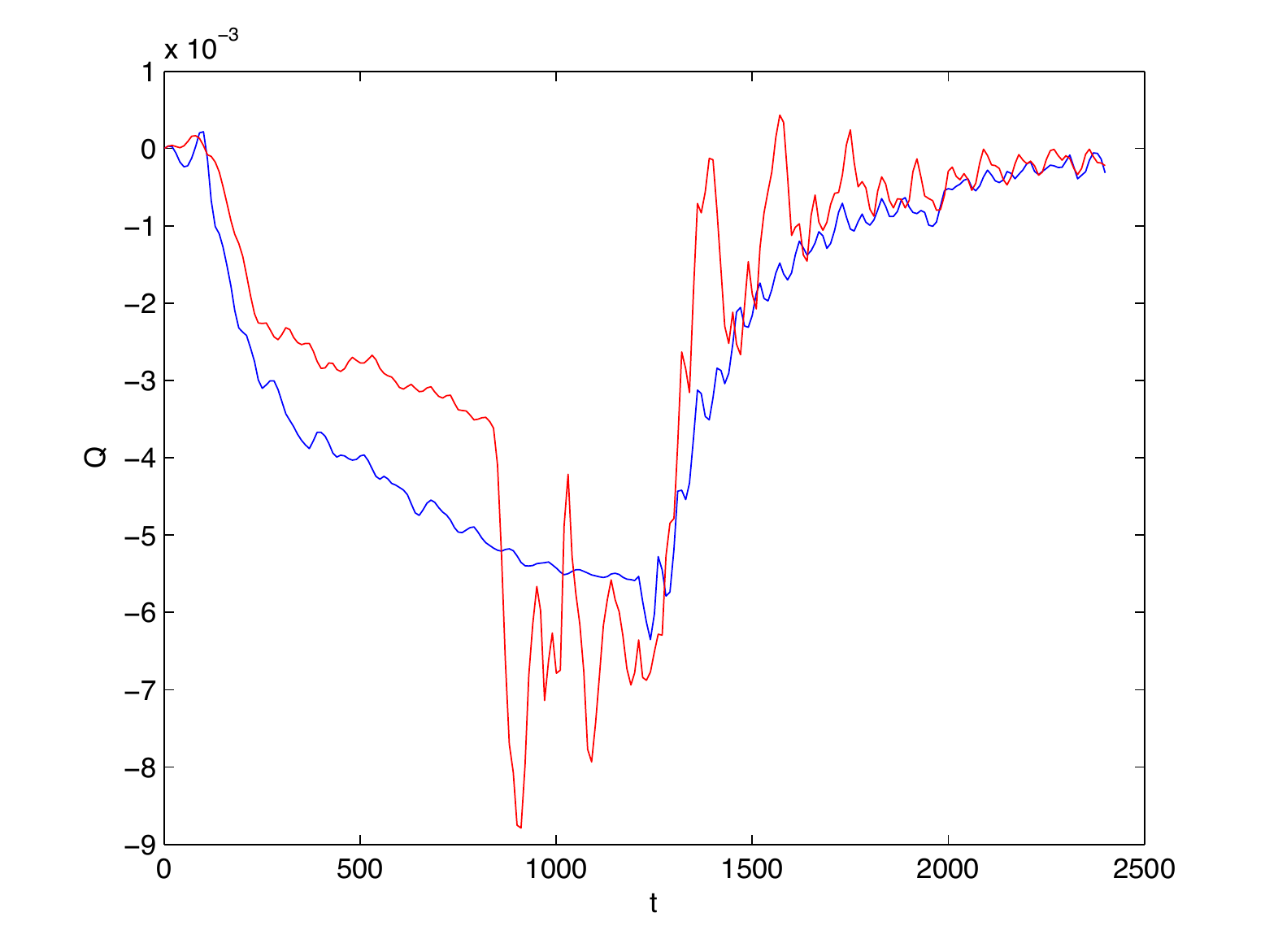}
\caption{The flow from the surface to the sewer system for all manholes (left). The comparison of the outflow from the free end in sewer sewer system (right): In case of the overload (red) and with the enlarged interceptor sewer (blue).
\label{Numerics:fig:Trunk_CompareOutflow}}
\end{figure}

\begin{figure}[htpb]
\includegraphics[width=7cm]{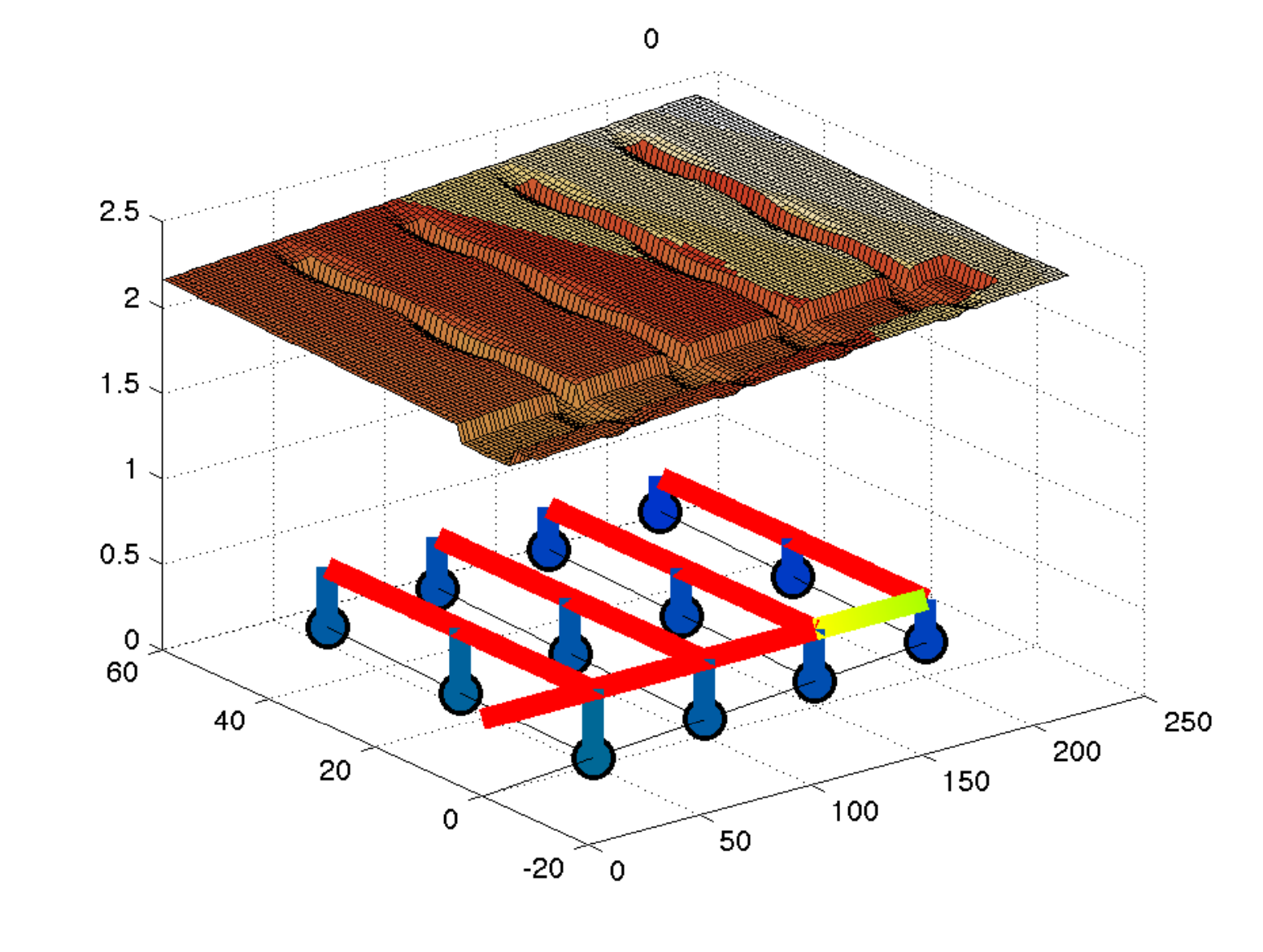}
\includegraphics[width=7cm]{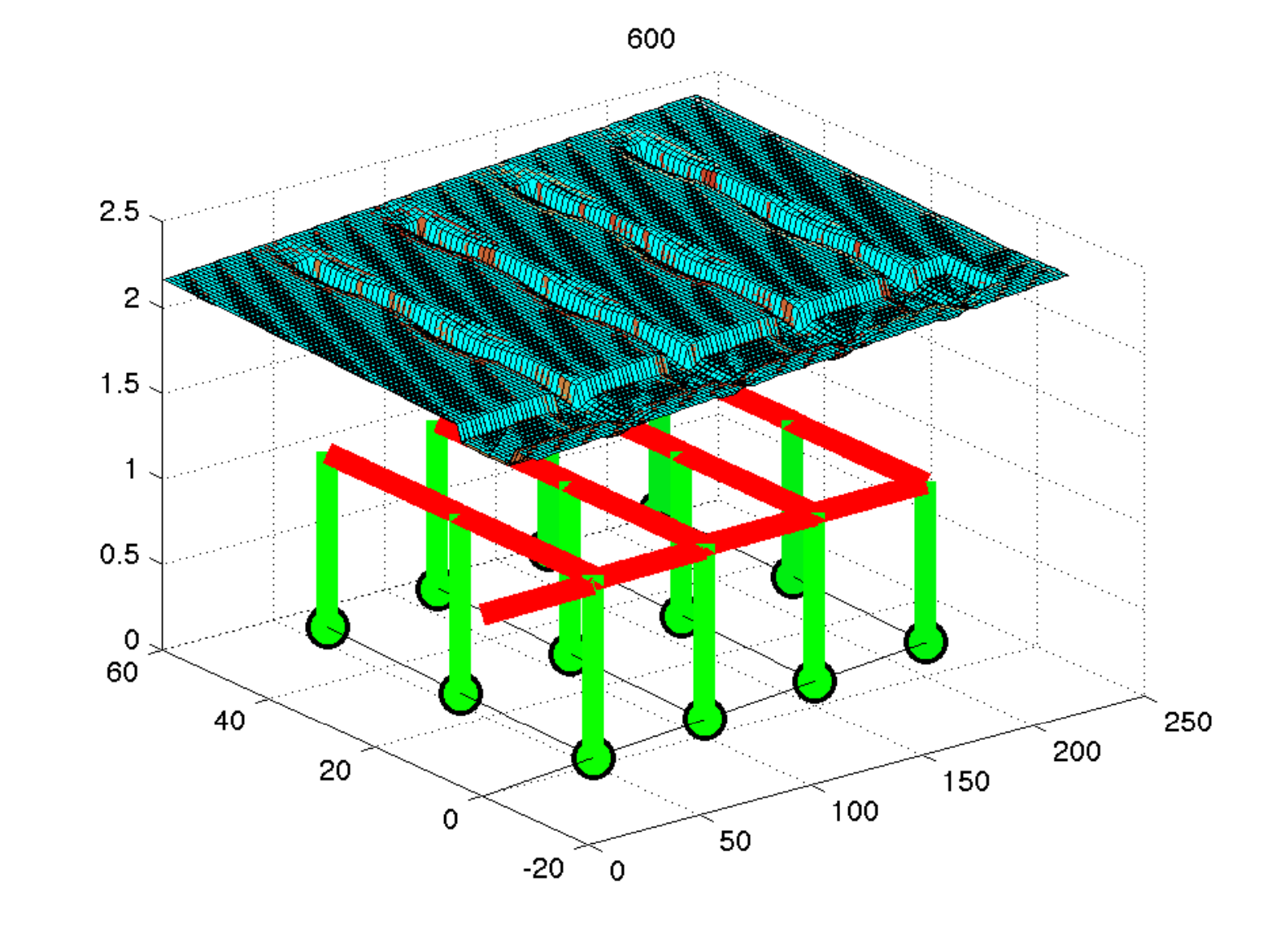}\\
\includegraphics[width=7cm]{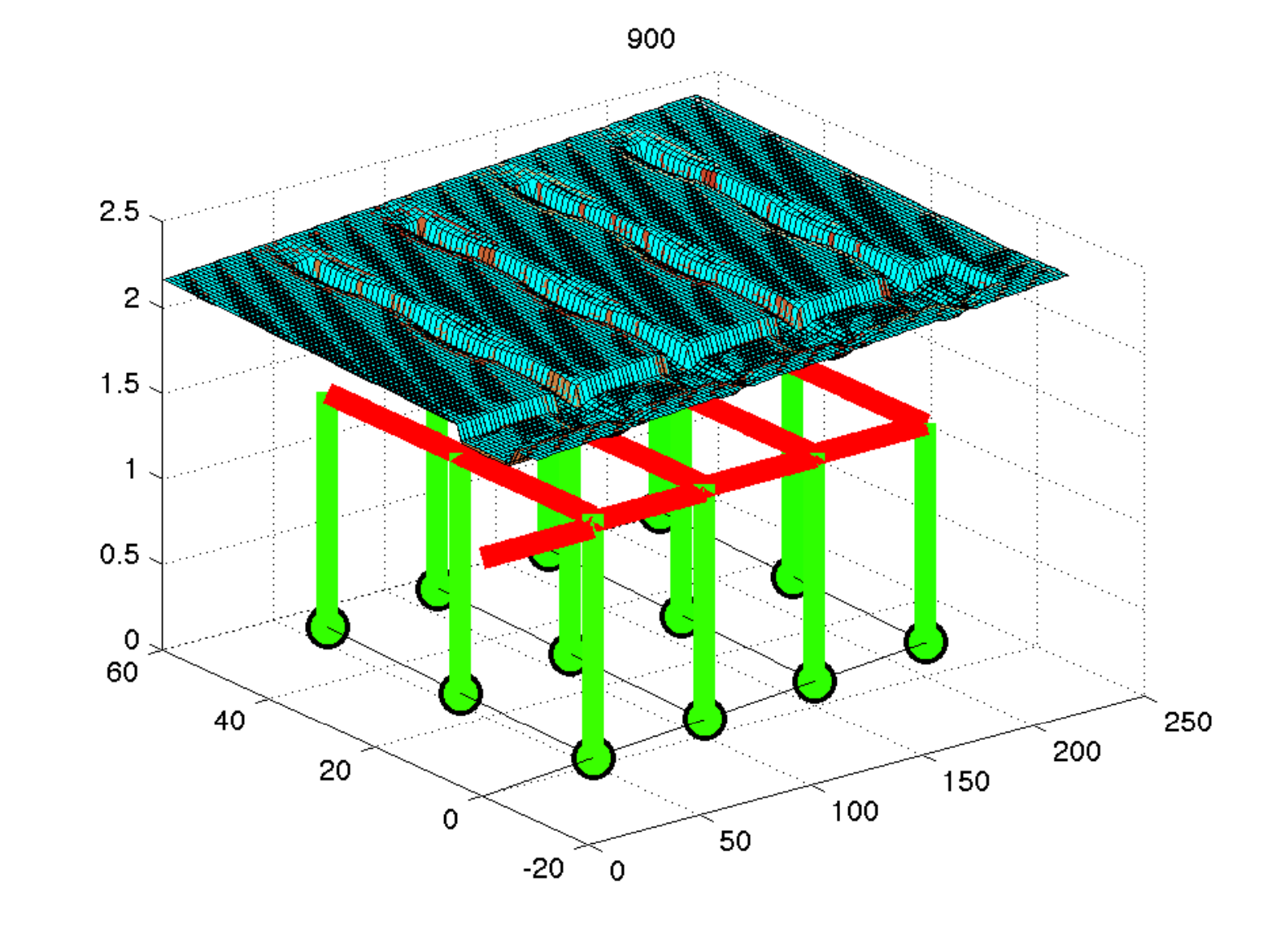}
\includegraphics[width=7cm]{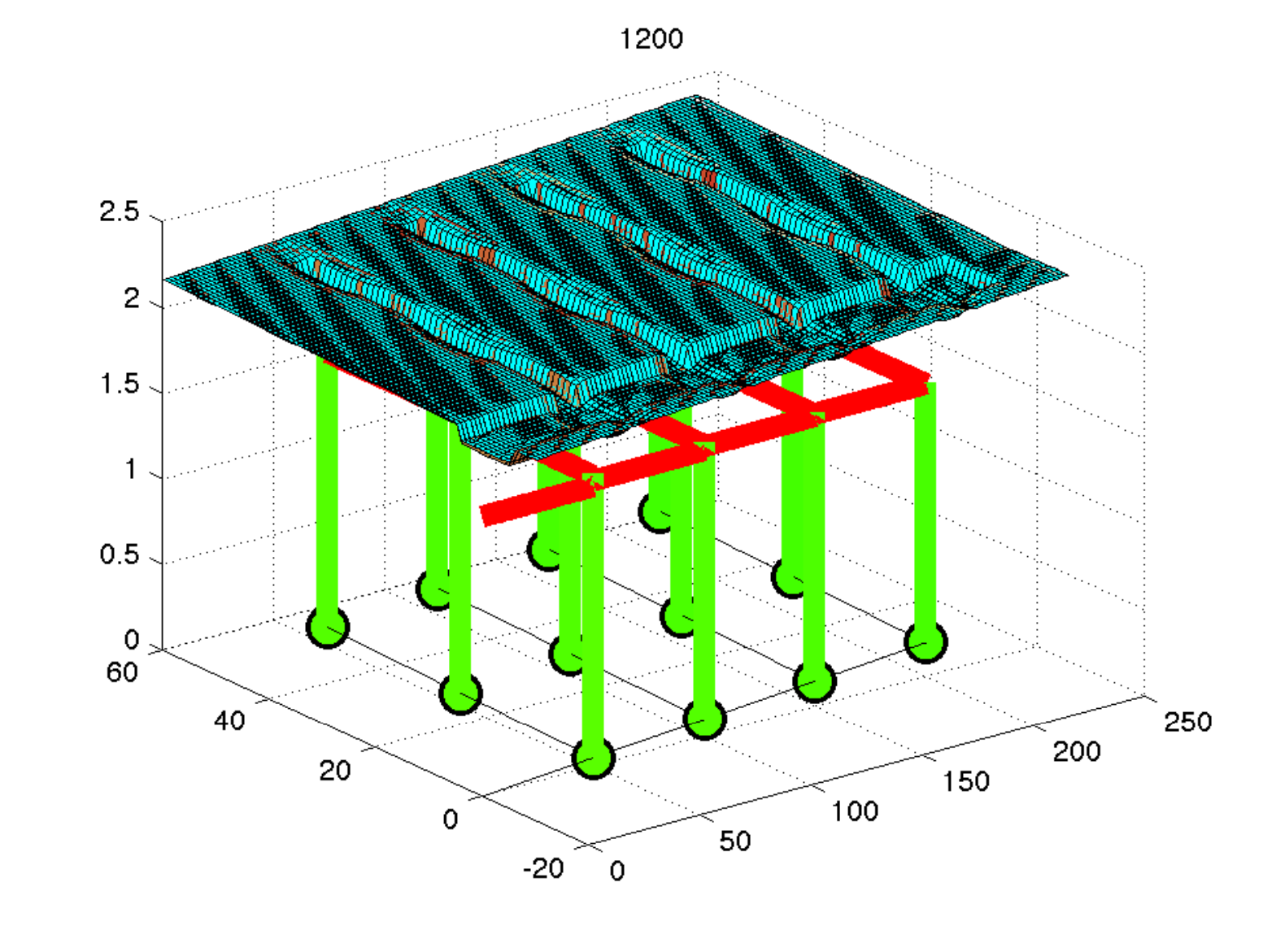}\\
\includegraphics[width=7cm]{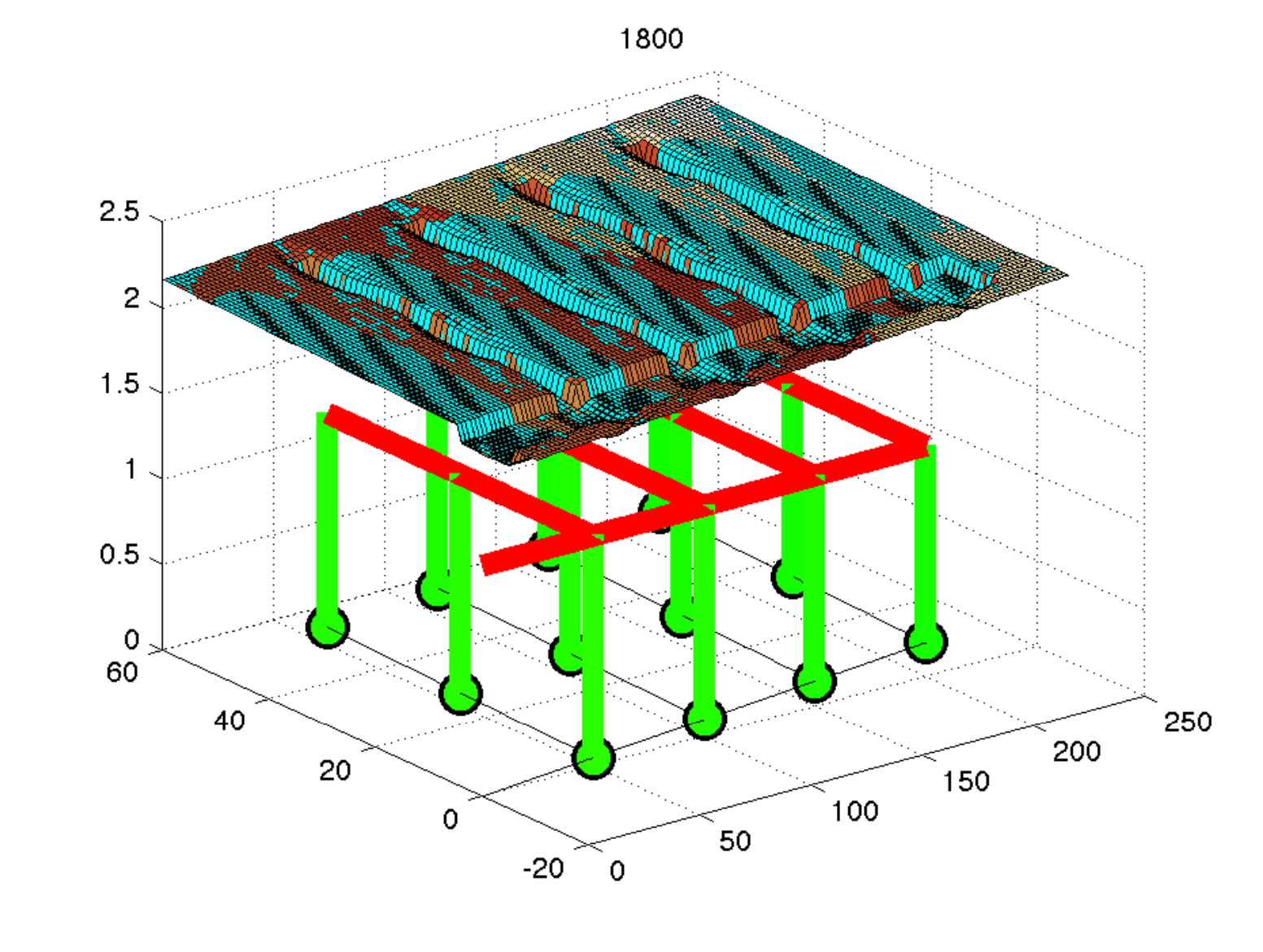}
\includegraphics[width=7cm]{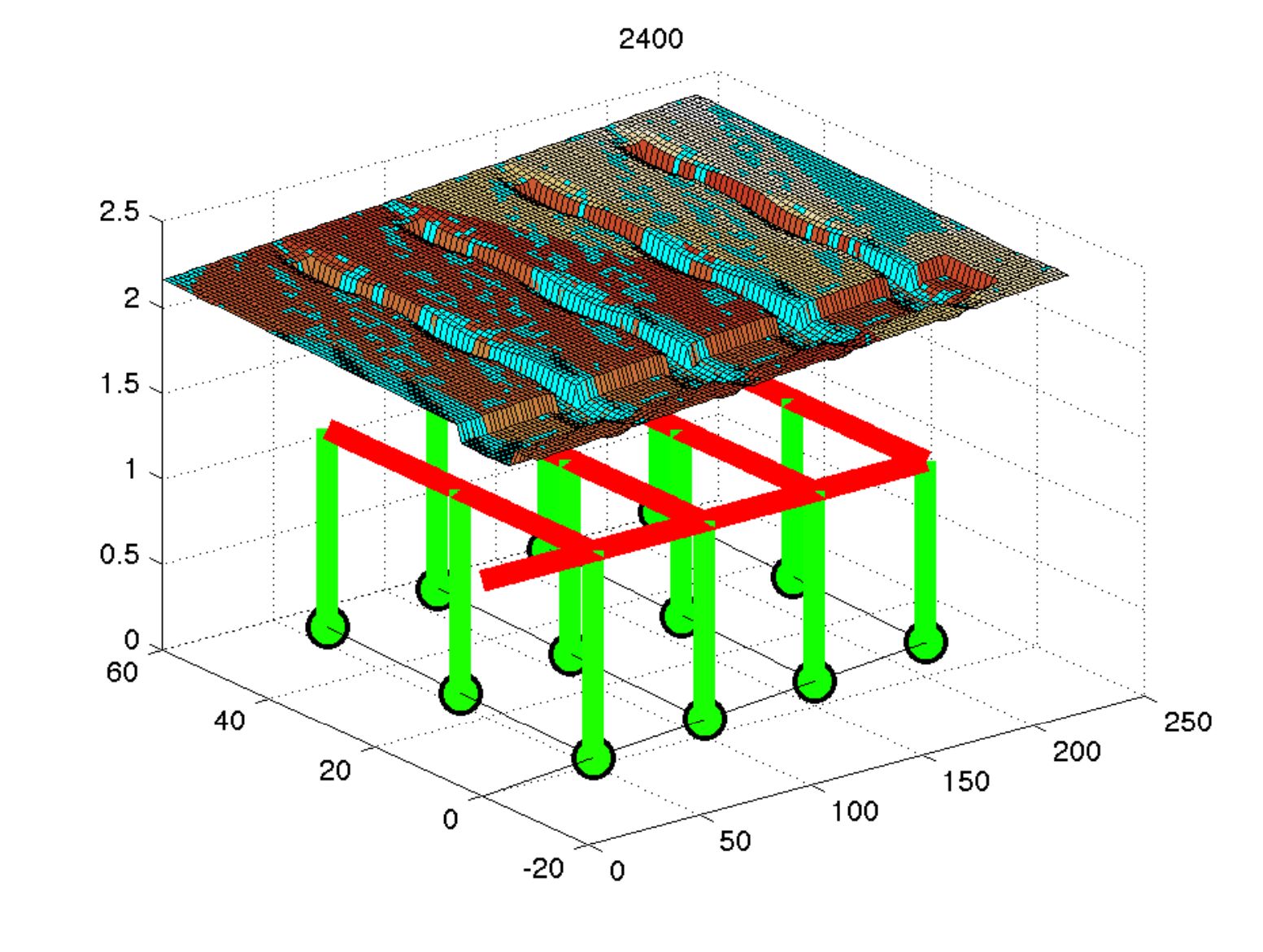}
\caption{The full system at the times $t=0\,,600\,,900\,,1200\,,1800\,,2400$.
\label{Numerics:fig:Trunk2_System}}
\end{figure}

{\bf An enlarged interceptor sewer}\\
{\noindent} 
In the second test case we consider the same scenario, but the interceptor sewer is enlarged to a diameter of $0.3$.
The intensity of the rain and thus the amount of added water is the same as in the previous example.
The total amount of water considered in the full system only differs by the additional water, which is needed to fill the larger tubes up to the initial water level.
This implies an almost identical behavior at the beginning of the computation.
The dynamics on the surface are not influenced by the sewers at this point and so the same inflow establishes, figure \ref{Numerics:fig:Trunk_CompareOutflow} (left).
The water level in the network rises due to the incoming water.
As shown in figure \ref{Numerics:fig:Trunk_CompareOutflow} (right), the outflow at the free end of the sewer system is slightly larger than in the previous configuration.
Thus, more water is transported out of the system and the water level does not reach the previous heights.
This avoids any surcharging of the manholes, as shown in figure \ref{Numerics:fig:Trunk_CompareOutflow} (left).
In figure \ref{Numerics:fig:Trunk2_System} we can observe that the water level in the network with enlarged interceptor returns faster to a normal state.
In both scenarios the states on the surface differ only next to the manholes, as the surcharging in the first test case is only moderate.
From figure \ref{Numerics:fig:Trunk_CompareOutflow} (right) we note that the surcharging in the first test case increased the output immediately, but also causes strong oscillations.
If no manholes surcharge the flow behaves much more regular.
While the outflow reduces, when the rainfall stops, the oscillations only decay as more and more manholes return into their normal states.
 
{\small

\bibliographystyle{siam}\bibliography{StormSewer}
}

\end{document}